\documentclass[twocolumn]{aastex7}
\usepackage{hyperref}
\usepackage{changepage}
\hypersetup{
    colorlinks=true,
    linkcolor=blue,
    filecolor=magenta,      
    urlcolor=cyan,
}

\usepackage{xspace}
\usepackage{float}
\usepackage{ucs}
\usepackage[version=4]{mhchem}
\usepackage{csvsimple}

\usepackage[caption=false]{subfig}
\usepackage[T1]{fontenc}
\newcommand{\ms}{\ensuremath{\mathrm{m}\,\mathrm{s}^{-1}}\xspace}

\newcommand{\Rsun}{\ensuremath{R_{\odot}}\xspace }
\newcommand{\Msun}{\ensuremath{M_{\odot}}\xspace}

\newcommand{\MEarth}{\ensuremath{M_{\oplus}}\xspace}
\newcommand{\REarth}{\ensuremath{R_{\oplus}}\xspace}

\defcitealias{Kuzuhara2024}{K24}
\defcitealias{Dholakia2024}{D24}

\begin{document}

\title{An Earth-like Density for the Temperate Earth-sized Planet GJ\,12\,b}

\correspondingauthor{Madison Brady}
\email{mtbrady@uchicago.edu}

\author[0000-0003-2404-2427]{Madison Brady}\email{mtbrady@uchicago.edu}
\affiliation{Department of Astronomy \& Astrophysics, University of Chicago, Chicago, IL 60637, USA}

\author[0000-0003-4733-6532]{Jacob L.\ Bean} \email{jacobbean@uchicago.edu}
\affiliation{Department of Astronomy \& Astrophysics, University of Chicago, Chicago, IL 60637, USA}

\author[0000-0003-4508-2436]{Ritvik Basant} \email{rbasant@uchicago.edu}
\affiliation{Department of Astronomy \& Astrophysics, University of Chicago, Chicago, IL 60637, USA}

\author[0009-0003-1142-292X]{Nina Brown} \email{ninabrown@uchicago.edu}
\affiliation{Department of Astronomy \& Astrophysics, University of Chicago, Chicago, IL 60637, USA}

\author[0009-0005-1486-8374]{Tanya Das} \email{tanyadas@uchicago.edu}
\affiliation{Department of Astronomy \& Astrophysics, University of Chicago, Chicago, IL 60637, USA}

\author[0000-0001-8236-5553]{Matthew Nixon} \email{mcnixon@umd.edu}
\affiliation{Department of Astronomy, University of Maryland, College Park, MD 20742, USA}

\author[0000-0002-4671-2957]{Rafael Luque} \email{rluque@uchicago.edu}
\affiliation{Department of Astronomy \& Astrophysics, University of Chicago, Chicago, IL 60637, USA}
\affiliation{NHFP Sagan Fellow}

\author[0000-0002-2875-917X]{Caroline Piaulet-Ghorayeb}\email{carolinepiaulet@uchicago.edu}
\affiliation{E. Margaret Burbridge Postdoctoral Fellow}
\affiliation{Department of Astronomy \& Astrophysics, University of Chicago, Chicago, IL 60637, USA}

\author[0000-0002-3328-1203]{Michael Radica} \email{radicamc@uchicago.edu}
\affiliation{Department of Astronomy \& Astrophysics, University of Chicago, Chicago, IL 60637, USA}
\affiliation{NSERC Postdoctoral Fellow}

\author[0000-0003-4526-3747]{Andreas Seifahrt} \email{andreas.seifahrt@noirlab.edu}
\affiliation{Gemini Observatory/NSF NOIRLab, 670 N. A'ohoku Place, Hilo, HI 96720, USA}

\author[0000-0002-4410-4712]{Julian St\"urmer} \email{stuermer@lsw.uni-heidelberg.de}
\affiliation{Landessternwarte, Zentrum f\"ur Astronomie der Universit\"at Heidelberg, K\"onigstuhl 12, D-69117 Heidelberg, Germany}

\author[0000-0002-3852-3590]{Lily Zhao}\email{lilylingzhao@uchicago.edu}
\affiliation{Department of Astronomy \& Astrophysics, University of Chicago, Chicago, IL 60637, USA}

%


\begin{abstract}

While \textit{JWST} has provided us with the opportunity to probe the atmospheres of potentially-habitable planets, observations of the TRAPPIST-1 system have shown us that active stars severely complicate efforts at studying their planets.  GJ\,12\,b is a newly-discovered temperate ($T_{\rm eq}\,\approx\,300$\,K), Earth-sized ($R_p\,=\,0.96\,\pm\,0.05$\,\REarth) planet orbiting an inactive M dwarf that might be a good alternate to the TRAPPIST-1 planets for atmospheric characterization.  In this paper, we use MAROON-X radial velocities to measure a mass of $0.71\,\pm\,0.12$\,\MEarth for GJ\,12\,b.  We also find moderate evidence that the planet has an eccentric ($e\,\approx\,0.16$) orbit.  GJ\,12\,b's mass results in a planetary density comparable to or less dense than Earth, possibly indicating the presence of water or a low bulk iron mass fraction. With its low mass, GJ\,12\,b is likely within reach of \textit{JWST} transmission spectroscopy observations, making it an excellent target for determining the location of the cosmic shoreline.  Its low mass may mean that the planet could have trouble retaining its primary atmosphere during the star's active pre-main-sequence phase.  However, if it has a heightened eccentricity, it may be able to sustain a secondary atmosphere through tidally-induced volcanism.

\end{abstract}


\section{Introduction}
\label{sec:intro}

The recent launch of \textit{JWST} has allowed for transformational developments in our understanding of exoplanet atmospheres \citep{kempton24}.  However, we have yet to answer a key question: which exoplanets are habitable?  The search for biosignatures on exoplanets is one of the high-priority research areas identified in the Astronomy 2020 decadal survey, but there is still much work to be done to pave the way for a habitable exoplanet discovery \citep{seager25}.

However, we may need to exercise caution when it comes to selecting targets for habitable planet searches.  While M dwarf exoplanets produce the largest-amplitude photometric signals (and thus are the best targets for time-intensive efforts at characterizing atmospheres), we have yet to find any definitive atmospheric signatures of rocky M dwarf planets with \textit{JWST}. 

Theoretically, atmospheres may be rarer on M dwarf planets.  Studies of photoevaporation and stellar evolution indicate that many transiting rocky M dwarf planets can lose their atmospheres due to X-ray/UV (XUV) activity from their host stars.  \cite{Zahnle2017} and \cite{Pass2025} show that, if we extrapolate the location of the cosmic shoreline (which is the delineation in planetary escape velocity versus lifetime incident XUV radiation space between planets with and without atmospheres) from solar system objects, the vast majority of currently-detected exoplanets are unlikely to possess atmospheres.  If this is true, many of the commonly-accepted targets for future biosignature searches (such as the habitable zone planets in the TRAPPIST-1 system) may be completely airless, making it necessary to reconsider how we prioritize targets for these \citep[potentially extremely time-intensive, see, e.g.,][]{Meadows2023} surveys.  Recent \textit{JWST} observations of several rocky M dwarf planets have struggled to find atmospheres, in apparent agreement with this theoretical expectation \citep{Greene2023, Zieba2023, Lustig-Yaeger2023, Kirk2024}.  However, \cite{Ji2025} has recently shown that the cosmic shoreline may be a transition region instead of a dividing line, and that it may be influenced by planets' initial volatile inventories.  We thus need further observational efforts to understand the precise nature of the shoreline for M dwarf planets.  

There are also numerous observational issues that complicate our efforts at studying M dwarf planet atmospheres.  Late-type stars have more heavily spotted surfaces than earlier-type stars, which can create a substantial amount of contamination in their planets' transmission spectra that can obscure or mimic atmospheric features \citep{Rackham2018, Canas2025}. In addition, M dwarfs flare more frequently than earlier stars \citep{Lin2019}.  Stellar flares during a \textit{JWST} observation can complicate our efforts at fitting the planet properties \citep[see, e.g.,][]{Lim2023, Howard2023, Ahrer2025}, making flaring stars risky \textit{JWST} targets.


TRAPPIST-1 is one of the coolest, lowest-mass stars known with a well-characterized system of transiting planets. In addition, at least one of its planets falls within the star's habitable zone \citep{Gillon2017}.  As TRAPPIST-1 is so small and nearby, its planets are the most amenable temperate rocky planets to \textit{JWST} observations.  This has made TRAPPIST-1 the topic of over 200\,hours of allocated \textit{JWST} time.  However, TRAPPIST-1 is an ill-behaved star.  Its short rotation period \citep{Dmitrienko2018} and extensive flare activity \citep{Howard2023} have contributed to planetary spectra that are dominated by stellar contamination \citep[see, e.g.,][]{Lim2023, Radica2025, Rathcke2025}.  Furthermore, its inner planets, TRAPPIST-1\,b \citep{Greene2023,ih23} and TRAPPIST-1\,c \citep{Zieba2023, lincowski23} appear to be completely airless.  The difficulty inherent to studying the TRAPPIST-1 planets has motivated the community to continue to search for new temperate planets around less active stars.

GJ\,12 (system parameters listed in Table~\ref{tab:host}) is a nearby ($d\approx 12$\,pc) mid-M dwarf with a $P_\mathrm{orb}\,=\,12.76$d transiting planet (parameters listed in Table~\ref{tab:planet}) recently discovered by \citet[][hereafter referred to as K24]{Kuzuhara2024} and \citet[][D24]{Dholakia2024}.  The planet, GJ\,12\,b, has a measured radius of around one Earth radius, meaning that it is very likely to be a rocky planet as opposed to a sub-Neptune with a thick gaseous envelope.  Even more interestingly, it has an equilibrium temperature of around 300\,K.  This places it significantly interior to the inner edge of the habitable zone (HZ) of its host star according to the HZ equations from \cite{Kopparapu2013}, but on the inner edge of the HZ assuming the planet is slowly rotating \citep[see][]{Yang2014}.  In addition, MEarth photometry of the system indicates that the host star has a rotation period $\approx$80\,d \citep{Irwin2011, Newton2016}, and it has a measured HARPS log$R'_{HK}$ value of $-5.23^{+0.10}_{-0.14}$ \citepalias{Kuzuhara2024} and HARPS-N log$R'_{HK}$ value of $-5.68\,\pm\,0.12$ \citepalias{Dholakia2024}.  This means that it is likely inactive, making it significantly less active than TRAPPIST-1.  Thus, GJ\,12\,b may be a promising alternative for the TRAPPIST-1 planets for future transmission spectroscopy measurements.

Obtaining a precise mass measurement for GJ\,12\,b is crucial to interpreting any future atmospheric measurements.  Radial velocity (RV) data from \citetalias{Kuzuhara2024} recovered a planetary mass of $1.57^{+0.78}_{-0.75}$\,\MEarth, a mass measurement with a 50\% uncertainty that is not sufficient to gain substantial insight into the planet's composition.  \cite{Batalha2019} showed that, due to a degeneracy between planetary surface gravity and atmosphere composition, a planet needs a 20\% mass measurement or better to have an accurate atmospheric characterization in transmission.  In addition, this mass measurement can help us determine whether or not transmission spectroscopy is even feasible in the first place- a high-mass planet will have a much lower atmospheric scale height, dampening the strength of its atmospheric features and complicating efforts at studying its transmission spectra. This mass will also allow us to determine where GJ\,12\,b lies on the cosmic shoreline plot, which will aid in future interpretations of its absence (or presence) of an atmosphere.

In this paper, we use a newly-collected dataset of RVs from the MAROON-X spectrograph \citep{seifahrt16, Seifahrt18, seifahrt20, Seifahrt22} to measure the mass of GJ\,12\,b.  In Section~\ref{sec:obs}, we describe the RVs we collected with MAROON-X that we use in our analysis to determine the mass in Section~\ref{sec:analysis}.  We then discuss the physical implications of our derived mass of GJ\,12\,b in Section~\ref{sec:discussion} and conclude in Section~\ref{sec:conclusions}.



\begin{table}[]
    \centering
    \begin{tabular}{|l|c|c|}
    \hline
    \textbf{Property}      & \textbf{Value} & \textbf{Reference} \\
    \hline
    RA (J2000)             & 00h 15m 49.24s & a \\
    Declination (J2000)    & +13$^o$ 33' 22.32" & a \\
    Distance (pc)          & 12.166 $\pm$ 0.005 & a \\
    Spectral Type          & M4V & b \\
    $J$ Mag                & 8.619 & c\\
    $K$ Mag                & 7.807 & c\\
    $V$ Mag                & 12.6  & d\\
    M$_\star$ (\Msun)      & $0.2414\,\pm\,0.0060$ & d\\
    R$_\star$ (\Rsun)      & $0.2616^{+0.0058}_{-0.0060}$ & e \\
    T$_\mathrm{eff}$ (K)   & $3296^{+48}_{-36}$   &  e \\
    log\,$g$               & 5.21 $\pm$ 0.07   & e \\ \relax
    [Fe/H]                 & -0.32$\pm$0.06    & e \\
    Age (Gyr)              & $7.0^{+2.8}_{-2.2}$ & f \\
    \hline
    \end{tabular}
    \caption{Properties of GJ\,12, the host star considered in this study.  The references are as follows: a) \cite{GaiaEDR3}, b) \cite{Newton2014}, c) \cite{Cutri2003}, d) \cite{Paegert2021}, e) \citetalias{Kuzuhara2024}, f) \citetalias{Dholakia2024}.}
    \label{tab:host}
\end{table}

\begin{table}[]
    \begin{center}
    \begin{tabular}{|l|c|c|}
    \hline
    \textbf{Property}      & \textbf{Value} & \textbf{Reference} \\
    \hline
    $R_p/R_\star$          & $0.0336\,\pm\,0.0015$ & a \\
    $b$                    & $0.758^{+0.049}_{-0.114}$ & a \\
    ln$\rho_\star$ ($\rho_\odot$) & $2.62\,\pm\,0.08$  & a \\
    $a/R_\star$            & $54.9\,\pm\,1.4$      & a \\
    $T_{\mathrm{eq}}$ (K)  & $314.6^{+6.0}_{-5.4}$ & a \\
    $S_\mathrm{eff}$ (S$_\oplus$) & $1.62^{+0.13}_{-0.11}$ & a \\
    $R_p$ (\REarth)        & $0.958\,\pm\,0.05$    & a \\
    $M_p$ (\MEarth)        & $0.71\,\pm\,0.12$    & b \\
    $e$                    & $0.16^{+0.14}_{-0.09}$    & b \\
    \hline
    \end{tabular}
    \caption{Properties of GJ\,12\,b.  The references are as follows: a) \citetalias{Kuzuhara2024} and b) this work.}
    \label{tab:planet}
    \end{center}

\end{table}

\section{Observations}
\label{sec:obs}

\subsection{MAROON-X}
We observed GJ\,12 using MAROON-X 51 times between August 2024 and January 2025 as part of programs GN-2024B-FT-104 (PI: Madison Brady) and GN-2024B-Q-124 (PI: Jacob Bean). MAROON-X is split into a red channel ($\lambda\,=\,650-920$\,nm) and a blue channel  ($\lambda\,=\,500-670$\,nm) that are exposed simultaneously, which means that there are two semi-independent RV measurements at each epoch.  Because the two channels have different wavelength coverages, and thus different sensitivities to stellar signals, we treat the RVs from them as if they are from two separate instruments in our subsequent analysis.

After collecting the data, we reduced it with a custom \texttt{Python3} pipeline that includes tools that were originally developed for CRIRES \citep{Bean10}.  We then used a version of \texttt{serval} \citep{Zechmeister2018} that we modified to work with MAROON-X data to calculate RVs from our dataset.  \texttt{serval} calculates the RVs of a system by comparing each individual spectrum to a template created by co-adding all of the science spectra together.  Our thirty-minute exposures had a median peak signal-to-noise ratio (SNR) of 334 (143) per-pixel in the red (blue) channel, resulting in a median RV internal uncertainty of 0.35 (0.48) \ms in our \texttt{serval} outputs. 

We note that one of our observations (taken on 9 October 2024 UT) has a low SNR  (about 10\% the SNR of the rest of our observations).  While we cannot pinpoint an obvious reason for the data being of such low quality (our observation plan requested fairly strict conditions), we omit this data from our following analysis to simplify our visualizations. After noticing some additional outliers in our data, we also removed those data points that were listed as either ``failed'' or ``usable'' in the MAROON-X observing logs, as these points were likely impacted by conditions or interruptions that may cause systematic RV deviations that could be a problem given the small ($<$1\,\ms) anticipated RV signal of GJ\,12's planet. 

There was one major instrumental issue over the course of the GJ\,12 observations. Starting on September 1, 2024, the instrument cooling system failed and caused an uncontrolled warming of MAROON-X. MAROON-X was returned to regular service on September 14th once this issue was fixed and the instrument had cooled down to close to its normal operating temperature. However, the instrument continued to settle through September 18th.  We therefore also exclude all data taken between September 1 and September 18, resulting in the exclusion of four additional observations from our analysis.  After excluding this data (in addition to the data described in the previous paragraphs), we performed our final analysis with 42 observations.

The MAROON-X temperature anomaly in September 2024 likely caused a change in the zero point of the observations because the instrument did not return to the exact same state as before. We measure a shift of $\sim$0.7 pixels in the primary dispersion direction and $\sim$0.1 pixels in the cross dispersion direction for both arms after the anomaly. We took flat and dark calibrations after the temperature anomaly, but we don't have calibration data specifically for the pre-anomaly data due to an oversight in our operations plan. The lack of well-matched darks is a particular problem because some of the etalon light from the calibration fiber contaminates the orders containing the science data. MAROON-X darks are taken with the etalon on, and the dark subtraction is designed to remove the etalon contamination from the science data. The etalon lines in our dark frames taken after the temperature anomaly are thus shifted relative to the pre-anomaly data in the same axis as stellar radial velocity variations. We have accounted for this shift in our reduction, but this correction may not be perfect since it requires sub-pixel interpolation. In addition, the etalon flux had varied up to 5\% between the original dark frames and the science observations. We have thus scaled the etalon flux from our dark frames to match that in our science frames. Also, the impact of the mismatched flats is difficult to predict. Therefore, the pre-anomaly data may have different systematics than the post-anomaly data, for which we have suitable calibrations.

Given the potential issues with calibration, we will consider analyses of the MAROON-X data both including and excluding the pre-anomaly data (which comprises nine observations), as well as an accompanying analysis of a MAROON-X calibrator (HD\,3651) with observations taken both pre- and post-anomaly.

\begin{figure}
    \centering
    \includegraphics[width=0.9\linewidth]{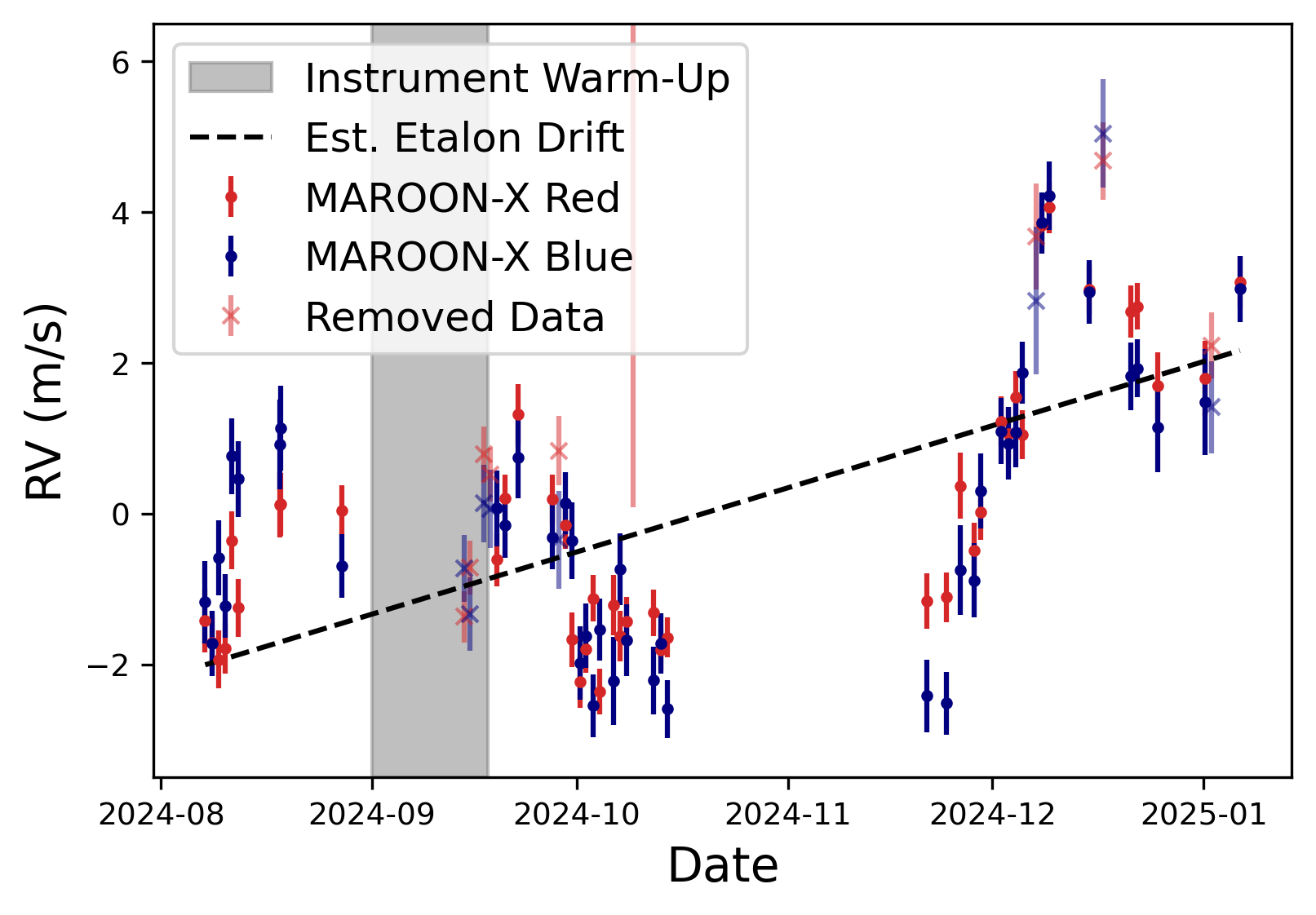}
    \caption{The MAROON-X RVs described in this paper.  The RVs that are not used in the final analysis are shown in lighter colors and indicated with 'x' markers.  The grayed-out region represents the time during the September 1-18 temperature anomaly.  We also include a dashed line indicative of the etalon drift slope as calculated for the calibrator star HD\,3651 (see Section~\ref{ssec:systematics} for more details).}
    \label{fig:GJ12_RV}
\end{figure}

Our final MAROON-X RVs are shown in Figure~\ref{fig:GJ12_RV}.  By eye, it is immediately obvious that there is some structure in the data, with at least one (if not several) signals at the 1-3\,\ms level.  The red-channel and blue-channel RVs appear to agree, indicating that the dominant signals in our RV data are achromatic.  

The MAROON-X data show an obvious slope that is consistent with the roughly 2-3\,cm\,s$^{-1}$\,d$^{-1}$ drift in the etalon that MAROON-X uses as the primary wavelength calibrator, described in \cite{Basant2025}.  In our fits of the MAROON-X RVs, we thus also fit a slope to the MAROON-X data.  We will discuss this in greater detail in Section~\ref{ssec:systematics}.

\subsection{Other RVs}

\begin{figure*}
    \centering
    \includegraphics[width=0.9\linewidth]{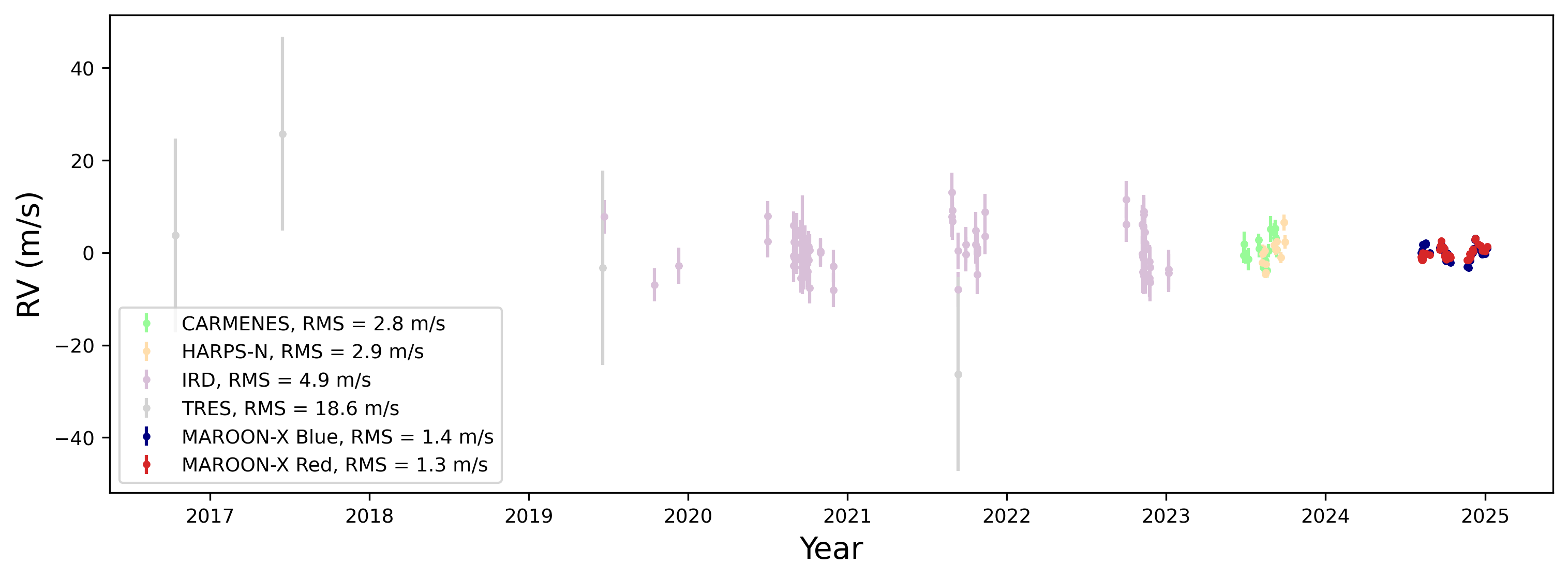}
    \caption{All of the RVs used in our analysis.  Each dataset shown is mean-subtracted to highlight the relative changes in RV of the star.  We have also subtracted the etalon drift slope and offsets (calibrated using those from HD\,3651, see Section~\ref{ssec:systematics} for more details) from the MAROON-X data.  Our MAROON-X data has a much smaller scatter than the data from other instruments, indicating that our MAROON-X data will provide much more constraining power than previous observations.}
    \label{fig:GJ12_RV_all}
\end{figure*}

In this section, we will briefly summarize all of the RVs collected on this target as a part of other surveys.  All of the RVs used in this paper are shown in Figure~\ref{fig:GJ12_RV_all}.

HARPS collected seven spectra on this target between 2003 and 2010.  While the HARPS RVs are available in \cite{Trifonov2020}, we decided against including them in our analysis given the substantial differences between the \texttt{serval}-calculated RV values (RMS\,=\,60\,\ms) and the DRS-calculated RVs (RMS\,=\,6\,\ms) for this target in \cite{Trifonov2020}.  Given the low quantity of data, as well as the substantial RV errors on the individual points, this is unlikely to dramatically influence our results.

\citetalias{Kuzuhara2024} observed GJ\,12 with CARMENES 17 times between June and September 2023, with a median RV error of 1.69\,\ms in the VIS channel.  They also collected 76 IRD spectra of GJ\,12 between 2019-2022, with median RV errors of 3.66\,\ms.  Despite their larger errors than those in our MAROON-X data, we include these RVs in our analysis to constrain any long-term and large-amplitude activity signals.  Further details on the data reduction of the CARMENES and IRD data are included in Section 3.3 of \citetalias{Kuzuhara2024}.

\citetalias{Dholakia2024} also collected 13 RVs on GJ\,12 with HARPS-N \citep{Cosentino2012} between August and October 2023.  They reduced the spectra using the traditional cross-correlation function (CCF) method as well as the line-by-line (LBL) method from \cite{Artigau2022}.  The CCF RVs have a median error of 2.64\,\ms and the LBL RVs have a lower median error of 1.14\,\ms.  We thus use the LBL-reduced RVs in our subsequent analysis.  We mean-subtracted these RVs ($\mu_{HARPS-N} = 51069.62$\,\ms) before using them in the analysis, as the MAROON-X RVs calculated with \texttt{serval} are relative values instead of absolute values.

There are also 4 TRES \citep{Furesz2008} RVs available for this system as a part of the \citet{Winters2021} survey, collected between 2016 and 2021.  Each has an RV error of 20\,\ms.  We include these RVs in our analysis primarily to probe the possibility of any long-term trends in the data.  We also mean-subtracted this dataset ($\mu_{TRES} = 51282.25$\,\ms) given our usage of relative (as opposed to absolute) RVs. 

\section{Analysis}
\label{sec:analysis}

\subsection{Rotation Period}
\label{ssec:rotper}

Understanding the rotation period of GJ\,12 is crucial for distinguishing between activity signals and bona-fide planets in our RV analysis.  There are already several different publications that attempted to constrain the rotation period of GJ\,12.  \cite{Irwin2011} and \cite{Newton2016} used MEarth photometry to find rotation period of $\approx 80$\,d.  \citetalias{Kuzuhara2024} and \citetalias{Dholakia2024} used additional RV and photometric data to constrain the stellar properties, but there is some disagreement between their rotation period measurements.  \citetalias{Kuzuhara2024} utilize public HARPS spectra and find log $R'_{HK}\,=\,-5.23^{+0.10}_{-0.14}$, while \citetalias{Dholakia2024} use their own HARPS-N spectra to find $R'_{HK}\,=\,-5.68\pm0.12$.  These correspond to rotation periods of roughly $60\,\pm\,20$d in the \citetalias{Kuzuhara2024} case \citep[using the relations from][]{Astudillo-Defru}, and $130\,\pm\,30$d in the \citetalias{Dholakia2024} case \citep[using the relations from][]{SuarezMascareno2018} -- numbers discrepant enough to be a cause for concern.  This discrepancy may be related to the relatively low SNRs of the HARPS and HARPS-N spectra, and we thus avoid using the analyses of $R'_{HK}$ to constrain the rotation period of GJ\,12.  

\citetalias{Kuzuhara2024} also attempted to constrain the rotation period of GJ\,12 using photometry from ASAS-SN \citep{Shappee2014, Kochanek17}, MEarth \citep{Berta2012}, and SuperWASP \citep{Pollacco2006}.  In these datasets, \citetalias{Kuzuhara2024} found photometric signals at periods ranging from 83 -- 108 days and identified the ASAS-SN signal at $\approx$85d as being the most reliable.  This is in agreement with the rotation signals identified in \cite{Irwin2011} and \cite{Newton2016}.

\begin{figure*}
    \centering
    \includegraphics[width=0.9\linewidth]{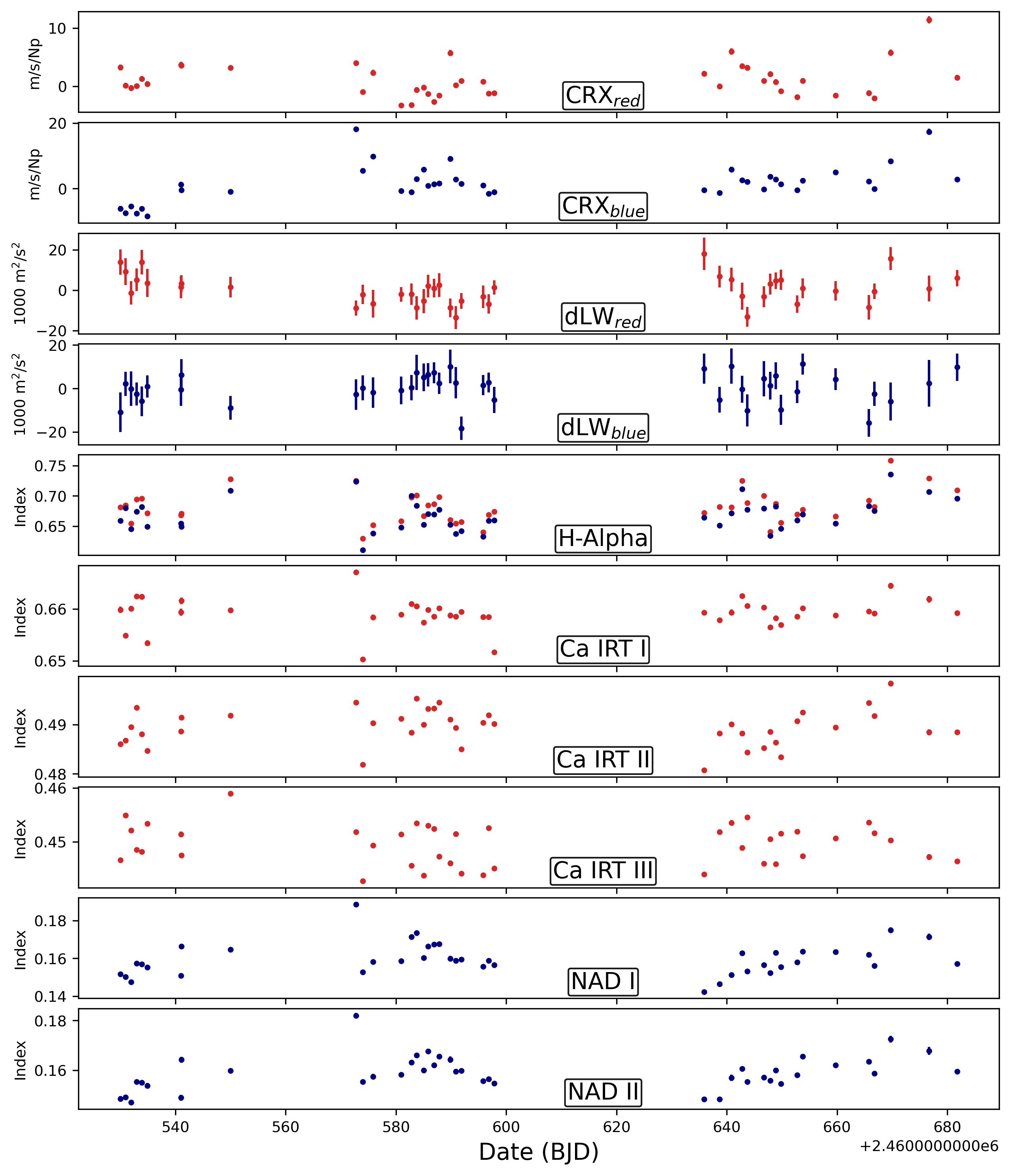}
    \caption{The \texttt{serval}-calculated activity indicators calculated from our MAROON-X data.}
    \label{fig:GJ12_act}
\end{figure*}

We examined our own \texttt{serval}-supplied spectral activity indicators (shown in Figure~\ref{fig:GJ12_act}) in order to provide further constraints on this number.  For this analysis, we included both the pre- and post-anomaly activity indicators to attempt to constrain the period, as we want to maintain as long a time baseline as possible given the star's long suspected rotation period.  We specifically examined the differential line-width and chromatic index \cite[as defined in ][]{Zechmeister2018}, as well as the hydrogen alpha, sodium doublet, and calcium infrared triplet line indices.

For the purposes of identifying any signals in our activity datasets, we performed a $\ell_1$ periodogram analysis \citep{Hara2017}.  The $\ell_1$ periodogram is useful for identifying signals in unevenly-sampled time series, and is less likely to identify signal aliases than the GLS periodogram, resulting a much more sparse final periodogram that only identifies the most significant signals.  For this analysis, we use \texttt{l1periodogram} code described in \cite{Hara2017} and implemented by the author on github\footnote{https://github.com/nathanchara/l1periodogram}.  As this code utilizes user-defined noise parameters to determine the covariance matrices in its analysis, we generated a grid of periodograms with varying values of red noise, white noise, calibration noise, and correlation times ($\tau$ = 1, 2, 4, 10, 20, 40, 60, 80, and 120\,d).  We allowed the noise parameters to vary between 0 and twice the standard deviation of the scatter for each activity parameter.  We then examined the 1\% of models for each activity indicator with the highest cross-validation values in order to identify the most robust signals for each indicator.

\begin{figure}
    \centering
    \includegraphics[width=0.9\linewidth]{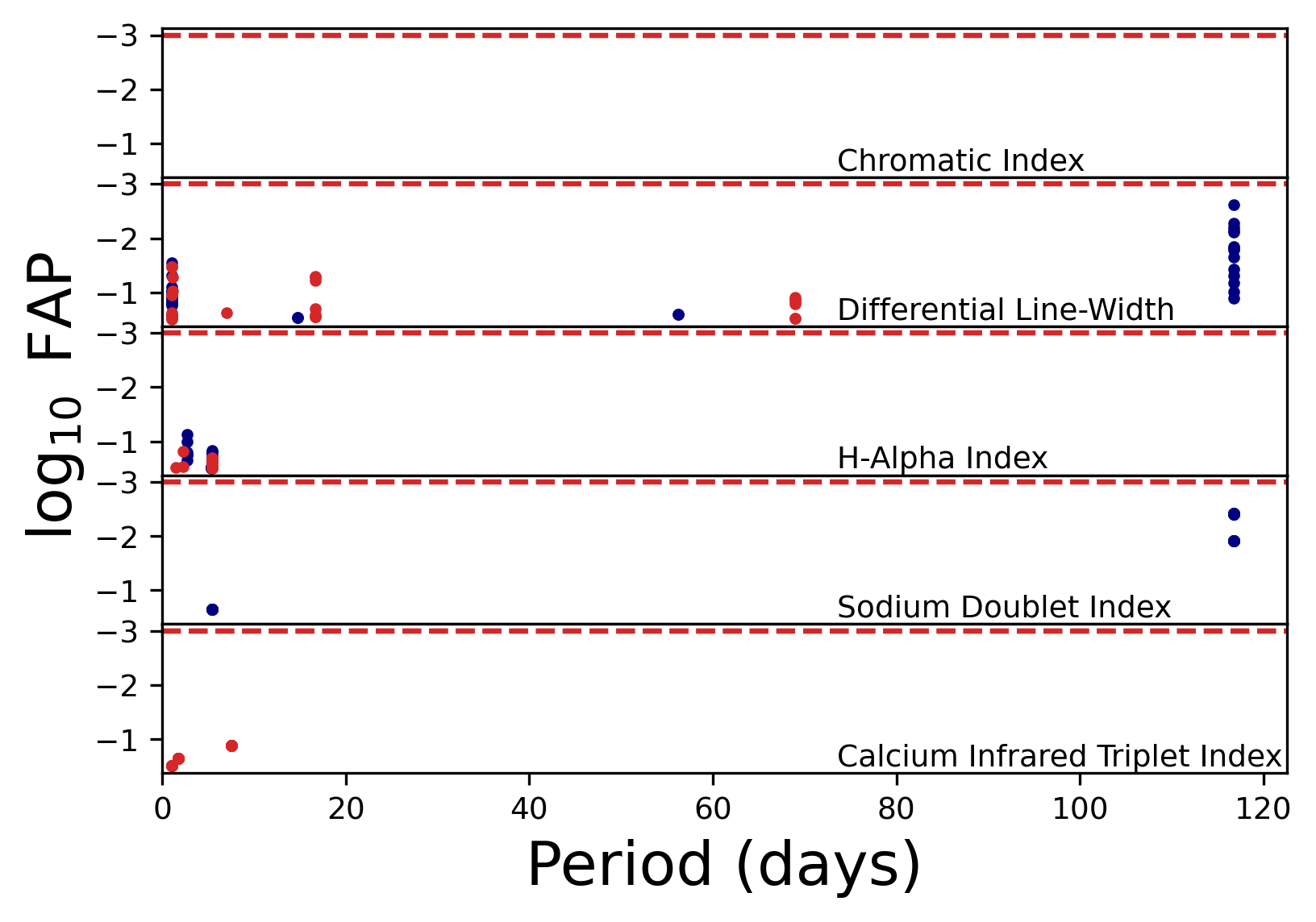}
    \caption{The \texttt{serval}-calculated activity indicators calculated from our MAROON-X data.}
    \label{fig:GJ12_activity_l1}
\end{figure}

As shown in Figure~\ref{fig:GJ12_activity_l1}, we were unable to recover any signals in our activity indicators at a log$_{10}$FAP level less than -3 (0.1\%).  The most significant signal that we find in the indicators is in our dLW (differential line-width) measurements in the blue channel, which have $\approx$1-10\% FAP signals at around 108d and 117d.  There is also a $\approx$1\% FAP signal in the sodium doublet index at 117d.  This signal could be related to the stellar activity, but we are hesitant to conclude the star has a rotation period of $\approx$120d given the low significance of the signal, as well as the discrepancy with the previously-quoted 80d signal from sources such as \cite{Irwin2011} and \cite{Newton2016}.  Our data do, however, support the theory that the star is relatively inactive and has a long rotation period.  Given the relatively short date range of our dataset (150\,days) and our unevenly sampled data, it is not a surprise that we have difficulty probing a 80 -- 120d signal.

\subsection{Characterizing The MAROON-X Systematics}
\label{ssec:systematics}

To constrain the instrumental systematics of MAROON-X in the GJ\,12 data, we examined the RV data of a calibrator star, HD\,3651, which was also observed over the same period.  HD\,3651 is useful as an RV calibrator given its massive, eccentric planetary companion, which \cite{Brewer2020} has shown is likely to have cleared its orbit of any additional planets.

\begin{figure}
    \centering
    \includegraphics[width=0.9\linewidth]{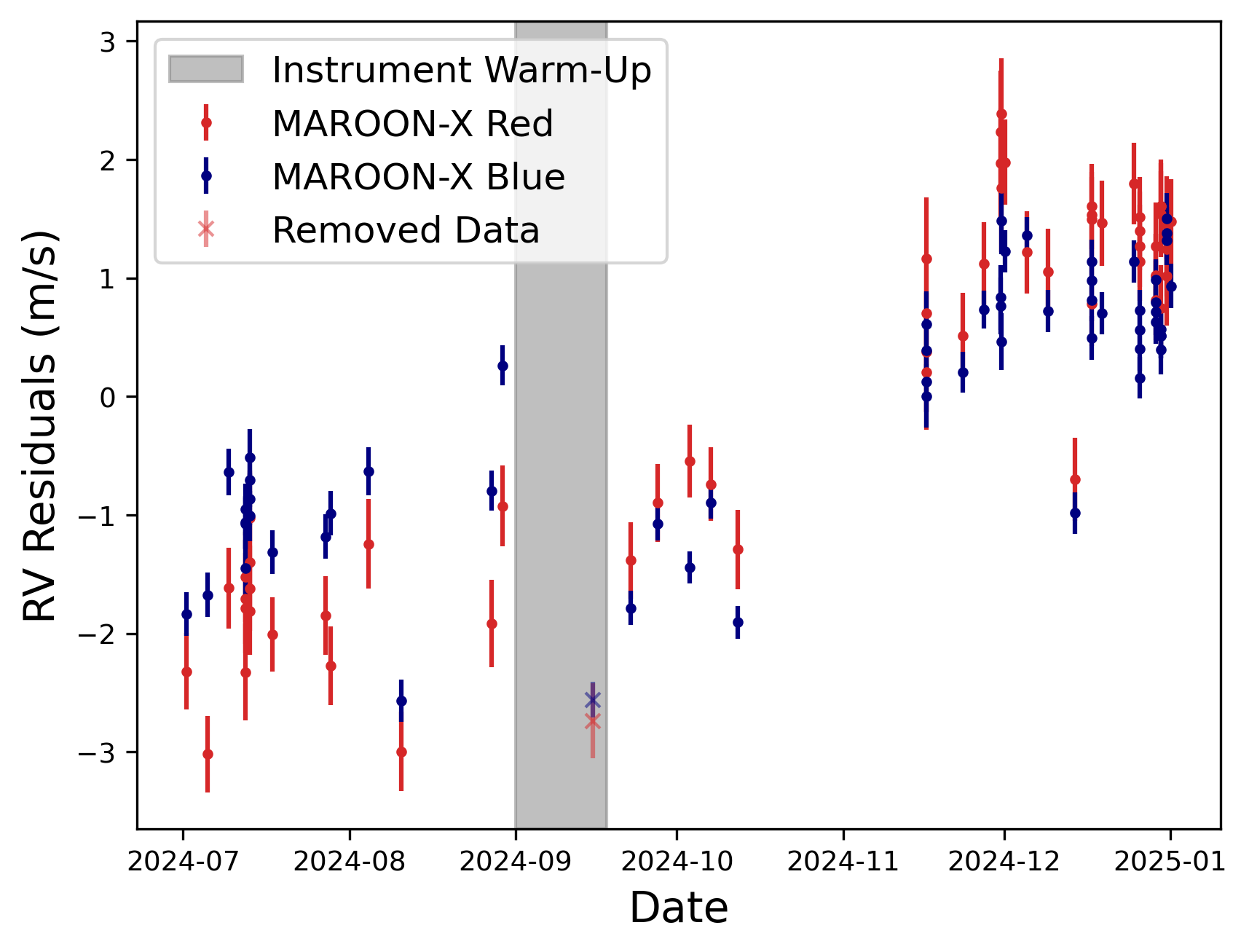}
    \caption{The MAROON-X RVs collected on calibrator star HD\,3651 in 2024, with the RV model for the eccentric planet from \cite{Basant2025} subtracted.  It is obvious that there is a slope in the RV residuals, as well as a potential offset around the time of the temperature anomaly.}
    \label{fig:HD3651_RV}
\end{figure}

Figure~\ref{fig:HD3651_RV} shows the RV data from HD\,3651 taken in 2024 (with data taken during the temperature anomaly omitted), with the Keplerian fit from \cite{Basant2025} subtracted to highlight the instrumental effects.  It is immediately obvious that there appears to be a slope in our RV data for HD\,3651, which is consistent with predictions from \cite{Basant2025}, who found that the MAROON-X etalon calibration causes a RV drift of roughly 2.2\,cm\,s$^{-1}$\,d$^{-1}$.  Given our long observing run for GJ\,12 ($>100$d), it is thus necessary to include a slope in the fits to the MAROON-X data.

Examining the HD\,3651 data more closely, we note the presence of a deviation between the red- and blue-channel RVs that indicates the offset during the temperature anomaly may be different between the two channels. We thus performed an RV analysis of the HD\,3651 data both with and without this offset to determine whether or not our calibrations statistically support including an offset in our GJ\,12 fits.

For the purposes of constraining the offset and slope, we included all of the MAROON-X data on HD\,3651 from 2024.  This includes 61 observations collected between 1 July 2024 and 1 January 2025, which is a similar time range as our GJ\,12 data.  We also included RV data from EXPRES \citep[from][]{Brewer2020}, APF \citep[from][]{Rosenthal2021}, HIRES \citep[from][]{Rosenthal2021}, and Lick \citep[from][]{Rosenthal2021}.  The inclusion of non-MAROON-X instruments will allow us to constrain the mass, eccentricity, and period of the eccentric planetary companion without the potential MAROON-X systematics.  We performed our fits using \texttt{juliet} \citep{Espinoza19}'s implementation of the nested sampling code \texttt{dynesty} \citep{Speagle2020, Koposov24} and the RV code \texttt{radvel} \citep{Fulton2018}.  We only included the known eccentric planet in our model, with a broad prior on the planet's eccentricity and mass.

\begin{figure}
    \centering
    \includegraphics[width=0.9\linewidth]{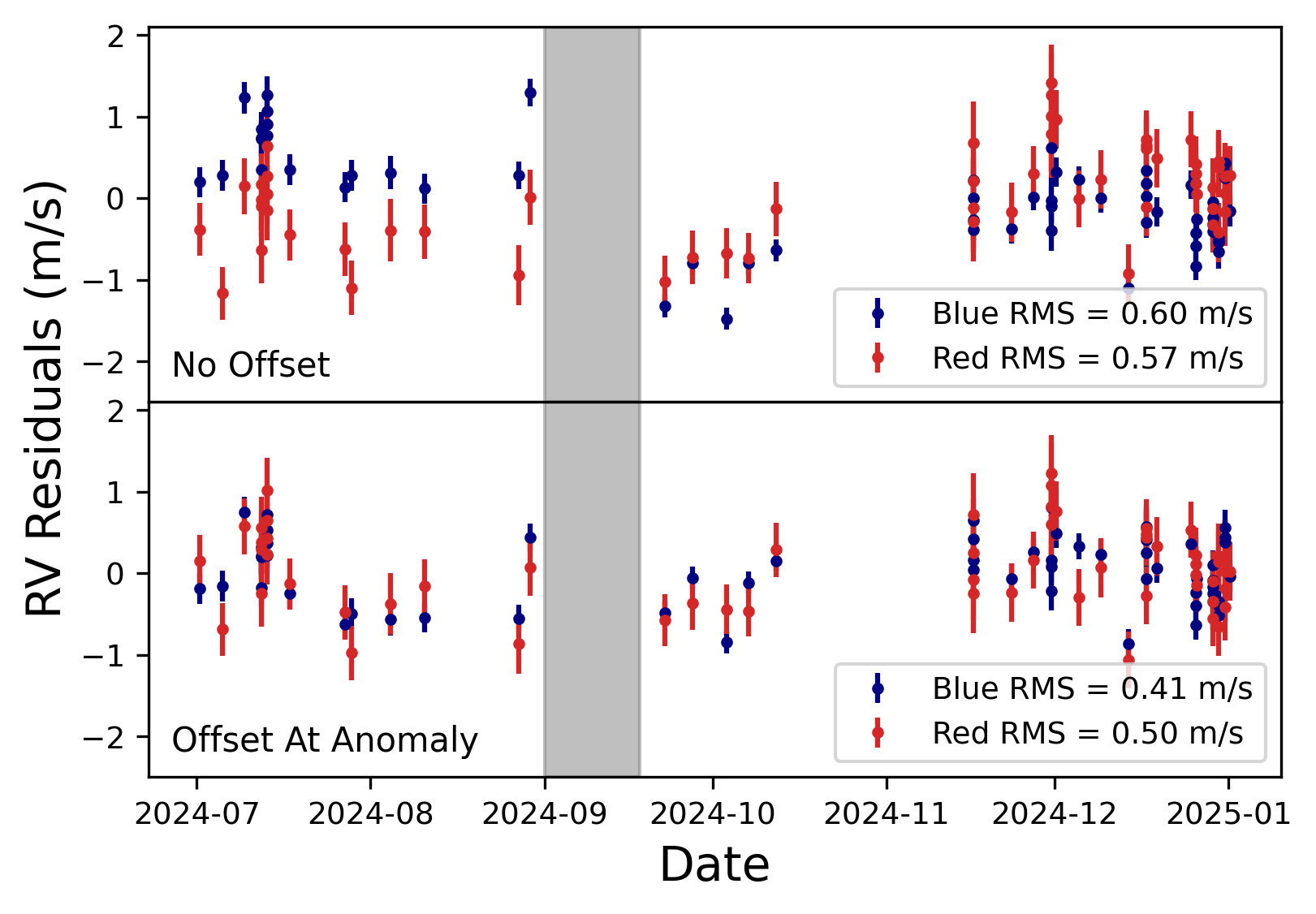}
    \caption{Residuals of our RV fits to the single eccentric planet in the HD\,3651 system.}
    \label{fig:HD3651_offsets}
\end{figure}

We found that, despite the increased model complexity, a model with a RV slope in the MAROON-X data and a chromatic offset placed at the temperature anomaly was preferred at very high confidence ($\Delta$ ln$Z$\,$>$\,15) over a model with just the etalon drift in the MAROON-X data.  We show the residuals for the model fits for the 2024 MAROON-X data in Figure~\ref{fig:HD3651_offsets}.  It is clear that including the offsets results in a substantial reduction in our model residuals. We thus include an offset in early September in our GJ\,12 analysis.  The value we recover for the etalon drift slope in 2024 is equal to $2.55^{+0.13}_{-0.17}$\,cm\,s$^{-1}$\,d$^{-1}$, which is comparable but not identical to the 2.2\,cm\,s$^{-1}$\,d$^{-1}$ value found in \cite{Basant2025} for pre-2024 MAROON-X data.  We also find an offset between the pre- and post-anomaly data of $-2.0^{+0.3}_{-0.2}$\,\ms in the blue channel and $-0.7^{+0.3}_{-0.2}$\,\ms in the red channel in addition to this slope.

Given the degeneracy between the etalon slope value and the offsets, as well as the fact that we are only examining a single calibrator, we are hesitant to place very strong constraints on the etalon drift slope or the offset values in our GJ\,12 analysis.  To be conservative, we thus allow for free uniform priors on the chromatic offsets between the pre- and post-anomaly MAROON-X data for GJ\,12, and a broad prior of $2.4\,\pm\,0.5$\,cm\,s$^{-1}$\,d$^{-1}$ on the etalon drift slope.  This broad slope prior captures both the value we fit for HD\,3651 in 2024 as well as the 2.2\,cm\,s$^{-1}$\,d$^{-1}$ value calculated in \cite{Basant2025} for data from earlier semesters.

\subsection{RV Model Selection}
\label{ssec:rv_models}

We performed a $\ell_1$ periodogram analysis \citep{Hara2017} to search for any signals in our RV dataset, including the transiting planet, stellar activity, and any other non-transiting planets. Similarly to our analysis in Section~\ref{ssec:rotper}, we performed a $\ell_1$ periodogram analysis with several values of noise parameters for the red, white and calibration noise ($\sigma$ = 0, 0.25, 0.5, 0.75, 1.0, 1.25, and 1.5\,\ms) as well as varying correlation times ($\tau$ = 1, 2, 4, 10, 20, 40, 60, 80, and 120\,d).  We considered offsets to be present in the data.  After generating a suite of periodogram results, we examined the 1\% of models with the highest cross-validation scores to identify potential RV signals.   As we could not simultaneously fit a slope to just the MAROON-X data, we also subtracted the 2.55\,cm\,s$^{-1}$\,d$^{-1}$ etalon drift signal calculated in Section~\ref{ssec:systematics} for HD\,3651 from the MAROON-X data before inputting it into the $\ell_1$ periodogram analysis.  In the cases where we included both the pre- and post-anomaly MAROON-X data, we also subtracted the offsets calculated in Section~\ref{ssec:systematics}.

\begin{figure}
    \centering
    \includegraphics[width=0.9\linewidth]{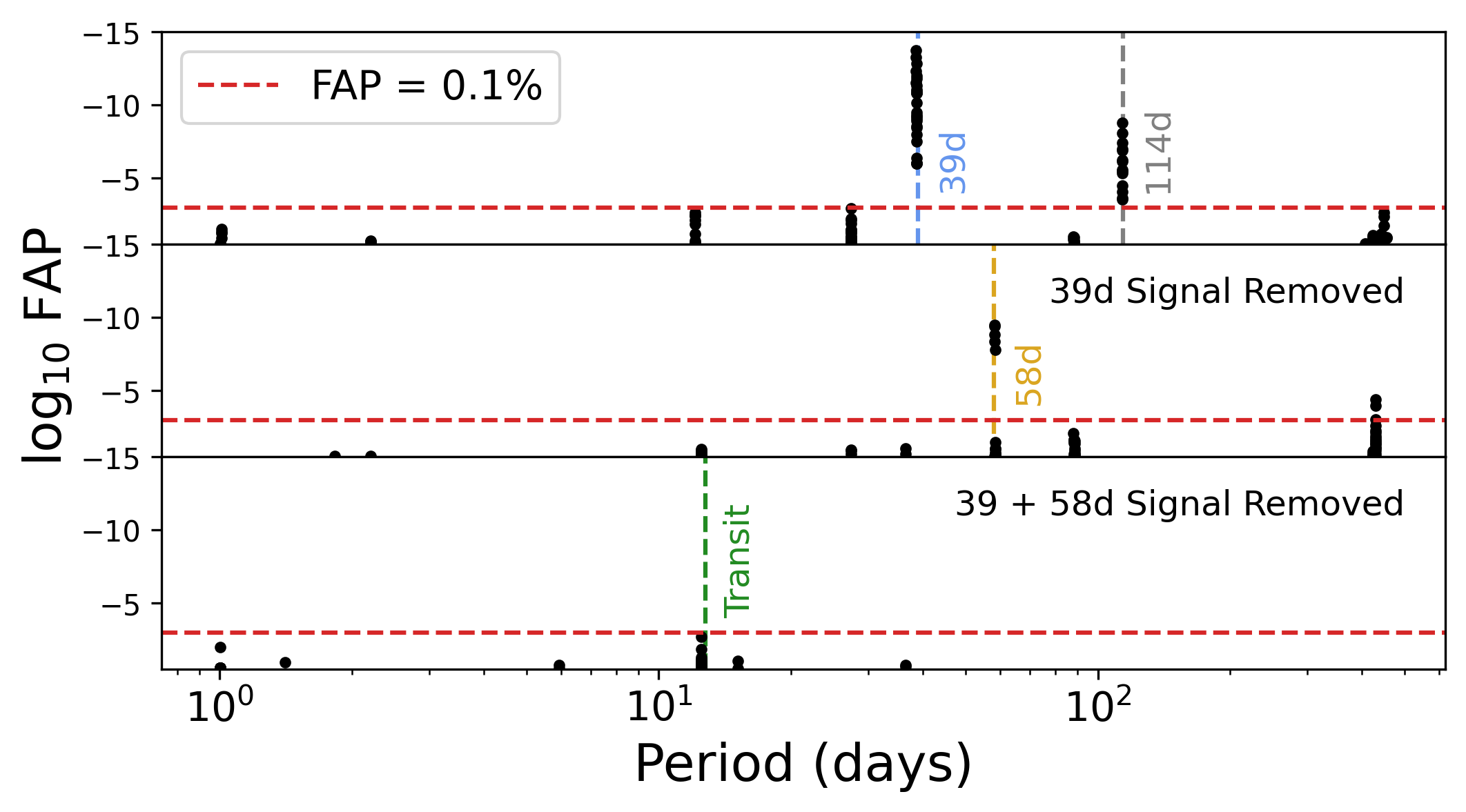}
    \caption{The results from our $\ell_1$ periodogram analysis of the RVs of GJ\,12.  The figure shows the identified significant periods of the top 1\% of noise models (in terms of cross-validation scores) for our RV data and identifies the FAPs that they are present in the data.  We include a red dashed line at FAP values of 0.1\% for reference.  We highlight and label the 39d, 58d, and 114d signals mentioned in the text, as well as the period of the transiting planet. \textbf{Top:} Our results with no unpenalized periods.  \textbf{Middle:} Our results with the 39d period unpenalized.  \textbf{Bottom}: Our results with both the 39d and 58d periods unpenalized.}
    \label{fig:GJ12_l1}
\end{figure}

After our first grid search, we found that, in most models, a very strong signal at around 39d is present, with a log$_{10}$FAP$\,<\,-6$.  There also appears to be another significant signal at around 114d that could be related to stellar rotation.  As noted in the $\ell_1$ periodogram code documentation, the $\ell_1$ periodogram can struggle to identify signals of vastly different amplitudes in data.  The recommended method to circumvent this is to unpenalize the period of a known strong signal in the data.  Thus, to search for any other signals, we then performed our grid search with the 39d signal unpenalized.  After removing the 39d signal, we recovered a 58d signal (which, given its period, is likely an alias of the 114d signal) with a log$_{10}$FAP that was as high as -9 in some models.  After unpenalizing this signal, we found no remaining signals in the periodogram with FAPs below -3.  However, we do note that several models recovered a 12.5d signal at a log$_{10}$FAP of around -1 to -2.7.  While this is not a high degree of significance for a blind search, the presence of this signal is promising as we know, from photometry, that there is a planet present in the system with a similar orbital period (12.8d).  We show the  results of our periodogram analysis in Figure~\ref{fig:GJ12_l1}.

In an effort to explore the presence of systematics in the pre-anomaly data, as well as any substantial deviations between our MAROON-X results and results from other instruments, we also performed several periodogram analyses of the MAROON-X data, including and excluding the data from August.  We found roughly similar results (with strong evidence for a signal at 40d, another signal at either 60d or 120d, and some evidence for the transiting planet at a lower significance), indicating that the MAROON-X data provides the majority of our constraining power.  We do note that including the August data resulted in a slightly reduced significance for the 40d signal and the transiting planet signal, but substantially higher evidence for the $\approx$120d signal compared to omitting the August data. It is unsurprising that including additional data increases the significance of the longer-period signal.

We note that the 39d signal is roughly consistent with being half of the supposed rotation period of the star ($\approx$80d according to past pubications using long-term photometry of the system), while the 58d (or 114d) signal could also be caused by the stellar activity (if the star's rotation period is $\approx$120d as suggested by some of our activity indicators).  We thus include several different models in our fits, with models including the transiting planet, Keplerians for a $58.2\,\pm\,2$d or a $38.5\,\pm\,2$d ``planet'' (these broad gaussian priors capture the periods identified by the $\ell_1$ analysis with some additional room for error), and stellar activity.

We performed our fits to the data using \texttt{juliet}, similarly to how we did for the HD\,3651 analysis.  For each fit, we set the number of live points used to be equal to the number of parameters squared or the number of parameters multiplied by forty (whichever was higher).  As recommended in the \texttt{juliet} code, we used a random walk method when the number of parameters was less than 20 and random slice sampling otherwise.  We also performed fits including and excluding the MAROON-X data from August, as we found that these two analyses produced different results.  

We modeled the rotation activity of the host star using gaussian processes (GPs) implemented in \texttt{juliet} with \texttt{celerite} \citep{celerite}, utilizing the double Simple Harmonic Oscillator GP (dSHO-GP) kernel described in \cite{Kossakowski2022}, which should be able to capture signals at both the rotation period of the star and half of the rotation period of the star.  For these GPs, we used a conservative gaussian prior on the rotation period of $100\,\pm\,30$d (which encompasses both the 80d and 120d rotation periods suggested via different methods) and included log-uniform priors on the GP amplitude of each instrument of between 0.01 and 10\,\ms, which seems to represent the amplitudes probed by our current dataset.  We allowed $f$, the fractional amplitude between the star's primary and secondary oscillations, to vary uniformly between 0 and 1, as is standard.  Finally, we allowed the quality factor of the secondary oscillation $Q_0$ and the difference in quality factors $dQ$ to vary log-uniformly between 10$^2$-10$^5$ and 10${-1}$-10$^{5}$, following the priors used in \citep{Kossakowski2022}.

We also included models with varying eccentricities and $\omega$ values in addition to models with eccentricities fixed at 0.  In our eccentric models, we allowed \textit{all} of the modeled planets to have varying eccentricities, as opposed to testing planetary eccentricities individually (which would have dramatically increased the number of models studied).  For our eccentric models, we defined our eccentricity priors based on the Beta distribution fit from \cite{Stevenson2025} for planets with masses between $1-20$\,\MEarth and periods between $10-300$\,days (which seems to roughly describe the anticipated mass and period range of these Keplerians).  We allowed the $\omega$ values to vary uniformly between -180 and 180.

For all of our models, we fit a slope (the same for both channels) in the MAROON-X data corresponding to the known etalon drift with a prior of $2.4\,\pm\,0.5$\,cm\,s$^{-1}$\,d$^{-1}$ (the derivation of this prior is discussed in Section~\ref{ssec:systematics}).  \texttt{juliet} handles RV slopes with the following parameterization:

\begin{equation}
    RV = slope * (t - 2458460) + intercept
\end{equation}

We did not fit the intercept value, as it would be completely degenerate with our offsets.  We found that fixing this value at -51\,\ms produced mean RVs per-instrument of around 0\,\ms, so we adopted -51\,\ms as the intercept value and fit for the slope.

\begin{table*}[]
\centering
\begin{tabular}{|l|c|c|c|c|}
\hline
\textbf{Model} & \textbf{$\Delta$ ln$Z$ (No Aug.} & \textbf{Mass$_{b}$ ($M_\oplus$)} & \textbf{$\Delta$ ln$Z$ (All)}  & \textbf{Mass$_{b}$ ($M_\oplus$)}\\
\hline
Offsets                                                & -62.65 &                        & -71.49 &   \\ 
GP                                                     & -27.70 &                        & -33.24 &   \\
Transiting Planet ($e=0$)                              & -64.81 & $0.69^{+0.28}_{-0.27}$ & -70.78  & $0.62^{+0.21}_{-0.20}$ \\
Transiting Planet ($e=0$), GP                          & -15.09 & $0.84^{+0.10}_{-0.11}$ & -23.97  & $0.72^{+0.12}_{-0.11}$ \\
Transiting Planet ($e\neq 0$)                          & -64.70 & $0.75^{+0.32}_{-0.30}$ & -72.41  & $0.65^{+0.22}_{-0.20}$ \\
Transiting Planet ($e\neq 0$), GP                      & -11.00 & $0.89^{+0.12}_{-0.12}$ & -20.87  & $0.82^{+0.12}_{-0.12}$ \\
Transiting Planet + 39d Keplerian ($e=0$)              & -27.99 & $0.94^{+0.16}_{-0.15}$ & -40.17  & $0.68^{+0.18}_{-0.18}$ \\
Transiting Planet + 39d Keplerian ($e=0$), GP          & -4.67  & $0.79^{+0.12}_{-0.10}$ & -10.40  & $0.69^{+0.10}_{-0.11}$ \\
Transiting Planet + 39d Keplerian ($e\neq 0$)          & -29.30 & $0.87^{+0.16}_{-0.16}$ & -37.37  & $0.84^{+0.17}_{-0.17}$ \\
Transiting Planet + 39d Keplerian ($e\neq 0$), GP      & -3.95  & $0.84^{+0.11}_{-0.12}$ & -7.69   & $0.80^{+0.10}_{-0.10}$ \\
Transiting Planet + 58d Keplerian ($e=0$)              & -56.47 & $0.41^{+0.25}_{-0.25}$ & -68.75  & $0.40^{+0.20}_{-0.19}$ \\
Transiting Planet + 58d Keplerian ($e=0$), GP          & -9.74  & $0.81^{+0.13}_{-0.13}$ & -16.20  & $0.66^{+0.11}_{-0.11}$ \\
Transiting Planet + 58d Keplerian ($e\neq 0$)          & -60.67 & $0.69^{+0.28}_{-0.30}$ & -69.91  & $0.41^{+0.23}_{-0.13}$ \\
Transiting Planet + 58d Keplerian ($e\neq 0$), GP      & -6.25  & $0.82^{+0.11}_{-0.11}$ & -9.17   & $0.79^{+0.10}_{-0.10}$ \\
Transiting Planet + 39d + 58d Keplerian ($e=0$)        & -5.09  & $0.76^{+0.12}_{-0.13}$ & -10.13  & $0.63^{+0.13}_{-0.13}$ \\
Transiting Planet + 39d + 58d Keplerian ($e=0$), GP    & 0      & $0.75^{+0.13}_{-0.12}$ & -3.52   & $0.63^{+0.10}_{-0.10}$ \\
Transiting Planet + 39d + 58d Keplerian ($e\neq 0$)    & -3.25  & $0.81^{+0.13}_{-0.13}$ & -6.76   & $0.70^{+0.13}_{-0.13}$ \\
Transiting Planet + 39d + 58d Keplerian ($e\neq 0$), GP& -0.28  & $0.74^{+0.11}_{-0.11}$ & 0       & $0.71^{+0.12}_{-0.12}$ \\
\hline
\end{tabular}
\caption{A comparison between several of the best-fit model likelihoods and masses for our GJ\,12 RVs, including and excluding the data from August.  We note that the $\Delta$ ln$Z$ values listed in each columns are only relative to other values in the same column, as the two datasets possess a significant offset in $\Delta$ ln$Z$ due to the differing amounts of data.}
\label{tab:model_GJ12}
\end{table*}

Our models are shown in Table~\ref{tab:model_GJ12}.  Overall, in both datasets we found very strong ($\Delta$ ln$Z$\,$>$\,10) evidence for a 39d signal in the data, even when also including a GP.  We note that, when \textit{not} including a Keplerian at 39d in our models, our GP fits would usually estimate rotation periods of around 72-80d (around two times the 39d period) supporting the preference of our models for this signal.  Without either this 39d signal or a GP included, there is no significant evidence for the transiting planet ($\Delta$ ln$Z$\,$<$\,1), which generally indicates that the 40/80d signal in both datasets completely overwhelms the signal of the transiting planet.  With the GP included, however, there is strong ($\Delta$ ln$Z$\,$>$\,10) evidence for a transiting planet.  

When the 39d signal is present, however, there is also a strong ($\Delta$ ln$Z$\,$>$\,10) preference for the models to also include either a 58d Keplerian or a GP.  This indicates that our dataset contains at least three signals: that of the transiting planet, a 40d signal, and a third signal around 60d, though it is somewhat ambiguous as to whether or not the 40d signal or the 60d signal is better fit as a Keplerian or as a GP.  In both the zero-eccentricity and variable-eccentricity cases, we find for both datasets that a three-Keplerian model is moderately to strongly preferred  (2.4\,$<$\,$\Delta$ ln$Z$\,$<$\,9.7) over a 60d Keplerian + GP model and inconclusively to moderately preferred (0.2\,$<$\,$\Delta$ ln$Z$\,$<$\,4.7) over a 40d Keplerian + GP model.  

Strangely enough, for both datasets, the models with the highest likelihoods include four signals- three Keplerians in addition to a GP signal.  As the star's expected rotation period is likely an alias of one of the longer-period Keplerians, a model that includes both the Keplerians and the rotation GP may be somewhat redundant.  However, \cite{Kossakowski2022} found similar results in their model fitting for AD Leonis, where there was a preference to fit the stellar rotation period with both a Keplerian \textit{and} a GP.  In their case, they concluded that there is a signal at roughly the rotation period of the star with both a stable component (the Keplerian) and a time-varying component (the GP) which could be indicative of either a planet in spin-orbit synchronization with the star or an unexpectedly stable rotation signal from the star.  These four-signal models are strongly ($\Delta$ ln$Z$\,$>$\,5) preferred over three-signal models with our full dataset but only moderately preferred (3\,$\lesssim$\,$\Delta$ ln$Z$\,$\lesssim$\,5) with the August data excluded. This makes sense given the increased time baseline we can access with the additional August data.  We do note that this could indicate that the MAROON-X August data possesses some kind of systematics, but we did not see any enhanced scatter in August from our analysis of HD\,3651 in Section~\ref{ssec:systematics}.  Given the likelihood preference, we thus choose to use the four-signal models in our final analyses, though we note that the final inferred masses for the transiting planets are not significantly ($>1\sigma$) affected by this decision.

When comparing models with similar numbers of signals, we also find that eccentric models are typically preferred over models with circular orbits, with a moderate-to-strong preference (2.7\,$<$\,$\Delta$ ln$Z$\,$<$\,7.0) in the full dataset and an inconclusive-to-moderate preference (0.3\,$<$\,$\Delta$ ln$Z$\,$<$\,3.5) in the dataset that excludes August.  Once again, we see a stronger preference for more complex models in the full dataset.  Given the long ($>$10d) orbital periods of the Keplerians in question, we have no reason to assume that the planets in the system must be on circular orbits, so we choose to quote these eccentric models in later analyses.  In addition, if the longer-period signals are related to stellar activity, it would also be unsurprising for them to be fit better by eccentric Keplerians, which are more flexible than circular Keplerians.

We note, in general, that, with the August data included, our inferred mass for GJ\,12\,b for most models is around 0.1\,\MEarth lower than it is with the August data excluded.  This may be some evidence that there is some source of systematics in the August data, but we currently have no other reason to definitely exclude this data from our analysis.  However, for the model with the highest likelihood with the full dataset (the rotation plus three eccentric Keplerians model), the mass inferred for the full dataset ($0.71\,\pm\,0.12$\,\MEarth) is almost identical to that inferred for the post-anomaly dataset ($0.74\,\pm\,0.11$\,\MEarth).  We thus choose to quote the full dataset in our final analysis.

We select the three-eccentric-Keplerian-plus-GP model as our final model for the data for the purposes of inferring the final planet mass, though we note that nearly every other model is within 1$\sigma$ of this model in terms of the inferred final mass.  This model is only moderately ($\Delta$ ln$Z$\,$\approx$\,3.5) preferred over a circular model, so we will also quote the circular model's parameters in our final fits and explore how those influence our results.

We note that, despite the data's preference for models that include at least three signals we cannot immediately infer the presence of additional planets.  As both the 39d and 58d Keplerians are close to harmonics of the rotation periods suggested by photometry and our activity indicators (respectively), we cannot conclude that they are planetary in origin, even though they are included in our most-preferred model.  It is not surprising that we have difficulty probing the nature of these signals given the relatively short (150\,days) timespan of our MAROON-X dataset. A quick search of the \textit{TESS} photometry on the system shows no clear signs of transits at their associated periods and $t_0$ values.  Further EPRV observations on the system that cover additional periods and phases of these signals may enable us to determine which signals in the dataset are likely to be planetary versus stellar.

\begin{figure*}
    \centering
    \includegraphics[width=0.9\linewidth]{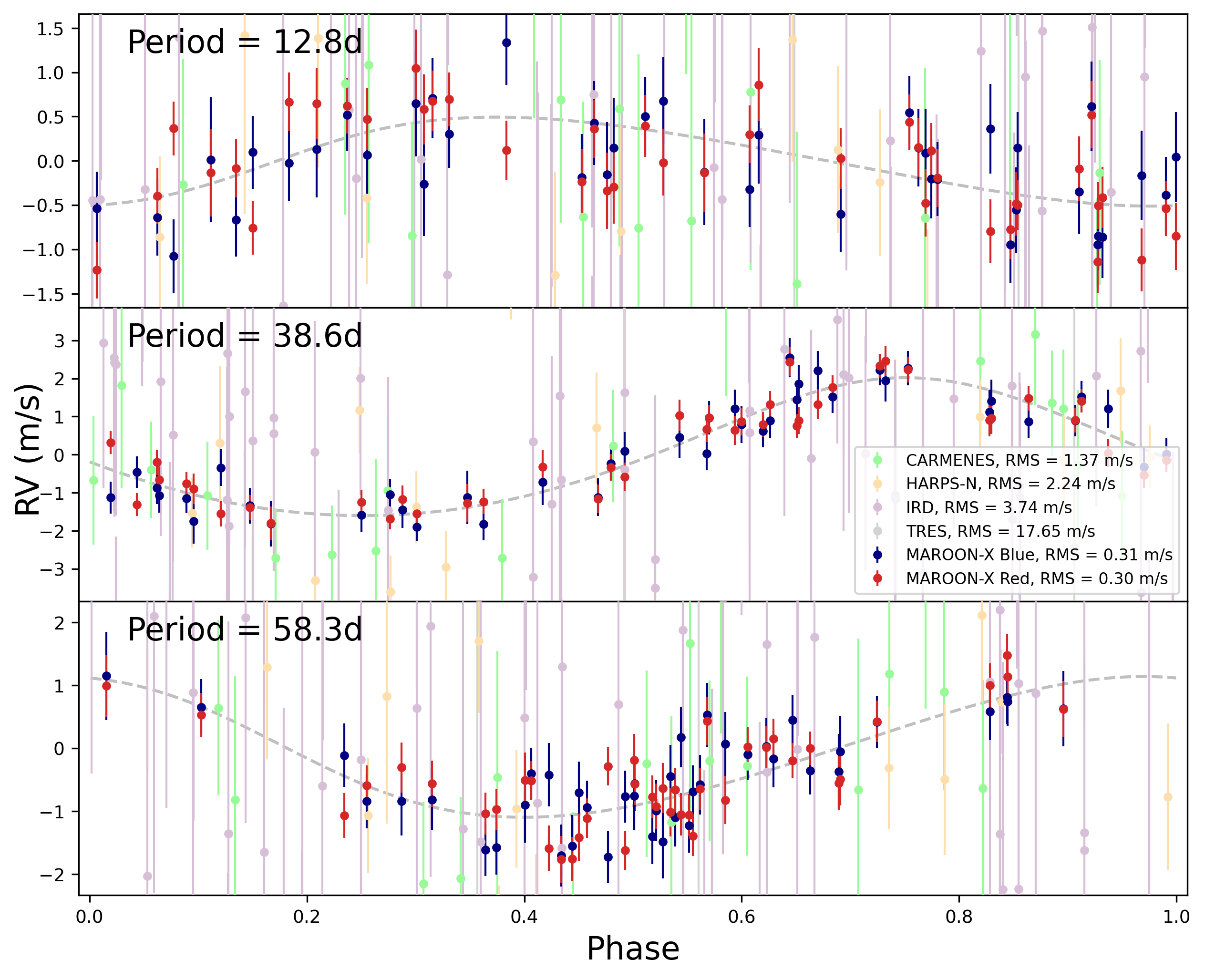}
    \caption{The phased RVs for the three Keplerians in our eccentric-orbit model for GJ\,12 with a rotation GP.}
    \label{fig:GJ12_RV_phase_ecc_act}
\end{figure*}

\begin{figure*}
    \centering
    \includegraphics[width=0.9\linewidth]{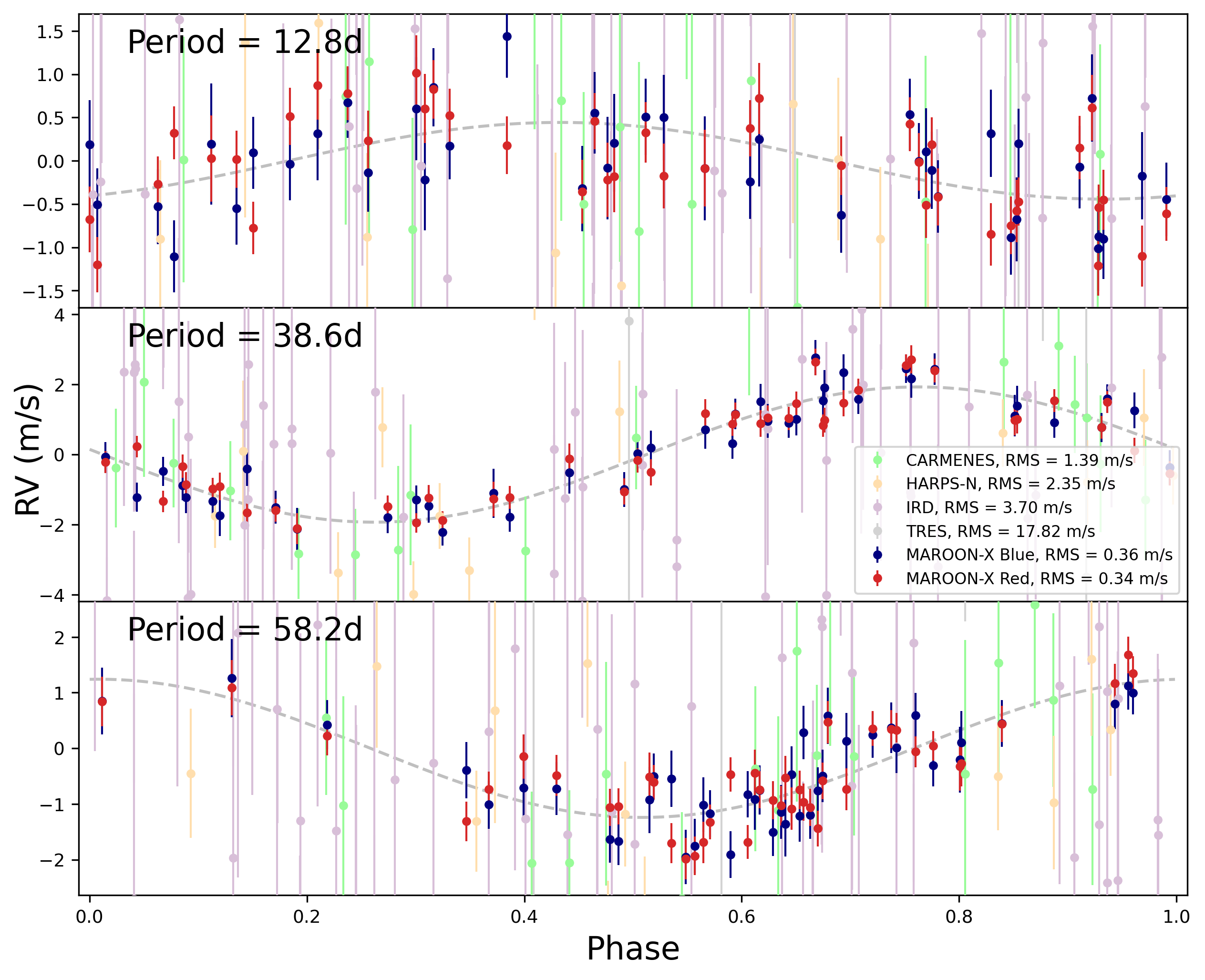}
    \caption{The phased RVs for the three Keplerians in our circular-orbit model for GJ\,12 with a rotation GP.}
    \label{fig:GJ12_RV_phase_act}
\end{figure*}

Thus, overall, we find a planet mass of $0.71\,\pm\,0.12$\,\MEarth, corresponding to a model that includes the MAROON-X etalon drift slope and a rotation GP, as well as eccentric Keplerians at the transiting planet's period, around 40d, and around 60d.  The phased RVs of this model are shown in Figure~\ref{fig:GJ12_RV_phase_ecc_act}, and the RVs of the similar model with the eccentricities fixed at zero are in Figure~\ref{fig:GJ12_RV_phase_act}.  The fit parameters (and priors) for both models are included in Table~\ref{tab:fit_results}.  We note that the model with circular orbits finds a lower planet mass of $0.63\,\pm\,0.10$\,\MEarth, but this is still within 1$\sigma$ of the eccentric model.  If we exclude the August data, the inferred planet masses are higher, at around $0.74\,\pm\,0.11$\,\MEarth and $0.75^{+0.13}_{-0.12}$\,\MEarth in the eccentric and circular cases, respectively.

\begin{table*}[]
    \centering
    \begin{tabular}{|l|c|c|c|}
    \hline
    \textbf{Parameter} & \textbf{Prior} & \textbf{Posterior (circular)} & \textbf{Posterior (eccentric)} \\
    \hline 
    \textbf{Transiting Planet} & & & \\
    $P$ (d) &  $\mathcal{N}$(12.761408, 0.000050) & $12.761394^{+0.000040}_{0.000041}$& $12.761416^{+0.000043}_{-0.000046}$ \\
    $t_0$ (BJD) & $\mathcal{N}$(2459497.1865, 0.0026)  & $2459497.1870^{+0.0023}_{-0.0025}$& $2459497.1871\,\pm\,0.0023$ \\   
    $K$ (\ms) & $\mathcal{U}$(0, 20)  & $0.44\,\pm\,0.07$ & $0.50\,\pm\,0.08$\\
    $e$       & $\beta$(1.78, 9.43)   &                   & $0.16^{+0.14}_{-0.09}$                      \\
    $\omega$ (degrees)  & $\mathcal{U}$(-180, 180)&                   &  $-99^{+14}_{-28}$                     \\
    $M$ (\MEarth)&                      &  $0.63\,\pm\,0.10$  & $0.71\,\pm\,0.12$                             \\
    \hline 
    \textbf{38d Keplerian} & & & \\
    $P$ (d) &  $\mathcal{N}$(38.5, 2) & $38.55^{+0.07}_{-0.06}$& $38.56^{+0.07}_{-0.08}$\\
    $t_0$ (BJD) & $\mathcal{U}$(2460530, 2460570)  & $2460547.20^{+0.48}_{-0.55}$&$2460546.26^{+0.69}_{-0.76}$ \\   
    $K$ (\ms) & $\mathcal{U}$(0, 20)  & $1.93^{0.21}_{-0.20}$& $1.81^{+0.24}_{-0.19}$\\  
    $e$       & $\beta$(1.78, 9.43)   &                   & $0.11\,\pm\,0.05$                      \\
    $\omega$(degrees) & $\mathcal{U}$(-180, 180)&                   & $0^{+25}_{-24}$                      \\
    $m$sin$i$ (\MEarth)&                      & $3.95^{+0.43}_{-0.42}$& $3.71^{+0.48}_{-0.40}$                   \\

    \hline 
    \textbf{59d Keplerian} & & & \\
    $P$ (d) &  $\mathcal{N}$(58.2, 2) & $58.24\,\pm\,0.25$ & $58.35^{+0.24}_{-0.25}$ \\
    $t_0$ (BJD) & $\mathcal{U}$(2460530, 2460590)  & $2460567.65^{+0.96}_{-0.94}$ & $2460569.59^{+1.86}_{-1.82}$\\   
    $K$ (\ms) & $\mathcal{U}$(0, 20)  & $1.24\,\pm\,0.16$ & $1.12\,\pm\,0.15$ \\    
    $e$       & $\beta$(1.78, 9.43)   &                   & $0.11^{+0.08}_{-0.06}$                      \\
    $\omega$ (degrees) & $\mathcal{U}$(-180, 180)&                   & $130^{+81}_{-66}$                      \\
    $m$sin$i$ (\MEarth)&                      & $2.91\,\pm\,0.41$ & $2.62\,\pm\,0.38$                        \\
    \hline
    \textbf{GP Parameters} & & & \\
    $\sigma_\mathrm{GP,TRES}$ (\ms)     & $\mathcal{LU}$(0.01, 10)          & $0.32^{+2.07}_{-0.28}$ & $0.46^{+3.08}_{-0.42}$ \\
    $\sigma_\mathrm{GP,IRD}$ (\ms)      & $\mathcal{LU}$(0.01, 10)          & $2.80^{+1.58}_{-0.93}$ & $2.86^{+1.55}_{-0.95}$ \\
    $\sigma_\mathrm{GP,CARMENES}$ (\ms) & $\mathcal{LU}$(0.01, 10)          & $0.20^{+0.66}_{-0.17}$ & $0.40^{+0.85}_{-0.34}$ \\
    $\sigma_\mathrm{GP,HARPS-N}$ (\ms)  & $\mathcal{LU}$(0.01, 10)          & $0.25^{+1.10}_{-0.21}$ & $0.67^{+1.15}_{-0.57}$ \\
    $\sigma_\mathrm{GP,MX Red}$ (\ms)   & $\mathcal{LU}$(0.01, 10)          & $0.47^{+0.40}_{-0.25}$ & $0.30^{+0.52}_{-0.23}$ \\
    $\sigma_\mathrm{GP,MX Blue}$ (\ms)  & $\mathcal{LU}$(0.01, 10)          & $0.40^{+0.26}_{-0.16}$ & $0.41^{+0.27}_{-0.18}$ \\
    $Q_0$                               & $\mathcal{LU}$(10$^2$, 10$^5$)    & $1008^{+6451}_{-788}$ & $1277^{+14484}_{-1049}$ \\
    $dQ$                                & $\mathcal{LU}$(10$^{-1}$, 10$^5$) & $37^{+3171}_{-36}$ & $55^{+6004}_{-55}$ \\
    $f$                                 & $\mathcal{U}$(0, 1)               & $0.66^{+0.23}_{-0.27}$ & $0.65^{+0.23}_{-0.25}$ \\
    $P$ (d)                             & $\mathcal{N}$(100, 30)            & $72.57^{+0.33}_{-0.39}$ & $72.58^{+0.32}_{-0.36}$ \\
    \hline
    \textbf{Instrument Parameters} & & & \\
    $\mu_{\mathrm{TRES}}$ (\ms) & $\mathcal{U}$(-30, 30)              & $1.18^{+9.38}_{-9.33}$ & $-0.08^{+9.61}_{-9.74}$\\
    $\mu_{\mathrm{IRD}}$ (\ms) & $\mathcal{U}$(-30, 30)               & $0.54^{+0.48}_{-0.49}$ & $0.42\,\pm\,0.51$ \\
    $\mu_{\mathrm{CARMENES}}$ (\ms) & $\mathcal{U}$(-30, 30)          & $0.68^{+0.41}_{-0.43}$ & $0.65^{+0.44}_{-0.43}$ \\
    $\mu_{\mathrm{HARPS-N}}$ (\ms) & $\mathcal{U}$(-30, 30)           & $0.67^{+0.61}_{-0.56}$ & $0.52^{+0.59}_{-0.57}$\\
    $\mu_{\mathrm{MX Red, Pre}}$ (\ms) & $\mathcal{U}$(-30, 30)       & $-1.29^{+4.19}_{-4.21}$ & $-5.42^{+3.70}_{-3.73}$\\
    $\mu_{\mathrm{MX Blue, Pre}}$ (\ms) & $\mathcal{U}$(-30, 30)      & $-1.19^{+4.14}_{-4.23}$ & $-5.23^{+3.78}_{-3.81}$\\
    $\mu_{\mathrm{MX Red, Post}}$ (\ms) & $\mathcal{U}$(-30, 30)      & $-1.04^{+4.41}_{-4.48}$ & $-5.58^{+3.93}_{-3.96}$\\
    $\mu_{\mathrm{MX Blue, Post}}$ (\ms) & $\mathcal{U}$(-30, 30)     & $-1.50^{+4.45}_{-4.43}$ & $-6.00^{+3.93}_{-3.96}$\\
    $\sigma_{\mathrm{TRES}}$ (\ms) & $\mathcal{LU}$(0.001, 20)        & $0.13^{+2.72}_{-0.13}$ & $0.11^{+2.56}_{-0.10}$ \\
    $\sigma_{\mathrm{IRD}}$ (\ms) & $\mathcal{LU}$(0.001, 20)         & $0.05^{+0.61}_{-0.06}$ & $0.12^{+0.87}_{-0.11}$\\
    $\sigma_{\mathrm{CARMENES}}$ (\ms) & $\mathcal{LU}$(0.001, 20)    & $0.04^{+0.21}_{-0.04}$  & $0.05^{+0.26}_{-0.04}$ \\
    $\sigma_{\mathrm{HARPS-N}}$ (\ms) & $\mathcal{LU}$(0.001, 20)     & $1.86^{+0.75}_{-0.57}$ & $1.54^{+0.80}_{-1.16}$\\
    $\sigma_{\mathrm{MX Red, Pre}}$ (\ms) & $\mathcal{LU}$(0.001, 20) & $0.43^{+0.31}_{-0.33}$ & $0.47^{+0.33}_{-0.41}$ \\
    $\sigma_{\mathrm{MX Blue, Pre}}$ (\ms) & $\mathcal{LU}$(0.001, 20)& $0.05^{+0.31}_{-0.05}$ & $0.23^{+0.30}_{-0.18}$ \\
    $\sigma_{\mathrm{MX Red, Post}}$ (\ms) & $\mathcal{LU}$(0.001, 20)& $0.32^{+0.10}_{-0.11}$ & $0.28^{+0.10}_{-0.09}$ \\
    $\sigma_{\mathrm{MX Blue, Post}}$ (\ms) & $\mathcal{LU}$(0.001, 20)&$0.05^{+0.16}_{-0.04}$ & $0.02^{+0.09}_{-0.02}$ \\
    Etalon Slope (cm\,s$^{-1}$\,d$^{-1}$) & $\mathcal{N}$(2.4, 0.5)   & $2.46\,\pm\,0.20$      & $2.66\,\pm\,0.18$\\
    Etalon Intercept (\ms) & Fixed & -51 & -51 \\
    \hline

    \end{tabular}
    \caption{Priors and posteriors of the circular (third column) and eccentric (fourth column) fit models for the RV data.  The listed $\omega$ values are shifted post-fitting by factors of 2$\pi$ as necessary so that they accurately represent the center and errors of the angular distributions.}
    \label{tab:fit_results}
\end{table*}

\section{Discussion}
\label{sec:discussion}

\subsection{The Composition of GJ\,12\,b}
\label{ssec:dynamics}

\begin{figure}
    \centering
    \includegraphics[width=0.9\linewidth]{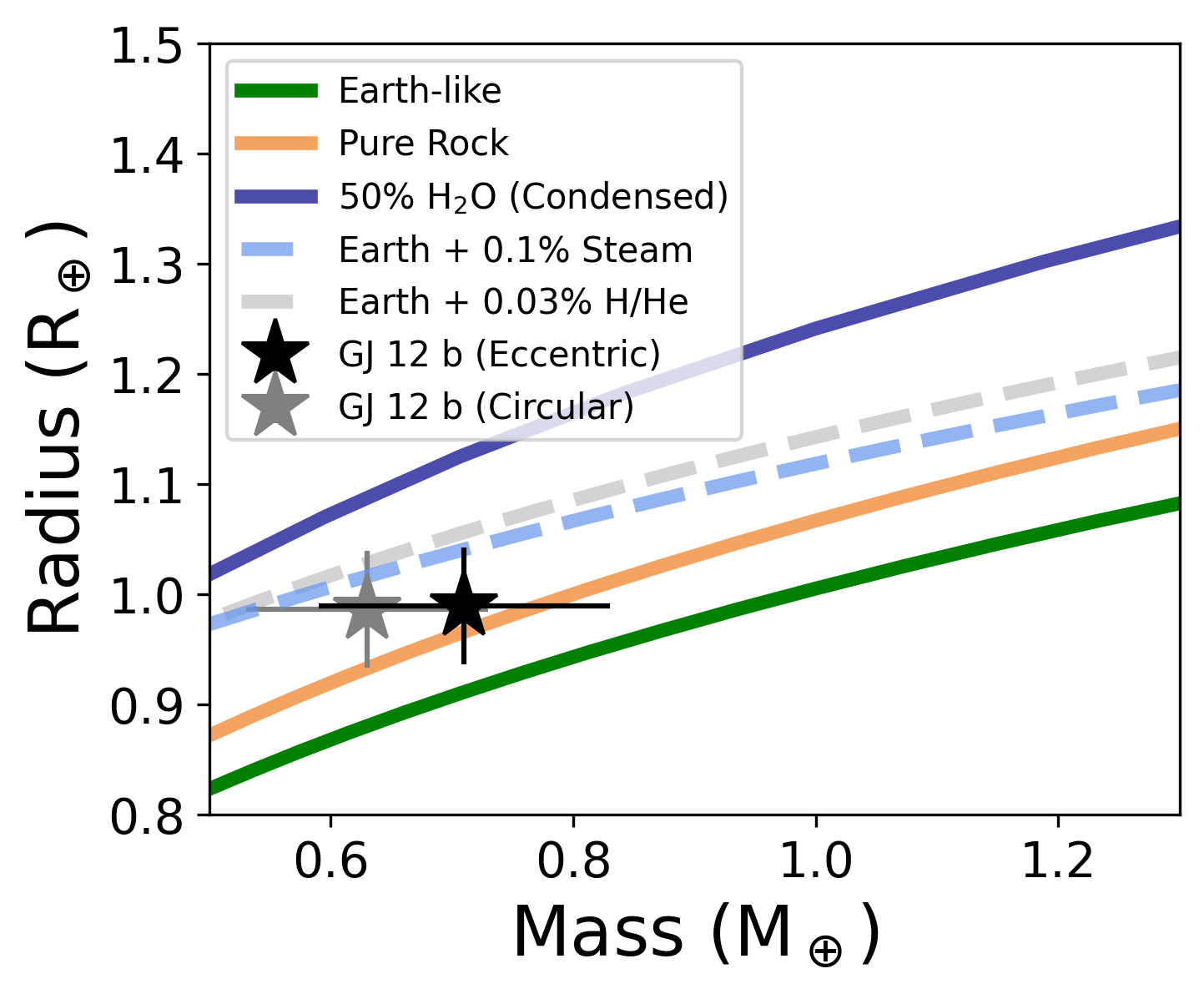}
    \caption{The mass and radius of our two models for GJ\,12\,b (circular and eccentric orbits) compared to various composition models from \cite{Zeng2019}, as well as a steam atmosphere model from \cite{Turbet2020} and a H/He atmosphere model from \cite{Nixon2021} and \cite{Nixon2024}.  We note that the 50\% water model, which is from \cite{Zeng2019}, both assumes a planet temperature of 300\,K, which is very close to the actual equilibrium temperature of this planet.}
    \label{fig:GJ12_intcomp}
\end{figure}

We show the placement of GJ\,12\,b on a mass-radius diagram in Figure~\ref{fig:GJ12_intcomp}.  With a mass of $0.71\,\pm\,0.12$\,\MEarth, GJ\,12\,b is one of the lowest-mass transiting planets with a mass measurement at the $5\sigma$ level.  However, its radius of $0.96\,\pm\,0.05$\,\REarth would make us anticipate a slightly larger mass for the planet. For example, the Earth-composition models from \cite{Zeng2019} predict a mass of $0.84\,\pm\,0.15$\,\MEarth for a planet of that radius.  If the planet has a pure-rock mass (with no iron core whatsoever), these same relations would predict a mass of $0.68\,\pm\,0.12$\,\MEarth for the target. Given our errors, it thus appears that GJ\,12\,b has a density somewhere between that of a pure-rock planet and a planet with an Earth-like composition.  Our circular-orbit model, with a mass of $0.63\,\pm\,0.10$\,\MEarth, is $>1\sigma$ less dense than an Earth composition and appears to be slightly less dense than a planet of pure rock.  In the circular orbit case, this could be indicative of a water-rich composition, pure-rock composition, or the presence of an atmosphere.  However, given the ambiguity due to both the model selection and the inclusion or exclusion of the August data from our RV analysis, we cannot robustly conclude that GJ\,12\,b has an atmospheric or volatile component.

\subsubsection{A Nonzero Eccentricity for GJ\,12\,b}

Given our moderate preference for an eccentric model to describe the orbit of GJ\,12\,b, it is necessary to consider whether or not such a model is feasible.  We can first calculate the eccentricity circularization timescale of GJ\,12\,b using the equation from \cite{Dobbs-Dixon04}.  If we assume an Earth-like modified tidal quality factor of 1500 \citep[see][]{Lainey2016}, which is reasonable given our calculated mass and radius for the planet, we calculate a circularization timescale of $>80$\,Gyr, which is significantly longer than the age of the system.  It thus seems perfectly feasible for GJ\,12\,b to maintain a heightened eccentricity.

It is next important to consider the stability of the system.  To examine this, we ran a suite of N-body simulations using the code \texttt{REBOUND} \citep{rebound}, using the WHFast integrator.  For our time step, we used a non-integer multiple of the shortest orbital period in the system (that of GJ\,12\,b), more specifically $P_b/\pi$.  We checked the stability of the system for both our circular and eccentric models with edge-on inclinations, assuming that either 1) all three of the Keplerians in the system are true planets, 2) the 39d signal is caused by stellar activity, or 3) the 58d signal is caused by stellar activity.  We found that all of our models are stable out to one billion orbits of GJ\,12\,b \textit{except} for the three-eccentric-Keplerian model, in which the planets are ejected within 2\,Myr (significantly shorter than the age of the system).  Thus, our eccentric model is only stable if either the 39d signal or the 58d signal in our RVs is \textit{not} caused by a planet and is instead caused by stellar rotation.  As both of these signals are consistent with harmonics of the estimates of the stellar rotation period, it still appears possible that an eccentric GJ\,12\,b, would be stable.  Thus, overall, we cannot rule out an eccentric orbit for GJ\,12\,b on theoretical grounds.

\subsubsection{An Atmosphere?}

\begin{figure}
    \centering
    \includegraphics[width=0.9\linewidth]{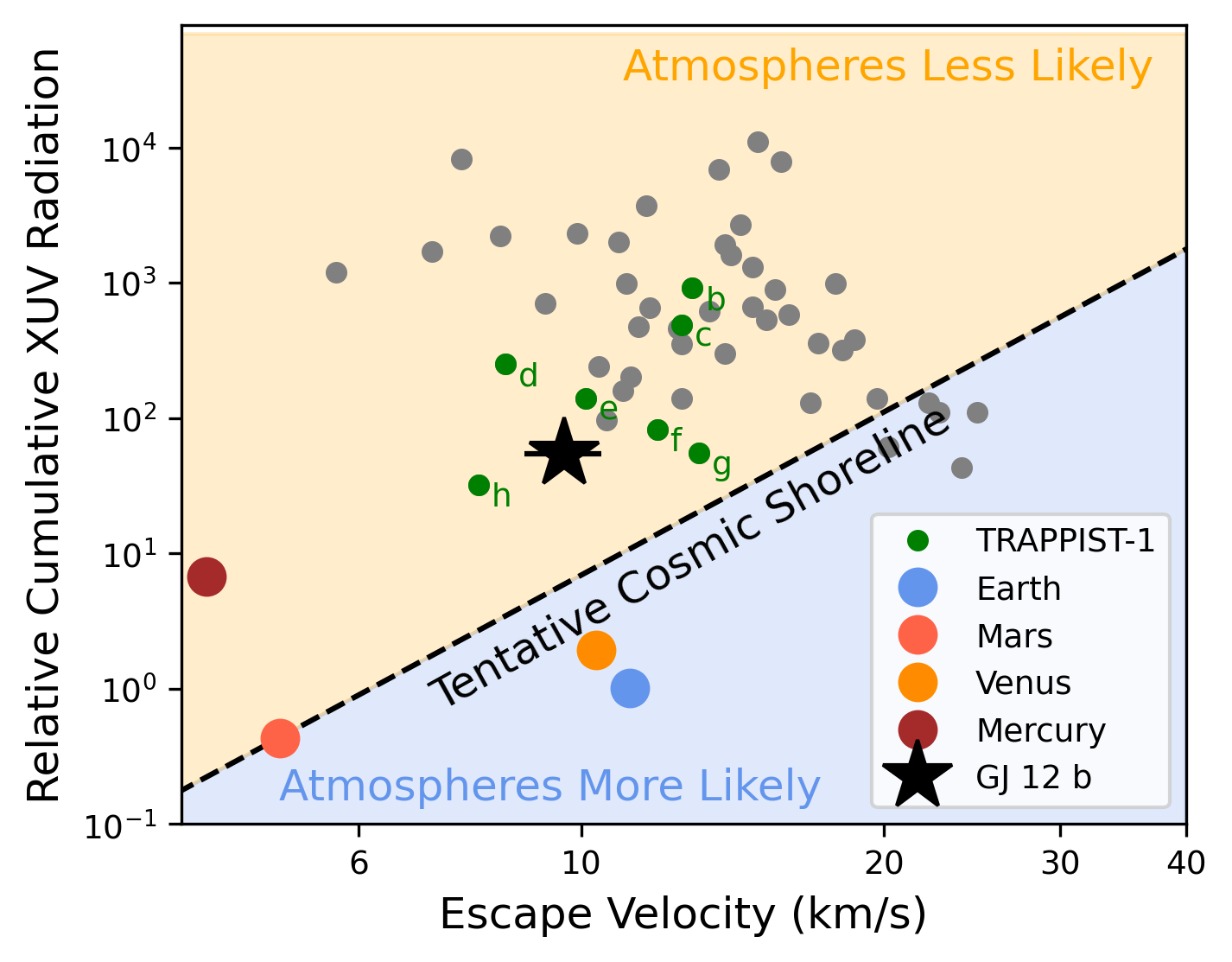}
    \caption{The cosmic shoreline formulation from \cite{Pass2025} for small planets orbiting mid-to-late M dwarfs within 50\,pc, showing our eccentric model for GJ\,12\,b (black star) in escape velocity versus XUV radiation space compared to the solar system planets (various colors), TRAPPIST-1 planets (green), and other planets with $R_p\,<\,1.8$\,\REarth (gray).  The location of the tentative shoreline is placed based on what we know from the solar system, with Mars on the shoreline while Mercury is airless.}
    \label{fig:shoreline}
\end{figure}

While our understanding of the XUV fluxes of M dwarfs is somewhat fragmentary (which poses a major issue when estimating photoevaporation timescales for their planets' atmospheres), we can compare GJ\,12\,b's characteristics to other planets to determine whether or not it is likely to retain an atmosphere.  Figure~\ref{fig:shoreline} shows the calculated XUV fluence versus escape velocity calculated in \cite{Pass2025} for planets orbiting mid-to-late M dwarfs within 50\,pc, calculated assuming both pre-main-sequence overluminosity and stellar flares.  The location of the shoreline (above which we expect planets to lose their atmospheres and below which we expect them to retain atmospheres) is identical that that quoted in \cite{Pass2025} and is extrapolated from solar system objects.  Using the formulas from \cite{Pass2025}, we calculated the XUV fluence of GJ\,12\,b (using linear interpolations based off of their calculations to estimate its overluminosity parameter) and placed it on the same plot.  It is clear that GJ\,12\,b sits near the shoreline, with a shoreline distance comparable to that of TRAPPIST-1\,e, which is a core target for \textit{JWST} habitability searches \citep[e.g.,][]{Morley2017, Claringbold2023}.  This is true regardless of whether we use the eccentric or circular orbit model for GJ\,12\,b.

However, there is no definitive atmospheric detection (or nondetection) of TRAPPIST-1\,e, so we still do not know the location of the shoreline relative to GJ\,12\,b.  TRAPPIST-1\,e's host is a much later M dwarf than GJ\,12, so it is likely much more active and thus may be less likely to host planets with atmospheres.  Studying GJ\,12\,b in detail and determining whether or not it has an atmosphere with \textit{JWST} may thus also be helpful for placing the cosmic shoreline and prioritizing future efforts on TRAPPIST-1\,e (and similar planets).  If GJ\,12\,b lacks an atmosphere, it may mean that planets such as TRAPPIST-1\,e also lack atmospheres (given their host star's higher modern activity levels), making them poor targets for habitability studies.  

We can use energy-limited photoevaporative mass-loss relations to get an estimate of the atmospheric mass-loss rate of GJ\,12\,b, which, when combined with the system's age, should give us a sense of how much atmospheric mass we expect the planet to have lost over its lifetime. To calculate the anticipated mass loss, we used the equations from \cite{Rogers2021}, which models the incident XUV flux on the planet with a broken power-law that depends on the host star mass.  Their model assumes that the flux from the star is constant up until the saturation time and decreases proportionally to $t^{-0.5}$ afterwards.  To determine the total loss, we had to first calculate the system's age.  We did this using the gyrochronology relationships from \cite{Engle2023}.  In order to be as conservative as possible (as we had little constraints on the system's rotation period from our own data), we quote the rotation period of the system as being uniform between 60 -- 120 days.  With that assumption, we find that the system's age is $>4$\,Gyr at a 99\% confidence level.  We then used the formulas from \cite{Rogers2021} to find that, if the planet formed in-situ, it would have lost around $5*10^{-3}$\,\MEarth ($7*10^{-3}$\,\MEarth in the circular-orbit case) of its atmosphere over its lifetime, which corresponds to about 0.7\% (1.1\%) of its current-day mass (which, given the relatively low amount of loss, is fairly similar to its initial mass).  This is only an estimate on the true amount of mass loss, however.  If GJ\,12\,b formed further out in the disk and later migrated inwards, it would have less incident radiation and thus would have less mass loss.  However, if GJ\,12 had energetic flares in its early stages, there could have been more loss than expected.  We thus only recommend using this value as an estimate.

We can next set limits on the current atmospheric thickness of GJ\,12\,b.  Examining the steam atmosphere models from \cite{Turbet2020}, which specifically describe the steam atmospheres of planets hotter than the runaway greenhouse limit \citep[which is true for GJ\,12\,b, see, e.g., the equations from][]{Kopparapu2013}, we found that, even in the circular orbit (lower-mass) case, if GJ\,12\,b has a steam atmosphere, it must make up less than 0.1\% of the mass of the planet given its radius constraints.  

We also considered internal structure models with H/He atmospheres described in \citet{Nixon2021} and \citet{Nixon2024}.  In these models, we assumed the core had an Earth-like iron-to-silicate ratio for the core and mantle and studied a range of H/He mass fractions ranging from 0.01-0.10\% (see Figure~\ref{fig:GJ12_intcomp}).  We used an isothermal temperature profile at the equlibrium temperature of GJ\,12\,b at pressures below 10\,bar, transitioning to an adiabat at higher pressures.  Overall, we found an upper limit of 0.03\% on the mass fraction of a H/He atmosphere in the circular-orbit case (with the higher-mass eccentric-orbit model having even more restrictive limits).  Thus, given the planet's current radius, its atmosphere must make up significantly less than 1\% of its mass today.  The planet retaining such a small proportion of its primordial atmosphere after significant photoevaporation represents a massive fine-tuning problem, and thus we conclude that GJ\,12\,b is unlikely to have retained any of its primordial H/He atmosphere. 

Our prior calculations cannot eliminate the possibility that GJ\,12\,b still has an atmosphere, as it may have outgassed a secondary atmosphere (possibly including the water sequestered in its interior) over time.  \cite{VanLooveren2024} found that none of the TRAPPIST-1 planets could sustain a secondary atmosphere long-term unless they were continually replenished by volcanism on $\approx$1\,Myr timescales.  While GJ\,12\,b has a similar instellation to the TRAPPIST-1 planets, it may be able to sustain volcanism via tidal heating if it has an enhanced eccentricity.  

To quantify the planet's tidal heating, we evaluated the state of GJ\,12\,b's interior using the open-source code \texttt{melt}\footnote{\url{https://github.com/cpiaulet/melt}}.  The energy balance model implemented in \texttt{melt} is described in detail in \citet{Peterson2023} and based on theoretical frameworks developed to study Io \citep{Fischer1990,Moore2003,Henning2009,Dobos2015,Barr2018}, and we summarize it briefly here.  The model calculates both the tidal energy dissipation flux $F_\mathrm{tidal}$ caused by the forced eccentricity of the planet \citep{Segatz1988}, as well as the convective flux $F_\mathrm{conv}$ describing the energy transport towards the surface \citep{Solomatov2000,Barr2008,Barr2018} over a range of mantle temperatures from 1400 to 1800 K.  For the tidal flux, we followed the Maxwell model of viscoelastic rheology for the estimate of the Love number \citep{Fischer1990,Moore2003,Henning2009,Dobos2015,Barr2018}. The shear modulus and viscosity followed the prescriptions described in previous work (\citealp{Peterson2023}; see also \citealp{Fischer1990,Henning2009,Renner2000,Moore2003}).  At temperatures beyond the solidus (set at $T_s = 1600$\,K), the rock viscosity has an exponential dependence on the melt fraction ($\propto \exp(-Bf)$) with the melt fraction coefficient experimentally constrained between 10 and 40. We therefore considered models over this full range of $B$ values to account for this uncertainty in material properties.  

We provided as inputs to \texttt{melt} the median planet mass, orbital period, and eccentricity for the eccentric model in Table~\ref{tab:fit_results}, and the median planet radius of $0.96 R_\oplus$ from \citetalias{Kuzuhara2024}.  

\begin{figure}
\centering
\includegraphics[width=0.5\textwidth]{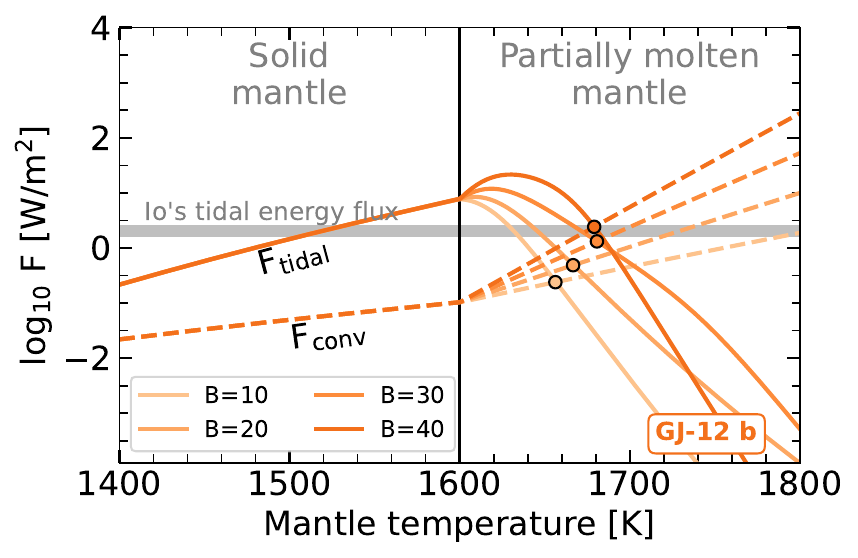}
\caption{\label{fig:energy_balance_tides} Illustration of the internal energy balance calculation with \texttt{melt} for GJ-12 b. We illustrate the calculations of both $F_\mathrm{conv}$ (dashed) and $F_\mathrm{tidal}$ (solid lines) as a function of mantle temperature, with equilibria  highlighted for each $B$ value with black circles. The constraints on tidal heating fluxes on Jupiter's moon Io are shown as a gray shaded region \citep{Veeder2012}.}
\end{figure}

Our \texttt{melt} results are shown in Figure~\ref{fig:energy_balance_tides}.  Over the full range of values for $B$, we see that the equilibria occur beyond the solidus, with stable mantle temperatures in the range 1660-1680\,K. These conditions are associated with a partially molten mantle, with inferred melt fractions between 14 and 20\%, corresponding to heat surface fluxes ranging from 0.2 to 2.4~W/m$^2$.  We note that our model ignores the contribution of the fluid molten rock to the tidal heat flux \citep{Farhat2025}, hence the values we quote can be considered to be lower limits (see also \citealp{Gkouvelis2025}).  These fluxes are about one order of magnitude larger than Earth's surface heat flux \citep{Davies2010} and are similar to Io's \citep{Veeder2012}.  Given the presence of extensive volcanism on the surface of Io, it is thus possible that GJ\,12\,b is undergoing surface outgassing as long as its orbit is eccentric.

Thus, we conclude that, while it is unlikely for GJ\,12\,b to have a thick primordial H/He atmosphere, it may have a secondary atmosphere due to tidally-induced volcanism as long as it has an eccentric orbit.  The detection of a substantial atmosphere with a high mean molecular weight on GJ\,12\,b could possibly be evidence of a secondary atmosphere formed via volcanism.

\subsubsection{A Water Layer?}

Next, we consider the planet's composition if its lower density is the result of a water/ice layer.   If GJ\,12\,b possesses a large water fraction, it could have formed beyond the water ice-line and then migrated inwards.  We can check to see if GJ\,12\,b has a density consistent with this formation scenario. 

In general, we expect that water-rich planets, if formed via pebble accretion \citep{Brugger2020}, would have a composition that is roughly 50\% ice given the composition of icy planetesimals beyond the water ice-line \citep{Marboef2014, Thiabaud2014}.  If formed via planetesimal accretion, the planet could have a composition between 0-50\% ice, given the slower formation of planetesimals over time \citep{Brugger2020}. 

Figure \ref{fig:GJ12_intcomp} shows that GJ\,12\,b is significantly too dense to be explained by a 50\% condensed water composition, even in the circular-orbit case.  We estimated limits to its internal water contents using \texttt{smint} \citep{Piaulet2021, Piaulet-Ghorayeb2024}'s implementation of the models for condensed-water planets from \cite{Zeng2016} and irradiated, water-rich planets from \cite{Aguichine2021}. 

\begin{figure}
    \centering
    \includegraphics[width=0.9\linewidth]{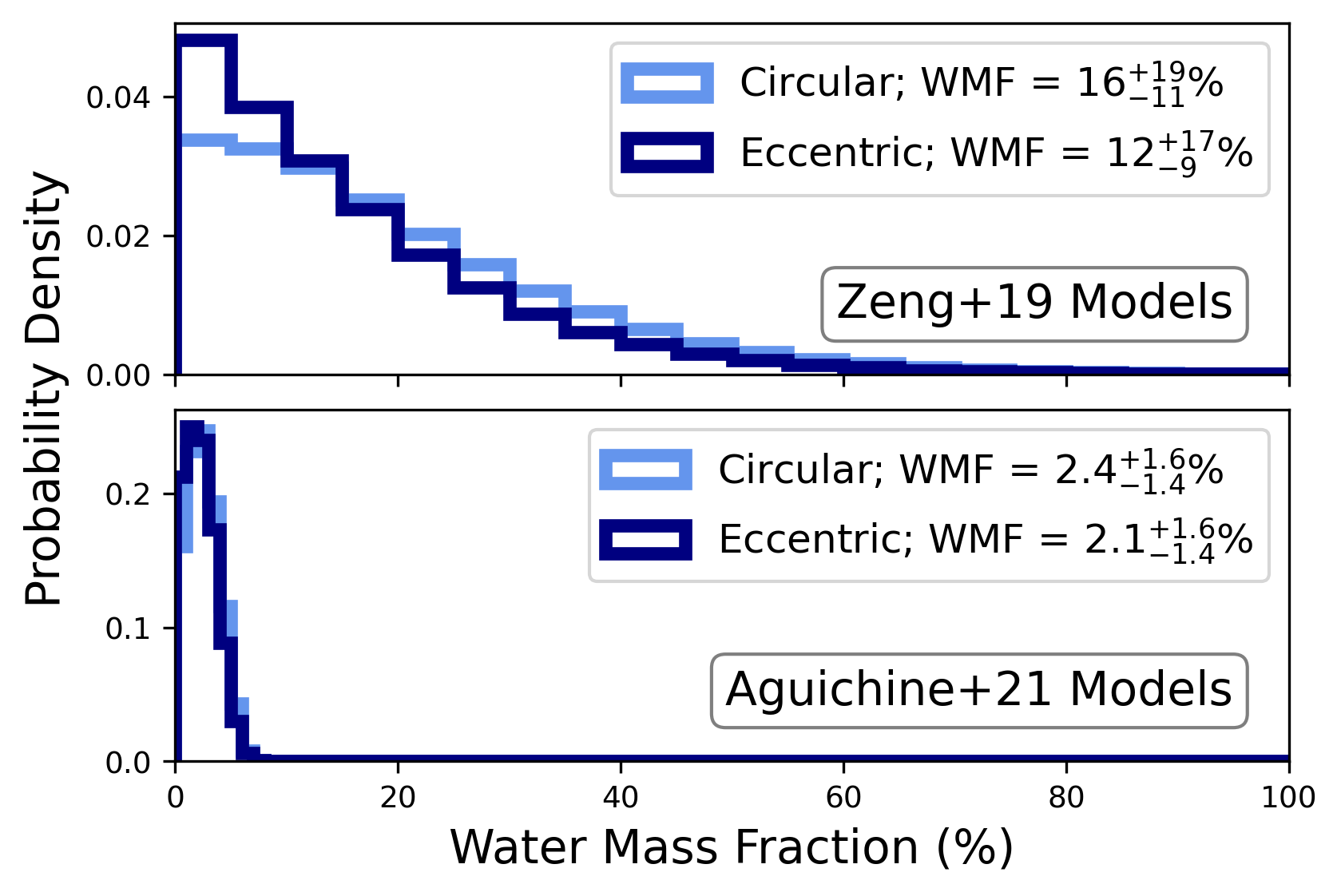}
    \caption{The inferred water contents of both our circular-orbit and eccentric-orbit GJ\,12\,b models using the \cite{Zeng2019} models (top) and \cite{Aguichine2021} models (bottom).}
    \label{fig:GJ12_watercontent}
\end{figure}

Our \texttt{smint} results are shown in Figure~\ref{fig:GJ12_watercontent}.  The \cite{Zeng2016} models, which assume a condensed water layer on top of a core with an Earth-like composition, find that GJ\,12\,b has a water mass fraction (WMF) of $12^{+17}_{-9}$\% in the eccentric case and $16^{+19}_{-11}$\% in the circular case.  In either scenario, a 50\% water composition is ruled out at a $>95\%$ level, meaning that this planet is unlikely to have been formed via by the pebble accretion scenario.  However, we note that these condensed models may not necessarily be reasonable given the planet's heavy irradiation.  When using the \cite{Aguichine2021} models, which considers a more physical equation of state for water and marginalizes over the core iron fraction, we instead recover a much lower water mass fraction, of around $2\,\pm\,2$\% in either scenario. Unfortunately, the \cite{Aguichine2021} model grid does not include any models with temperatures below 400\,K, and our planet temperature is 300\,K, so these results (which we generated assuming a temperature of 400\,K) may feature some systematic errors.  As GJ\,12\,b is slightly cooler, it might have a slightly higher water mass fraction than is represented by Figure~\ref{fig:GJ12_watercontent}.  However, these results still show that it is unlikely that GJ\,12\,b has a water mass fraction $>10\%$, and that it is possible that GJ\,12\,b has no water at all.

We note that studies such as \cite{Luo2024} have shown that the mass-radius relations typically used to describe planets with substantial bulk water contents do not properly account for the effects of water mixing with the planet's mantle and core, which can have a substantial influence on planet radius even at relatively low mass fractions.  In general, the mixing of water within the planet's interior will result in less dramatic increases in planet radius than water present entirely on the surface as a steam atmosphere, so it is possible that the upper limit of the volatile fraction of the planet is larger than described by our \texttt{smint} analysis.

\subsubsection{Rocky Composition}

\begin{figure}
    \centering
    \includegraphics[width=0.9\linewidth]{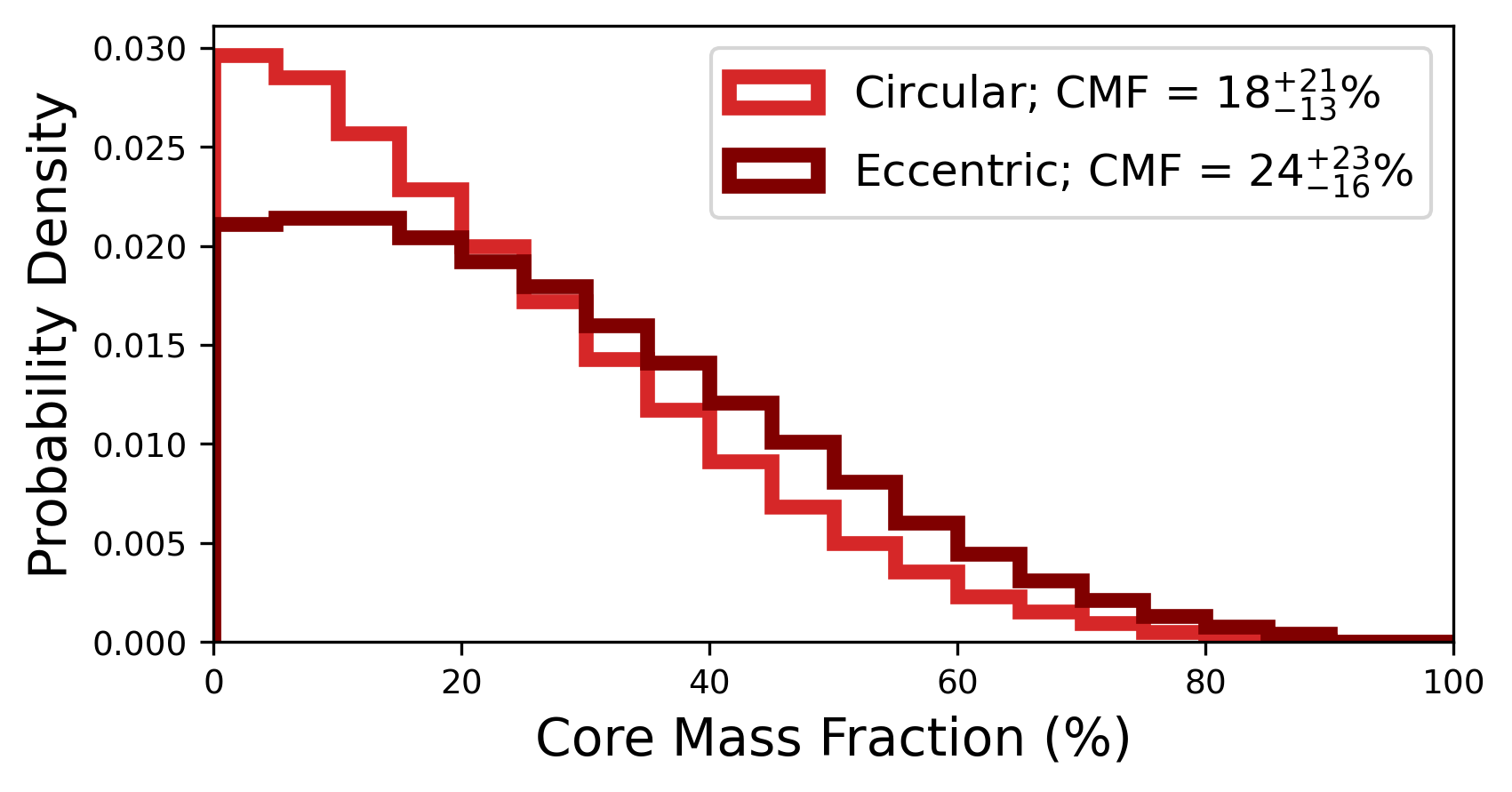}
    \caption{The inferred CMF of both our circular-orbit and eccentric-orbit GJ\,12\,b models using the \cite{Zeng2019} models, assuming the planet lacks both water and an atmosphere.}
    \label{fig:GJ12_ironcontent}
\end{figure}

We can also use \texttt{smint} to infer constraints on the core mass fraction (CMF) of GJ\,12\,b.  To do this, we will use the models from \cite{Zeng2016}, assuming that the planets have no atmospheres and no water.  We note that this will thus produce a lower limit on the CMF.  Our results are shown in Figure~\ref{fig:GJ12_ironcontent}.  The CMFs for the circular-orbit and eccentric-orbit models are 1$\sigma$ consistent with one another, though (as expected) the lower-mass circular-orbit model has a lower CMF.  Both models appear to have a somewhat sub-Earth CMF, but are still 1$\sigma$ consistent with Earth's CMF \citep[0.33, see][]{Szurgot2015}.

In addition, given their mass and radius errors, both models are also consistent with a 0\% iron composition if the planet is volatile- or atmosphere-free.  Thus, while we cannot rule out an Earth-like composition for GJ\,12\,b with our current dataset, it seems possible that the planet has a lower CMF than that of Earth.  A higher-precision mass for GJ\,12\,b will be necessary in order to definitively constrain its density.

While recent surveys have shown that the majority of terrestrial planets appear to have Earth-like compositions \citep{Luque2022, Brinkman2024}, rocky bodies with lower CMFs are certainly capable of being formed. The moon, with a CMF of less than 5\% \citep{Szurgot2015}, is an obvious example of such an object. The moon's low density has been explained as a result of a past collision with a proto-Earth \citep{Hartmann1975} that resulted in the formation of an iron-poor body.  It is possible that some similar mechanism occurred in the past of GJ\,12\,b.  An exoplanet that could potentially host a CMF $<0.1$ is TOI-561\,b \citep{Brinkman2023}, though its density could also be explained by an atmosphere.  

Another potential explanation for an iron-poor composition on GJ\,12\,b is related to its host star.  GJ\,12 has a somewhat sub-solar metallicity, with a metallicity of $-0.32\,\pm\,0.06$ quoted in \citetalias{Kuzuhara2024} and a metallicity of $-0.29\,\pm\,0.09$ quoted in \cite{Maldonado2020}.  K2-11\,b, which orbits a similarly metal-poor star, was found by \cite{Mortier2020} to have a low CMF ($13\%$) and the authors theorized that the lower iron abundance of the star could influence the final planet's composition.  Studies such as \cite{Adebekyan2021} have looked into the connection between stellar and planetary iron abundances.  However, more recent studies have shown that the relationship between stellar metallicities and planet CMFs is complex and likely heavily affected by the process of planet formation \citep{Brinkman2024b}. 

\subsection{Accessibility with \textit{JWST}}

Two useful metrics for identifying targets amenable to \textit{JWST} observations are the TSM (transmission spectroscopy metric) and the ESM (emission spectroscopy metric), both described in \cite{Kempton2018}.  In general, \cite{Kempton2018} describes planets with a TSM\,$>$\,12 as being amenable to \textit{JWST} characterization in transmission.  Given our calculated mass, we find that GJ\,12\,b has a TSM of $20\,\pm\,5$ ($22\,\pm\,5$ in the circular case), making it highly observable.  As a point of reference, TRAPPIST-1\,e, which falls at a similar location on the cosmic shoreline (see Figure~\ref{fig:shoreline}), has a comparable TSM of $20\,\pm\,1$.  Thus, GJ\,12\,b not only orbits a less active host star than the TRAPPIST-1 planets, it also has a similar TSM.  This, combined with its low temperature, makes it an ideal target for follow-up.

In addition, our current mass of GJ\,12\,b is at a 17\% precision, which is sufficient for transmission spectroscopy of small planets \citep[as described by][]{Batalha2019}.  A precise mass supports transmission observations, as a planet's surface gravity is degenerate with the atmospheric composition when it comes to determining the atmosphere's scale height.  While there are some ambiguities between the circular-orbit model and the eccentric-orbit model, the masses are still within 1$\sigma$ of one another and thus our model is sufficient for planning upcoming observations.  Given the high-priority nature of this target, we thus recommend planning for \textit{JWST} transmission spectroscopy on this target in the upcoming cycle and collecting additional RVs in the meantime to allow for enhanced insight into the planet's internal composition, as well as constraining the planet's eccentricity.

While GJ\,12\,b is on the side of the cosmic shoreline that implies a lack of atmosphere, the current location of the cosmic shoreline is largely the result of extrapolation from solar system objects, which all orbit a star that is very different (in terms of mass, temperature, and activity lifetime) from the host stars of the planets typically accessible with \textit{JWST} observations. It is thus impossible to conclude definitively that the GJ\,12\,b \textit{does not} have an atmosphere, even though it sits to the left of the cosmic shoreline. In addition, its low equilibrium temperature ($T_{\rm eq}\,\approx\,300$\,K) places it outside the typical temperature range for atmospheric haze formation \citep{Morley2015, Brande2024}, which has severely complicated attempts at observing the atmospheres of other planets in transmission.   The system's longer rotation period and low flare rate also make it an attractive target for follow-up \citepalias{Dholakia2024, Kuzuhara2024}.  GJ\,12\,b thus may be one of the best rocky targets for future \textit{JWST} transmission spectroscopy.

Unfortunately, however, with an ESM$\,<\,1$, GJ\,12\,b is unsuitable for observations in emission.  Given our low precision on the planet's eccentricity, the planet would be challenging to observe in emission anyway, as a planet's secondary eclipse timing is heavily influenced by its eccentricity.  Using the equation for secondary eclipse timing from \cite{Winn2010}, we find that our model's secondary eclipse timing has a standard deviation of 8.5\,hours, complicating efforts at scheduling observations.

\section{Summary and Conclusions}
\label{sec:conclusions}

GJ\,12 is a nearby M4 dwarf with a recently-discovered transiting Earth-sized companion.  The planet's long orbital period ($\approx\,13$d) means that it likely has a temperate ($\approx\,300$\,K) surface temperature.  Given its inactive host star, GJ\,12\,b may be one of the most promising targets for cool rocky planet studies with \textit{JWST}.

In this paper, we observed the system with MAROON-X over forty times in 2024 to measure the planet's mass.  We found evidence of additional Keplerian signals in our RV data at 39d and 58d.  However, given the uncertain rotation period of the host star, one or the other of these signals may be a product of stellar activity and we thus do not claim them as additional planets.  While we were not able to independently confirm the presence of the 12.8d transiting planet with RV data alone, we were able to us our data in combination with available data from other instruments to constrain its mass at the $5\sigma$ level, finding it to be $0.71\,\pm\,0.12$\,\MEarth.  We also found moderate evidence that the planet has a heightened eccentricity, which is feasible given its long eccentricity circularization timescale.  However, we currently only have moderate evidence, so we emphasize that future observations of the system may be necessary to properly constrain the planet's eccentricity.  We also note that our inferred model is influenced by our inclusion or exclusion of MAROON-X data collected during a time that lacked high-quality calibrations, further underscoring the need for further observations.

Regardless of the model chosen, GJ\,12\,b  appears to possess an Earth-like or sub-Earth density.  A low density could be explained if the planet has a volatile component (such as water mixed within the mantle), an atmosphere, a low iron mass fraction, or some combination of these factors.  Its low mass (and thus low surface gravity) may result in it having a diffuse atmosphere that would be highly amenable to \textit{JWST} observations, but it is also possible that the planet has too low a mass to retain an atmosphere long-term unless it is replenished regularly by tidally-induced volcanism.

\vskip 5.8mm plus 1mm minus 1mm
\vskip1sp

The University of Chicago group acknowledges funding for the MAROON-X project from the David and Lucile Packard Foundation, the Heising-Simons Foundation, the Gordon and Betty Moore Foundation, the Gemini Observatory, the NSF (award number 2108465), and NASA (grant number 80NSSC22K0117). The Gemini observations are associated with programs GN-2024A-FT-105 and GN-2024B-Q-124.  We also made use of MAROON-X calibration observations of HD\,3651, which are a part of programs GN-2024A-CAL-201 and GN-2024B-CAL-201.

Support for this work was provided by NASA through the NASA Hubble Fellowship grant \#HST-HF2-51559.001-A awarded by the Space Telescope Science Institute, which is operated by the Association of Universities for Research in Astronomy, Inc., for NASA, under contract NAS5-26555. C.P.-G. acknowledges support from the E. Margaret Burbidge Prize Postdoctoral Fellowship from the Brinson Foundation.



This research has also made use of NASA's Astrophysics Data System Bibliographic Services.

\software{Astropy \citep{Astropy1, Astropy2, Astropy3}, batman \citep{batman}, celerite \citep{celerite}, dynesty \citep{Skilling2004, Skilling2006, Speagle2020, Koposov24}, emcee \citep{emcee}, juliet \citep{Espinoza19}, lightkurve \citep{lightkurve}, Matplotlib \citep{matplotlib}, melt, Numpy \citep{numpy}, PyAstronomy \citep{PyAstronomy}, radvel \cite{Fulton2018}, REBOUND \citep{rebound}, scipy \citep{2020SciPy-NMeth}, smint \citep{Piaulet2021, Piaulet-Ghorayeb2024}}

\facility {Exoplanet Archive, Gemini-N (MAROON-X)}

\bibliography{manuscript}

\begin{thebibliography}{}
\expandafter\ifx\csname natexlab\endcsname\relax\def\natexlab#1{#1}\fi
\providecommand{\url}[1]{\href{#1}{#1}}
\providecommand{\dodoi}[1]{doi:~\href{http://doi.org/#1}{\nolinkurl{#1}}}
\providecommand{\doeprint}[1]{\href{http://ascl.net/#1}{\nolinkurl{http://ascl.net/#1}}}
\providecommand{\doarXiv}[1]{\href{https://arxiv.org/abs/#1}{\nolinkurl{https://arxiv.org/abs/#1}}}

\bibitem[{V. {Adibekyan} {et~al.}(2021){Adibekyan}, {Dorn}, {Sousa}, {Santos}, {Bitsch}, {Israelian}, {Mordasini}, {Barros}, {Delgado Mena}, {Demangeon}, {Faria}, {Figueira}, {Hakobyan}, {Oshagh}, {Soares}, {Kunitomo}, {Takeda}, {Jofr{\'e}}, {Petrucci}, \& {Martioli}}]{Adebekyan2021}
{Adibekyan}, V., {Dorn}, C., {Sousa}, S.~G., {et~al.} 2021, \bibinfo{title}{{A compositional link between rocky exoplanets and their host stars},} Science, 374, 330, \dodoi{10.1126/science.abg8794}

\bibitem[{A. {Aguichine} {et~al.}(2021){Aguichine}, {Mousis}, {Deleuil}, \& {Marcq}}]{Aguichine2021}
{Aguichine}, A., {Mousis}, O., {Deleuil}, M., \& {Marcq}, E. 2021, \bibinfo{title}{{Mass-Radius Relationships for Irradiated Ocean Planets},} \apj, 914, 84, \dodoi{10.3847/1538-4357/abfa99}

\bibitem[{E.-M. {Ahrer} {et~al.}(2025){Ahrer}, {Radica}, {Piaulet-Ghorayeb}, {Raul}, {Wiser}, {Welbanks}, {Acu{\~n}a}, {Allart}, {Coulombe}, {Louca}, {MacDonald}, {Saidel}, {Evans-Soma}, {Benneke}, {Christie}, {Beatty}, {Cadieux}, {Cloutier}, {Doyon}, {Fortney}, {Gagnebin}, {Gapp}, {Innes}, {Knutson}, {Komacek}, {Krissansen-Totton}, {Miguel}, {Pierrehumbert}, {Roy}, \& {Schlichting}}]{Ahrer2025}
{Ahrer}, E.-M., {Radica}, M., {Piaulet-Ghorayeb}, C., {et~al.} 2025, \bibinfo{title}{{Escaping Helium and a Highly Muted Spectrum Suggest a Metal-enriched Atmosphere on Sub-Neptune GJ 3090 b from JWST Transit Spectroscopy},} \apjl, 985, L10, \dodoi{10.3847/2041-8213/add010}

\bibitem[{{\'E}. {Artigau} {et~al.}(2022){Artigau}, {Cadieux}, {Cook}, {Doyon}, {Vandal}, {Donati}, {Moutou}, {Delfosse}, {Fouqu{\'e}}, {Martioli}, {Bouchy}, {Parsons}, {Carmona}, {Dumusque}, {Astudillo-Defru}, {Bonfils}, \& {Mignon}}]{Artigau2022}
{Artigau}, {\'E}., {Cadieux}, C., {Cook}, N.~J., {et~al.} 2022, \bibinfo{title}{{Line-by-line Velocity Measurements: an Outlier-resistant Method for Precision Velocimetry},} \aj, 164, 84, \dodoi{10.3847/1538-3881/ac7ce6}

\bibitem[{ {Astropy Collaboration} {et~al.}(2013){Astropy Collaboration}, {Robitaille}, {Tollerud}, {Greenfield}, {Droettboom}, {Bray}, {Aldcroft}, {Davis}, {Ginsburg}, {Price-Whelan}, {Kerzendorf}, {Conley}, {Crighton}, {Barbary}, {Muna}, {Ferguson}, {Grollier}, {Parikh}, {Nair}, {Unther}, {Deil}, {Woillez}, {Conseil}, {Kramer}, {Turner}, {Singer}, {Fox}, {Weaver}, {Zabalza}, {Edwards}, {Azalee Bostroem}, {Burke}, {Casey}, {Crawford}, {Dencheva}, {Ely}, {Jenness}, {Labrie}, {Lim}, {Pierfederici}, {Pontzen}, {Ptak}, {Refsdal}, {Servillat}, \& {Streicher}}]{Astropy1}
{Astropy Collaboration}, {Robitaille}, T.~P., {Tollerud}, E.~J., {et~al.} 2013, \bibinfo{title}{{Astropy: A community Python package for astronomy},} \aap, 558, A33, \dodoi{10.1051/0004-6361/201322068}

\bibitem[{ {Astropy Collaboration} {et~al.}(2018){Astropy Collaboration}, {Price-Whelan}, {Sip{\H{o}}cz}, {G{\"u}nther}, {Lim}, {Crawford}, {Conseil}, {Shupe}, {Craig}, {Dencheva}, {Ginsburg}, {VanderPlas}, {Bradley}, {P{\'e}rez-Su{\'a}rez}, {de Val-Borro}, {Aldcroft}, {Cruz}, {Robitaille}, {Tollerud}, {Ardelean}, {Babej}, {Bach}, {Bachetti}, {Bakanov}, {Bamford}, {Barentsen}, {Barmby}, {Baumbach}, {Berry}, {Biscani}, {Boquien}, {Bostroem}, {Bouma}, {Brammer}, {Bray}, {Breytenbach}, {Buddelmeijer}, {Burke}, {Calderone}, {Cano Rodr{\'\i}guez}, {Cara}, {Cardoso}, {Cheedella}, {Copin}, {Corrales}, {Crichton}, {D'Avella}, {Deil}, {Depagne}, {Dietrich}, {Donath}, {Droettboom}, {Earl}, {Erben}, {Fabbro}, {Ferreira}, {Finethy}, {Fox}, {Garrison}, {Gibbons}, {Goldstein}, {Gommers}, {Greco}, {Greenfield}, {Groener}, {Grollier}, {Hagen}, {Hirst}, {Homeier}, {Horton}, {Hosseinzadeh}, {Hu}, {Hunkeler}, {Ivezi{\'c}}, {Jain}, {Jenness}, {Kanarek}, {Kendrew}, {Kern}, {Kerzendorf}, {Khvalko}, {King}, {Kirkby}, {Kulkarni},
  {Kumar}, {Lee}, {Lenz}, {Littlefair}, {Ma}, {Macleod}, {Mastropietro}, {McCully}, {Montagnac}, {Morris}, {Mueller}, {Mumford}, {Muna}, {Murphy}, {Nelson}, {Nguyen}, {Ninan}, {N{\"o}the}, {Ogaz}, {Oh}, {Parejko}, {Parley}, {Pascual}, {Patil}, {Patil}, {Plunkett}, {Prochaska}, {Rastogi}, {Reddy Janga}, {Sabater}, {Sakurikar}, {Seifert}, {Sherbert}, {Sherwood-Taylor}, {Shih}, {Sick}, {Silbiger}, {Singanamalla}, {Singer}, {Sladen}, {Sooley}, {Sornarajah}, {Streicher}, {Teuben}, {Thomas}, {Tremblay}, {Turner}, {Terr{\'o}n}, {van Kerkwijk}, {de la Vega}, {Watkins}, {Weaver}, {Whitmore}, {Woillez}, {Zabalza}, \& {Astropy Contributors}}]{Astropy2}
{Astropy Collaboration}, {Price-Whelan}, A.~M., {Sip{\H{o}}cz}, B.~M., {et~al.} 2018, \bibinfo{title}{{The Astropy Project: Building an Open-science Project and Status of the v2.0 Core Package},} \aj, 156, 123, \dodoi{10.3847/1538-3881/aabc4f}

\bibitem[{ {Astropy Collaboration} {et~al.}(2022){Astropy Collaboration}, {Price-Whelan}, {Lim}, {Earl}, {Starkman}, {Bradley}, {Shupe}, {Patil}, {Corrales}, {Brasseur}, {N{\"o}the}, {Donath}, {Tollerud}, {Morris}, {Ginsburg}, {Vaher}, {Weaver}, {Tocknell}, {Jamieson}, {van Kerkwijk}, {Robitaille}, {Merry}, {Bachetti}, {G{\"u}nther}, {Aldcroft}, {Alvarado-Montes}, {Archibald}, {B{\'o}di}, {Bapat}, {Barentsen}, {Baz{\'a}n}, {Biswas}, {Boquien}, {Burke}, {Cara}, {Cara}, {Conroy}, {Conseil}, {Craig}, {Cross}, {Cruz}, {D'Eugenio}, {Dencheva}, {Devillepoix}, {Dietrich}, {Eigenbrot}, {Erben}, {Ferreira}, {Foreman-Mackey}, {Fox}, {Freij}, {Garg}, {Geda}, {Glattly}, {Gondhalekar}, {Gordon}, {Grant}, {Greenfield}, {Groener}, {Guest}, {Gurovich}, {Handberg}, {Hart}, {Hatfield-Dodds}, {Homeier}, {Hosseinzadeh}, {Jenness}, {Jones}, {Joseph}, {Kalmbach}, {Karamehmetoglu}, {Ka{\l}uszy{\'n}ski}, {Kelley}, {Kern}, {Kerzendorf}, {Koch}, {Kulumani}, {Lee}, {Ly}, {Ma}, {MacBride}, {Maljaars}, {Muna}, {Murphy}, {Norman},
  {O'Steen}, {Oman}, {Pacifici}, {Pascual}, {Pascual-Granado}, {Patil}, {Perren}, {Pickering}, {Rastogi}, {Roulston}, {Ryan}, {Rykoff}, {Sabater}, {Sakurikar}, {Salgado}, {Sanghi}, {Saunders}, {Savchenko}, {Schwardt}, {Seifert-Eckert}, {Shih}, {Jain}, {Shukla}, {Sick}, {Simpson}, {Singanamalla}, {Singer}, {Singhal}, {Sinha}, {Sip{\H{o}}cz}, {Spitler}, {Stansby}, {Streicher}, {{\v{S}}umak}, {Swinbank}, {Taranu}, {Tewary}, {Tremblay}, {de Val-Borro}, {Van Kooten}, {Vasovi{\'c}}, {Verma}, {de Miranda Cardoso}, {Williams}, {Wilson}, {Winkel}, {Wood-Vasey}, {Xue}, {Yoachim}, {Zhang}, {Zonca}, \& {Astropy Project Contributors}}]{Astropy3}
{Astropy Collaboration}, {Price-Whelan}, A.~M., {Lim}, P.~L., {et~al.} 2022, \bibinfo{title}{{The Astropy Project: Sustaining and Growing a Community-oriented Open-source Project and the Latest Major Release (v5.0) of the Core Package},} \apj, 935, 167, \dodoi{10.3847/1538-4357/ac7c74}

\bibitem[{N. {Astudillo-Defru} {et~al.}(2017){Astudillo-Defru}, {Delfosse}, {Bonfils}, {Forveille}, {Lovis}, \& {Rameau}}]{Astudillo-Defru}
{Astudillo-Defru}, N., {Delfosse}, X., {Bonfils}, X., {et~al.} 2017, \bibinfo{title}{{Magnetic activity in the HARPS M dwarf sample. The rotation-activity relationship for very low-mass stars through R'$_{HK}$},} \aap, 600, A13, \dodoi{10.1051/0004-6361/201527078}

\bibitem[{A.~C. Barr(2008)Barr}]{Barr2008}
Barr, A.~C. 2008, \bibinfo{title}{Mobile lid convection beneath {Enceladus}' south polar terrain,} Journal of Geophysical Research (Planets), 113, E07009, \dodoi{10.1029/2008JE003114}

\bibitem[{A.~C. Barr {et~al.}(2018)Barr, Dobos, \& Kiss}]{Barr2018}
Barr, A.~C., Dobos, V., \& Kiss, L.~L. 2018, \bibinfo{title}{Interior structures and tidal heating in the {TRAPPIST}-1 planets,} Astronomy and Astrophysics, 613, A37, \dodoi{10.1051/0004-6361/201731992}

\bibitem[{R. {Basant} {et~al.}(2025){Basant}, {Das}, {Bean}, {Luque}, {Seifahrt}, {Brady}, {Brown}, {St{\"u}rmer}, {Kasper}, \& {Stef{\'a}nsson}}]{Basant2025}
{Basant}, R., {Das}, T., {Bean}, J.~L., {et~al.} 2025, \bibinfo{title}{{Calibrating the Instrumental Drift in MAROON-X Using an Ensemble Analysis},} \aj, 169, 253, \dodoi{10.3847/1538-3881/adba48}

\bibitem[{N.~E. Batalha {et~al.}(2019)Batalha, Lewis, Fortney, Batalha, Kempton, Lewis, \& Line}]{Batalha2019}
Batalha, N.~E., Lewis, T., Fortney, J.~J., {et~al.} 2019, \bibinfo{title}{The Precision of Mass Measurements Required for Robust Atmospheric Characterization of Transiting Exoplanets,} \apjl, 885, L25, \dodoi{10.3847/2041-8213/ab4909}

\bibitem[{J.~L. {Bean} {et~al.}(2010){Bean}, {Seifahrt}, {Hartman}, {Nilsson}, {Wiedemann}, {Reiners}, {Dreizler}, \& {Henry}}]{Bean10}
{Bean}, J.~L., {Seifahrt}, A., {Hartman}, H., {et~al.} 2010, \bibinfo{title}{{The CRIRES Search for Planets Around the Lowest-mass Stars. I. High-precision Near-infrared Radial Velocities with an Ammonia Gas Cell},} \apj, 713, 410, \dodoi{10.1088/0004-637X/713/1/410}

\bibitem[{Z.~K. {Berta} {et~al.}(2012){Berta}, {Irwin}, {Charbonneau}, {Burke}, \& {Falco}}]{Berta2012}
{Berta}, Z.~K., {Irwin}, J., {Charbonneau}, D., {Burke}, C.~J., \& {Falco}, E.~E. 2012, \bibinfo{title}{{Transit Detection in the MEarth Survey of Nearby M Dwarfs: Bridging the Clean-first, Search-later Divide},} \aj, 144, 145, \dodoi{10.1088/0004-6256/144/5/145}

\bibitem[{J. {Brande} {et~al.}(2024){Brande}, {Crossfield}, {Kreidberg}, {Morley}, {Barman}, {Benneke}, {Christiansen}, {Dragomir}, {Fortney}, {Greene}, {Hardegree-Ullman}, {Howard}, {Knutson}, {Lothringer}, \& {Mikal-Evans}}]{Brande2024}
{Brande}, J., {Crossfield}, I. J.~M., {Kreidberg}, L., {et~al.} 2024, \bibinfo{title}{{Clouds and Clarity: Revisiting Atmospheric Feature Trends in Neptune-size Exoplanets},} \apjl, 961, L23, \dodoi{10.3847/2041-8213/ad1b5c}

\bibitem[{J.~M. {Brewer} {et~al.}(2020){Brewer}, {Fischer}, {Blackman}, {Cabot}, {Davis}, {Laughlin}, {Leet}, {Ong}, {Petersburg}, {Szymkowiak}, {Zhao}, {Henry}, \& {Llama}}]{Brewer2020}
{Brewer}, J.~M., {Fischer}, D.~A., {Blackman}, R.~T., {et~al.} 2020, \bibinfo{title}{{EXPRES. I. HD 3651 as an Ideal RV Benchmark},} \aj, 160, 67, \dodoi{10.3847/1538-3881/ab99c9}

\bibitem[{C.~L. {Brinkman} {et~al.}(2024{\natexlab{a}}){Brinkman}, {Polanski}, {Huber}, {Weiss}, {Valencia}, \& {Plotnykov}}]{Brinkman2024b}
{Brinkman}, C.~L., {Polanski}, A.~S., {Huber}, D., {et~al.} 2024{\natexlab{a}}, \bibinfo{title}{{Revisiting the Relationship Between Rocky Exoplanet and Stellar Compositions: Reduced Evidence for a Super-Mercury Population},} \aj, 168, 281, \dodoi{10.3847/1538-3881/ad82eb}

\bibitem[{C.~L. {Brinkman} {et~al.}(2023){Brinkman}, {Weiss}, {Dai}, {Huber}, {Kite}, {Valencia}, {Bean}, {Beard}, {Behmard}, {Blunt}, {Brady}, {Fulton}, {Giacalone}, {Howard}, {Isaacson}, {Kasper}, {Lubin}, {MacDougall}, {Akana Murphy}, {Plotnykov}, {Polanski}, {Rice}, {Seifahrt}, {Stef{\'a}nsson}, \& {St{\"u}rmer}}]{Brinkman2023}
{Brinkman}, C.~L., {Weiss}, L.~M., {Dai}, F., {et~al.} 2023, \bibinfo{title}{{TOI-561 b: A Low-density Ultra-short-period ``Rocky'' Planet around a Metal-poor Star},} \aj, 165, 88, \dodoi{10.3847/1538-3881/acad83}

\bibitem[{C.~L. {Brinkman} {et~al.}(2024{\natexlab{b}}){Brinkman}, {Weiss}, {Huber}, {Lee}, {Kolecki}, {Tenn}, {Zhang}, {Narayanan}, {Polanski}, {Dai}, {Bean}, {Beard}, {Brady}, {Brodheim}, {Brown}, {Deich}, {Edelstein}, {Fulton}, {Giacalone}, {Gibson}, {Gilbert}, {Halverson}, {Handley}, {Hill}, {Holcomb}, {Holden}, {Householder}, {Howard}, {Isaacson}, {Kaye}, {Laher}, {Lanclos}, {Ong}, {Payne}, {Petigura}, {Pidhorodetska}, {Poppett}, {Roy}, {Rubenzahl}, {Saunders}, {Schwab}, {Seifahrt}, {Shaum}, {Sirk}, {Smith}, {Smith}, {Stef{\'a}nsson}, {St{\"u}rmer}, {Thorne}, {Turtelboom}, {Tyler}, {Valliant}, {Van Zandt}, {Walawender}, {Yee}, {Yeh}, \& {Zink}}]{Brinkman2024}
{Brinkman}, C.~L., {Weiss}, L.~M., {Huber}, D., {et~al.} 2024{\natexlab{b}}, \bibinfo{title}{{The Compositions of Rocky Planets in Close-in Orbits Tend to be Earth-Like},} arXiv e-prints, arXiv:2410.00213, \dodoi{10.48550/arXiv.2410.00213}

\bibitem[{N. {Br{\"u}gger} {et~al.}(2020){Br{\"u}gger}, {Burn}, {Coleman}, {Alibert}, \& {Benz}}]{Brugger2020}
{Br{\"u}gger}, N., {Burn}, R., {Coleman}, G.~A.~L., {Alibert}, Y., \& {Benz}, W. 2020, \bibinfo{title}{{Pebbles versus planetesimals. The outcomes of population synthesis models},} \aap, 640, A21, \dodoi{10.1051/0004-6361/202038042}

\bibitem[{C.~I. {Ca{\~n}as} {et~al.}(2025){Ca{\~n}as}, {Lustig-Yaeger}, {Tsai}, {M{\"u}ller}, {Helled}, {Louie}, {Guzm{\'a}n Caloca}, {Kanodia}, {Gao}, {Libby-Roberts}, {Hardegree-Ullman}, {Col{\'o}n}, {Czekala}, {Delamer}, {Han}, {Lin}, {Mahadevan}, {May}, {Ninan}, {Piette}, {Stef{\'a}nsson}, {Stevenson}, {Teske}, \& {Wallack}}]{Canas2025}
{Ca{\~n}as}, C.~I., {Lustig-Yaeger}, J., {Tsai}, S.-M., {et~al.} 2025, \bibinfo{title}{{GEMS JWST: Transmission spectroscopy of TOI-5205b reveals significant stellar contamination and a metal-poor atmosphere},} arXiv e-prints, arXiv:2502.06966, \dodoi{10.48550/arXiv.2502.06966}

\bibitem[{A.~B. {Claringbold} {et~al.}(2023){Claringbold}, {Rimmer}, {Rugheimer}, \& {Shorttle}}]{Claringbold2023}
{Claringbold}, A.~B., {Rimmer}, P.~B., {Rugheimer}, S., \& {Shorttle}, O. 2023, \bibinfo{title}{{Prebiosignature Molecules Can Be Detected in Temperate Exoplanet Atmospheres with JWST},} \aj, 166, 39, \dodoi{10.3847/1538-3881/acdacc}

\bibitem[{R. {Cosentino} {et~al.}(2012){Cosentino}, {Lovis}, {Pepe}, {Collier Cameron}, {Latham}, {Molinari}, {Udry}, {Bezawada}, {Black}, {Born}, {Buchschacher}, {Charbonneau}, {Figueira}, {Fleury}, {Galli}, {Gallie}, {Gao}, {Ghedina}, {Gonzalez}, {Gonzalez}, {Guerra}, {Henry}, {Horne}, {Hughes}, {Kelly}, {Lodi}, {Lunney}, {Maire}, {Mayor}, {Micela}, {Ordway}, {Peacock}, {Phillips}, {Piotto}, {Pollacco}, {Queloz}, {Rice}, {Riverol}, {Riverol}, {San Juan}, {Sasselov}, {Segransan}, {Sozzetti}, {Sosnowska}, {Stobie}, {Szentgyorgyi}, {Vick}, \& {Weber}}]{Cosentino2012}
{Cosentino}, R., {Lovis}, C., {Pepe}, F., {et~al.} 2012, in Society of Photo-Optical Instrumentation Engineers (SPIE) Conference Series, Vol. 8446, Ground-based and Airborne Instrumentation for Astronomy IV, ed. I.~S. {McLean}, S.~K. {Ramsay}, \& H.~{Takami}, 84461V, \dodoi{10.1117/12.925738}

\bibitem[{R.~M. {Cutri} {et~al.}(2003){Cutri}, {Skrutskie}, {van Dyk}, {Beichman}, {Carpenter}, {Chester}, {Cambresy}, {Evans}, {Fowler}, {Gizis}, {Howard}, {Huchra}, {Jarrett}, {Kopan}, {Kirkpatrick}, {Light}, {Marsh}, {McCallon}, {Schneider}, {Stiening}, {Sykes}, {Weinberg}, {Wheaton}, {Wheelock}, \& {Zacarias}}]{Cutri2003}
{Cutri}, R.~M., {Skrutskie}, M.~F., {van Dyk}, S., {et~al.} 2003, \bibinfo{title}{{VizieR Online Data Catalog: 2MASS All-Sky Catalog of Point Sources (Cutri+ 2003)},}, VizieR On-line Data Catalog: II/246. Originally published in: University of Massachusetts and Infrared Processing and Analysis Center, (IPAC/California Institute of Technology) (2003)

\bibitem[{S. {Czesla} {et~al.}(2019){Czesla}, {Schr{\"o}ter}, {Schneider}, {Huber}, {Pfeifer}, {Andreasen}, \& {Zechmeister}}]{PyAstronomy}
{Czesla}, S., {Schr{\"o}ter}, S., {Schneider}, C.~P., {et~al.} 2019, \bibinfo{title}{{PyA: Python astronomy-related packages},} \doeprint{1906.010}

\bibitem[{J.~H. Davies \& D.~R. Davies(2010)Davies \& Davies}]{Davies2010}
Davies, J.~H., \& Davies, D.~R. 2010, \bibinfo{title}{Earth's surface heat flux,} Solid Earth, 1, 5, \dodoi{10.5194/se-1-5-2010}

\bibitem[{S. {Dholakia} {et~al.}(2024){Dholakia}, {Palethorpe}, {Venner}, {Mortier}, {Wilson}, {Huang}, {Rice}, {Van Eylen}, {Nabbie}, {Cloutier}, {Boschin}, {Ciardi}, {Delrez}, {Dransfield}, {Ducrot}, {Essack}, {Everett}, {Gillon}, {Hooton}, {Kunimoto}, {Latham}, {L{\'o}pez-Morales}, {Li}, {Li}, {McDermott}, {Murphy}, {Murray}, {Seager}, {Timmermans}, {Triaud}, {Turner}, {Twicken}, {Vanderburg}, {Wang}, {Wittenmyer}, \& {Wright}}]{Dholakia2024}
{Dholakia}, S., {Palethorpe}, L., {Venner}, A., {et~al.} 2024, \bibinfo{title}{{Gliese 12 b, a temperate Earth-sized planet at 12 parsecs discovered with TESS and CHEOPS},} \mnras, 531, 1276, \dodoi{10.1093/mnras/stae1152}

\bibitem[{E.~S. {Dmitrienko} \& I.~S. {Savanov}(2018){Dmitrienko} \& {Savanov}}]{Dmitrienko2018}
{Dmitrienko}, E.~S., \& {Savanov}, I.~S. 2018, \bibinfo{title}{{Activity of the M8 Dwarf TRAPPIST-1},} Astronomy Reports, 62, 412, \dodoi{10.1134/S1063772918060033}

\bibitem[{I. {Dobbs-Dixon} {et~al.}(2004){Dobbs-Dixon}, {Lin}, \& {Mardling}}]{Dobbs-Dixon04}
{Dobbs-Dixon}, I., {Lin}, D.~N.~C., \& {Mardling}, R.~A. 2004, \bibinfo{title}{{Spin-Orbit Evolution of Short-Period Planets},} \apj, 610, 464, \dodoi{10.1086/421510}

\bibitem[{V. Dobos \& E.~L. Turner(2015)Dobos \& Turner}]{Dobos2015}
Dobos, V., \& Turner, E.~L. 2015, \bibinfo{title}{Viscoelastic {Models} of {Tidally} {Heated} {Exomoons},} The Astrophysical Journal, 804, 41, \dodoi{10.1088/0004-637X/804/1/41}

\bibitem[{S.~G. {Engle} \& E.~F. {Guinan}(2023){Engle} \& {Guinan}}]{Engle2023}
{Engle}, S.~G., \& {Guinan}, E.~F. 2023, \bibinfo{title}{{Living with a Red Dwarf: The Rotation-Age Relationships of M Dwarfs},} \apjl, 954, L50, \dodoi{10.3847/2041-8213/acf472}

\bibitem[{N. {Espinoza} {et~al.}(2019){Espinoza}, {Kossakowski}, \& {Brahm}}]{Espinoza19}
{Espinoza}, N., {Kossakowski}, D., \& {Brahm}, R. 2019, \bibinfo{title}{{juliet: a versatile modelling tool for transiting and non-transiting exoplanetary systems},} \mnras, 490, 2262, \dodoi{10.1093/mnras/stz2688}

\bibitem[{M. {Farhat} {et~al.}(2025){Farhat}, {Auclair-Desrotour}, {Bou{\'e}}, {Lichtenberg}, \& {Laskar}}]{Farhat2025}
{Farhat}, M., {Auclair-Desrotour}, P., {Bou{\'e}}, G., {Lichtenberg}, T., \& {Laskar}, J. 2025, \bibinfo{title}{{Tides on Lava Worlds: Application to Close-in Exoplanets and the Early Earth{\textendash}Moon System},} \apj, 979, 133, \dodoi{10.3847/1538-4357/ad9b93}

\bibitem[{G. {F{\H{u}}r{\'e}sz} {et~al.}(2008){F{\H{u}}r{\'e}sz}, {Szentgyorgyi}, \& {Meibom}}]{Furesz2008}
{F{\H{u}}r{\'e}sz}, G., {Szentgyorgyi}, A.~H., \& {Meibom}, S. 2008, in Precision Spectroscopy in Astrophysics, ed. N.~C. {Santos}, L.~{Pasquini}, A.~C.~M. {Correia}, \& M.~{Romaniello}, 287--290, \dodoi{10.1007/978-3-540-75485-5_68}

\bibitem[{H.-J. Fischer \& T. Spohn(1990)Fischer \& Spohn}]{Fischer1990}
Fischer, H.-J., \& Spohn, T. 1990, \bibinfo{title}{Thermal-orbital histories of viscoelastic models of {Io} ({J1}),} Icarus, 83, 39, \dodoi{10.1016/0019-1035(90)90005-T}

\bibitem[{D. {Foreman-Mackey} {et~al.}(2017){Foreman-Mackey}, {Agol}, {Angus}, \& {Ambikasaran}}]{celerite}
{Foreman-Mackey}, D., {Agol}, E., {Angus}, R., \& {Ambikasaran}, S. 2017, \bibinfo{title}{Fast and scalable Gaussian process modeling with applications to astronomical time series,} ArXiv.
\newblock \url{https://arxiv.org/abs/1703.09710}

\bibitem[{D. {Foreman-Mackey} {et~al.}(2013){Foreman-Mackey}, {Hogg}, {Lang}, \& {Goodman}}]{emcee}
{Foreman-Mackey}, D., {Hogg}, D.~W., {Lang}, D., \& {Goodman}, J. 2013, \bibinfo{title}{{emcee: The MCMC Hammer},} \pasp, 125, 306, \dodoi{10.1086/670067}

\bibitem[{B.~J. {Fulton} {et~al.}(2018){Fulton}, {Petigura}, {Blunt}, \& {Sinukoff}}]{Fulton2018}
{Fulton}, B.~J., {Petigura}, E.~A., {Blunt}, S., \& {Sinukoff}, E. 2018, \bibinfo{title}{{RadVel: The Radial Velocity Modeling Toolkit},} \pasp, 130, 044504, \dodoi{10.1088/1538-3873/aaaaa8}

\bibitem[{ {Gaia Collaboration} {et~al.}(2021){Gaia Collaboration}, {Brown}, {Vallenari}, {Prusti}, {de Bruijne}, {Babusiaux}, {Biermann}, {Creevey}, {Evans}, {Eyer}, {Hutton}, {Jansen}, {Jordi}, {Klioner}, {Lammers}, {Lindegren}, {Luri}, {Mignard}, {Panem}, {Pourbaix}, {Randich}, {Sartoretti}, {Soubiran}, {Walton}, {Arenou}, {Bailer-Jones}, {Bastian}, {Cropper}, {Drimmel}, {Katz}, {Lattanzi}, {van Leeuwen}, {Bakker}, {Cacciari}, {Casta{\~n}eda}, {De Angeli}, {Ducourant}, {Fabricius}, {Fouesneau}, {Fr{\'e}mat}, {Guerra}, {Guerrier}, {Guiraud}, {Jean-Antoine Piccolo}, {Masana}, {Messineo}, {Mowlavi}, {Nicolas}, {Nienartowicz}, {Pailler}, {Panuzzo}, {Riclet}, {Roux}, {Seabroke}, {Sordo}, {Tanga}, {Th{\'e}venin}, {Gracia-Abril}, {Portell}, {Teyssier}, {Altmann}, {Andrae}, {Bellas-Velidis}, {Benson}, {Berthier}, {Blomme}, {Brugaletta}, {Burgess}, {Busso}, {Carry}, {Cellino}, {Cheek}, {Clementini}, {Damerdji}, {Davidson}, {Delchambre}, {Dell'Oro}, {Fern{\'a}ndez-Hern{\'a}ndez}, {Galluccio}, {Garc{\'\i}a-Lario},
  {Garcia-Reinaldos}, {Gonz{\'a}lez-N{\'u}{\~n}ez}, {Gosset}, {Haigron}, {Halbwachs}, {Hambly}, {Harrison}, {Hatzidimitriou}, {Heiter}, {Hern{\'a}ndez}, {Hestroffer}, {Hodgkin}, {Holl}, {Jan{\ss}en}, {Jevardat de Fombelle}, {Jordan}, {Krone-Martins}, {Lanzafame}, {L{\"o}ffler}, {Lorca}, {Manteiga}, {Marchal}, {Marrese}, {Moitinho}, {Mora}, {Muinonen}, {Osborne}, {Pancino}, {Pauwels}, {Petit}, {Recio-Blanco}, {Richards}, {Riello}, {Rimoldini}, {Robin}, {Roegiers}, {Rybizki}, {Sarro}, {Siopis}, {Smith}, {Sozzetti}, {Ulla}, {Utrilla}, {van Leeuwen}, {van Reeven}, {Abbas}, {Abreu Aramburu}, {Accart}, {Aerts}, {Aguado}, {Ajaj}, {Altavilla}, {{\'A}lvarez}, {{\'A}lvarez Cid-Fuentes}, {Alves}, {Anderson}, {Anglada Varela}, {Antoja}, {Audard}, {Baines}, {Baker}, {Balaguer-N{\'u}{\~n}ez}, {Balbinot}, {Balog}, {Barache}, {Barbato}, {Barros}, {Barstow}, {Bartolom{\'e}}, {Bassilana}, {Bauchet}, {Baudesson-Stella}, {Becciani}, {Bellazzini}, {Bernet}, {Bertone}, {Bianchi}, {Blanco-Cuaresma}, {Boch}, {Bombrun}, {Bossini},
  {Bouquillon}, {Bragaglia}, {Bramante}, {Breedt}, {Bressan}, {Brouillet}, {Bucciarelli}, {Burlacu}, {Busonero}, {Butkevich}, {Buzzi}, {Caffau}, {Cancelliere}, {C{\'a}novas}, {Cantat-Gaudin}, {Carballo}, {Carlucci}, {Carnerero}, {Carrasco}, {Casamiquela}, {Castellani}, {Castro-Ginard}, {Castro Sampol}, {Chaoul}, {Charlot}, {Chemin}, {Chiavassa}, {Cioni}, {Comoretto}, {Cooper}, {Cornez}, {Cowell}, {Crifo}, {Crosta}, {Crowley}, {Dafonte}, {Dapergolas}, {David}, \& {David}}]{GaiaEDR3}
{Gaia Collaboration}, {Brown}, A.~G.~A., {Vallenari}, A., {et~al.} 2021, \bibinfo{title}{{Gaia Early Data Release 3. Summary of the contents and survey properties},} \aap, 649, A1, \dodoi{10.1051/0004-6361/202039657}

\bibitem[{M. {Gillon} {et~al.}(2017){Gillon}, {Triaud}, {Demory}, {Jehin}, {Agol}, {Deck}, {Lederer}, {de Wit}, {Burdanov}, {Ingalls}, {Bolmont}, {Leconte}, {Raymond}, {Selsis}, {Turbet}, {Barkaoui}, {Burgasser}, {Burleigh}, {Carey}, {Chaushev}, {Copperwheat}, {Delrez}, {Fernandes}, {Holdsworth}, {Kotze}, {Van Grootel}, {Almleaky}, {Benkhaldoun}, {Magain}, \& {Queloz}}]{Gillon2017}
{Gillon}, M., {Triaud}, A. H.~M.~J., {Demory}, B.-O., {et~al.} 2017, \bibinfo{title}{{Seven temperate terrestrial planets around the nearby ultracool dwarf star TRAPPIST-1},} \nat, 542, 456, \dodoi{10.1038/nature21360}

\bibitem[{L. {Gkouvelis} {et~al.}(2025){Gkouvelis}, {Pozuelos}, {Drant}, {Farhat}, {Tian}, \& {Ak{\i}n}}]{Gkouvelis2025}
{Gkouvelis}, L., {Pozuelos}, F.~J., {Drant}, T., {et~al.} 2025, \bibinfo{title}{{Interior redox state effects on the stability of secondary atmospheres and observational manifestations: LP 791-18 d as a case study for outgassing rocky exoplanets},} arXiv e-prints, arXiv:2506.02188, \dodoi{10.48550/arXiv.2506.02188}

\bibitem[{T.~P. {Greene} {et~al.}(2023){Greene}, {Bell}, {Ducrot}, {Dyrek}, {Lagage}, \& {Fortney}}]{Greene2023}
{Greene}, T.~P., {Bell}, T.~J., {Ducrot}, E., {et~al.} 2023, \bibinfo{title}{{Thermal emission from the Earth-sized exoplanet TRAPPIST-1 b using JWST},} \nat, 618, 39, \dodoi{10.1038/s41586-023-05951-7}

\bibitem[{N.~C. {Hara} {et~al.}(2017){Hara}, {Bou{\'e}}, {Laskar}, \& {Correia}}]{Hara2017}
{Hara}, N.~C., {Bou{\'e}}, G., {Laskar}, J., \& {Correia}, A.~C.~M. 2017, \bibinfo{title}{{Radial velocity data analysis with compressed sensing techniques},} \mnras, 464, 1220, \dodoi{10.1093/mnras/stw2261}

\bibitem[{C.~R. Harris {et~al.}(2020)Harris, Millman, van~der Walt, Gommers, Virtanen, Cournapeau, Wieser, Taylor, Berg, Smith, Kern, Picus, Hoyer, van Kerkwijk, Brett, Haldane, del R{\'{i}}o, Wiebe, Peterson, G{\'{e}}rard-Marchant, Sheppard, Reddy, Weckesser, Abbasi, Gohlke, \& Oliphant}]{numpy}
Harris, C.~R., Millman, K.~J., van~der Walt, S.~J., {et~al.} 2020, \bibinfo{title}{Array programming with {NumPy},} Nature, 585, 357, \dodoi{10.1038/s41586-020-2649-2}

\bibitem[{W.~K. {Hartmann} \& D.~R. {Davis}(1975){Hartmann} \& {Davis}}]{Hartmann1975}
{Hartmann}, W.~K., \& {Davis}, D.~R. 1975, \bibinfo{title}{{Satellite-Sized Planetesimals and Lunar Origin},} \icarus, 24, 504, \dodoi{10.1016/0019-1035(75)90070-6}

\bibitem[{W.~G. Henning {et~al.}(2009)Henning, O'Connell, \& Sasselov}]{Henning2009}
Henning, W.~G., O'Connell, R.~J., \& Sasselov, D.~D. 2009, \bibinfo{title}{Tidally {Heated} {Terrestrial} {Exoplanets}: {Viscoelastic} {Response} {Models},} The Astrophysical Journal, 707, 1000, \dodoi{10.1088/0004-637X/707/2/1000}

\bibitem[{W.~S. {Howard} {et~al.}(2023){Howard}, {Kowalski}, {Flagg}, {MacGregor}, {Lim}, {Radica}, {Piaulet}, {Roy}, {Lafreni{\`e}re}, {Benneke}, {Brown}, {Espinoza}, {Doyon}, {Coulombe}, {Johnstone}, {Cowan}, {Jayawardhana}, {Turner}, \& {Dang}}]{Howard2023}
{Howard}, W.~S., {Kowalski}, A.~F., {Flagg}, L., {et~al.} 2023, \bibinfo{title}{{Characterizing the Near-infrared Spectra of Flares from TRAPPIST-1 during JWST Transit Spectroscopy Observations},} \apj, 959, 64, \dodoi{10.3847/1538-4357/acfe75}

\bibitem[{J.~D. Hunter(2007)Hunter}]{matplotlib}
Hunter, J.~D. 2007, \bibinfo{title}{Matplotlib: A 2D graphics environment,} Computing in Science \& Engineering, 9, 90, \dodoi{10.1109/MCSE.2007.55}

\bibitem[{J. {Ih} {et~al.}(2023){Ih}, {Kempton}, {Whittaker}, \& {Lessard}}]{ih23}
{Ih}, J., {Kempton}, E. M.~R., {Whittaker}, E.~A., \& {Lessard}, M. 2023, \bibinfo{title}{{Constraining the Thickness of TRAPPIST-1 b's Atmosphere from Its JWST Secondary Eclipse Observation at 15 {\ensuremath{\mu}}m},} \apjl, 952, L4, \dodoi{10.3847/2041-8213/ace03b}

\bibitem[{J. {Irwin} {et~al.}(2011){Irwin}, {Berta}, {Burke}, {Charbonneau}, {Nutzman}, {West}, \& {Falco}}]{Irwin2011}
{Irwin}, J., {Berta}, Z.~K., {Burke}, C.~J., {et~al.} 2011, \bibinfo{title}{{On the Angular Momentum Evolution of Fully Convective Stars: Rotation Periods for Field M-dwarfs from the MEarth Transit Survey},} \apj, 727, 56, \dodoi{10.1088/0004-637X/727/1/56}

\bibitem[{X. {Ji} {et~al.}(2025){Ji}, {Chatterjee}, {Park Coy}, \& {Kite}}]{Ji2025}
{Ji}, X., {Chatterjee}, R.~D., {Park Coy}, B., \& {Kite}, E.~S. 2025, \bibinfo{title}{{The Cosmic Shoreline Revisited: A Metric for Atmospheric Retention Informed by Hydrodynamic Escape},} arXiv e-prints, arXiv:2504.19872, \dodoi{10.48550/arXiv.2504.19872}

\bibitem[{E.~M.~R. {Kempton} \& H.~A. {Knutson}(2024){Kempton} \& {Knutson}}]{kempton24}
{Kempton}, E. M.~R., \& {Knutson}, H.~A. 2024, \bibinfo{title}{{Transiting Exoplanet Atmospheres in the Era of JWST},} Reviews in Mineralogy and Geochemistry, 90, 411, \dodoi{10.2138/rmg.2024.90.12}

\bibitem[{E.~M.~R. {Kempton} {et~al.}(2018){Kempton}, {Bean}, {Louie}, {Deming}, {Koll}, {Mansfield}, {Christiansen}, {L{\'o}pez-Morales}, {Swain}, {Zellem}, {Ballard}, {Barclay}, {Barstow}, {Batalha}, {Beatty}, {Berta-Thompson}, {Birkby}, {Buchhave}, {Charbonneau}, {Cowan}, {Crossfield}, {de Val-Borro}, {Doyon}, {Dragomir}, {Gaidos}, {Heng}, {Hu}, {Kane}, {Kreidberg}, {Mallonn}, {Morley}, {Narita}, {Nascimbeni}, {Pall{\'e}}, {Quintana}, {Rauscher}, {Seager}, {Shkolnik}, {Sing}, {Sozzetti}, {Stassun}, {Valenti}, \& {von Essen}}]{Kempton2018}
{Kempton}, E. M.~R., {Bean}, J.~L., {Louie}, D.~R., {et~al.} 2018, \bibinfo{title}{{A Framework for Prioritizing the TESS Planetary Candidates Most Amenable to Atmospheric Characterization},} \pasp, 130, 114401, \dodoi{10.1088/1538-3873/aadf6f}

\bibitem[{J. {Kirk} {et~al.}(2024){Kirk}, {Stevenson}, {Fu}, {Lustig-Yaeger}, {Moran}, {Peacock}, {Alam}, {Batalha}, {Bennett}, {Gonzalez-Quiles}, {L{\'o}pez-Morales}, {Lothringer}, {MacDonald}, {May}, {Mayorga}, {Rustamkulov}, {Sing}, {Sotzen}, {Valenti}, \& {Wakeford}}]{Kirk2024}
{Kirk}, J., {Stevenson}, K.~B., {Fu}, G., {et~al.} 2024, \bibinfo{title}{{JWST/NIRCam Transmission Spectroscopy of the Nearby Sub-Earth GJ 341b},} \aj, 167, 90, \dodoi{10.3847/1538-3881/ad19df}

\bibitem[{C.~S. {Kochanek} {et~al.}(2017){Kochanek}, {Shappee}, {Stanek}, {Holoien}, {Thompson}, {Prieto}, {Dong}, {Shields}, {Will}, {Britt}, {Perzanowski}, \& {Pojma{\'n}ski}}]{Kochanek17}
{Kochanek}, C.~S., {Shappee}, B.~J., {Stanek}, K.~Z., {et~al.} 2017, \bibinfo{title}{{The All-Sky Automated Survey for Supernovae (ASAS-SN) Light Curve Server v1.0},} \pasp, 129, 104502, \dodoi{10.1088/1538-3873/aa80d9}

\bibitem[{S. Koposov {et~al.}(2024)Koposov, Speagle, Barbary, Ashton, Bennett, Buchner, Scheffler, Cook, Talbot, Guillochon, Cubillos, Ramos, Dartiailh, Ilya, Tollerud, Lang, Johnson, jtmendel, Higson, Vandal, Daylan, Angus, patelR, Cargile, Sheehan, Pitkin, Kirk, Leja, joezuntz, \& Goldstein}]{Koposov24}
Koposov, S., Speagle, J., Barbary, K., {et~al.} 2024, \bibinfo{title}{joshspeagle/dynesty: v2.1.4,}, v2.1.4 Zenodo, \dodoi{10.5281/zenodo.12537467}

\bibitem[{R.~K. {Kopparapu} {et~al.}(2013){Kopparapu}, {Ramirez}, {Kasting}, {Eymet}, {Robinson}, {Mahadevan}, {Terrien}, {Domagal-Goldman}, {Meadows}, \& {Deshpande}}]{Kopparapu2013}
{Kopparapu}, R.~K., {Ramirez}, R., {Kasting}, J.~F., {et~al.} 2013, \bibinfo{title}{{Habitable Zones around Main-sequence Stars: New Estimates},} \apj, 765, 131, \dodoi{10.1088/0004-637X/765/2/131}

\bibitem[{D. {Kossakowski} {et~al.}(2022){Kossakowski}, {K{\"u}rster}, {Henning}, {Trifonov}, {Caballero}, {Lafarga}, {Bauer}, {Stock}, {Kemmer}, {Jeffers}, {Amado}, {P{\'e}rez-Torres}, {B{\'e}jar}, {Cort{\'e}s-Contreras}, {Ribas}, {Reiners}, {Quirrenbach}, {Aceituno}, {Baroch}, {Cifuentes}, {Dreizler}, {Hatzes}, {Kaminski}, {Montes}, {Morales}, {Pavlov}, {Pena}, {Perdelwitz}, {Reffert}, {Revilla}, {Rodr{\'\i}guez Lopez}, {Rosich}, {Sadegi}, {Sanz-Forcada}, {Sch{\"o}fer}, {Schweitzer}, \& {Zechmeister}}]{Kossakowski2022}
{Kossakowski}, D., {K{\"u}rster}, M., {Henning}, T., {et~al.} 2022, \bibinfo{title}{{The CARMENES search for exoplanets around M dwarfs. Stable radial-velocity variations at the rotation period of AD Leonis: A test case study of current limitations to treating stellar activity},} \aap, 666, A143, \dodoi{10.1051/0004-6361/202243773}

\bibitem[{L. {Kreidberg}(2015){Kreidberg}}]{batman}
{Kreidberg}, L. 2015, \bibinfo{title}{{batman: BAsic Transit Model cAlculatioN in Python},} Publications of the Astronomical Society of the Pacific, 127, 1161, \dodoi{10.1086/683602}

\bibitem[{M. {Kuzuhara} {et~al.}(2024){Kuzuhara}, {Fukui}, {Livingston}, {Caballero}, {de Leon}, {Hirano}, {Kasagi}, {Murgas}, {Narita}, {Omiya}, {Orell-Miquel}, {Palle}, {Changeat}, {Esparza-Borges}, {Harakawa}, {Hellier}, {Hori}, {Ikuta}, {Ishikawa}, {Kodama}, {Kotani}, {Kudo}, {Morales}, {Mori}, {Nagel}, {Parviainen}, {Perdelwitz}, {Reiners}, {Ribas}, {Sanz-Forcada}, {Sato}, {Schweitzer}, {Tabernero}, {Takarada}, {Uyama}, {Watanabe}, {Zechmeister}, {Garc{\'\i}a}, {Aoki}, {Beichman}, {B{\'e}jar}, {Brandt}, {Calatayud-Borras}, {Carleo}, {Charbonneau}, {Collins}, {Currie}, {Doty}, {Dreizler}, {Fern{\'a}ndez-Rodr{\'\i}guez}, {Fukuda}, {Gal{\'a}n}, {Gerald{\'\i}a-Gonz{\'a}lez}, {Gonz{\'a}lez-Rodr{\'\i}guez}, {Hayashi}, {Hedges}, {Henning}, {Hodapp}, {Ikoma}, {Isogai}, {Jacobson}, {Janson}, {Jenkins}, {Kagetani}, {Kambe}, {Kawai}, {Kawauchi}, {Kokubo}, {Konishi}, {Korth}, {Krishnamurthy}, {Kurokawa}, {Kusakabe}, {Kwon}, {Laza-Ramos}, {Libotte}, {Luque}, {Madrigal-Aguado}, {Matsumoto}, {Mawet}, {McElwain}, {Meni
  Gallardo}, {Morello}, {Mu{\~n}oz Torres}, {Nishikawa}, {Nugroho}, {Ogihara}, {Pel{\'a}ez-Torres}, {Rapetti}, {S{\'a}nchez-Benavente}, {Schlecker}, {Seager}, {Serabyn}, {Serizawa}, {Stangret}, {Takahashi}, {Teng}, {Tamura}, {Terada}, {Ueda}, {Usuda}, {Vanderspek}, {Vievard}, {Watanabe}, {Winn}, \& {Zapatero Osorio}}]{Kuzuhara2024}
{Kuzuhara}, M., {Fukui}, A., {Livingston}, J.~H., {et~al.} 2024, \bibinfo{title}{{Gliese 12 b: A Temperate Earth-sized Planet at 12 pc Ideal for Atmospheric Transmission Spectroscopy},} \apjl, 967, L21, \dodoi{10.3847/2041-8213/ad3642}

\bibitem[{V. {Lainey}(2016){Lainey}}]{Lainey2016}
{Lainey}, V. 2016, \bibinfo{title}{{Quantification of tidal parameters from Solar System data},} Celestial Mechanics and Dynamical Astronomy, 126, 145, \dodoi{10.1007/s10569-016-9695-y}

\bibitem[{ {Lightkurve Collaboration} {et~al.}(2018){Lightkurve Collaboration}, {Cardoso}, {Hedges}, {Gully-Santiago}, {Saunders}, {Cody}, {Barclay}, {Hall}, {Sagear}, {Turtelboom}, {Zhang}, {Tzanidakis}, {Mighell}, {Coughlin}, {Bell}, {Berta-Thompson}, {Williams}, {Dotson}, \& {Barentsen}}]{lightkurve}
{Lightkurve Collaboration}, {Cardoso}, J.~V.~d.~M., {Hedges}, C., {et~al.} 2018, \bibinfo{title}{{Lightkurve: Kepler and TESS time series analysis in Python},}, Astrophysics Source Code Library \doeprint{1812.013}

\bibitem[{O. {Lim} {et~al.}(2023){Lim}, {Benneke}, {Doyon}, {MacDonald}, {Piaulet}, {Artigau}, {Coulombe}, {Radica}, {L'Heureux}, {Albert}, {Rackham}, {de Wit}, {Salhi}, {Roy}, {Flagg}, {Fournier-Tondreau}, {Taylor}, {Cook}, {Lafreni{\`e}re}, {Cowan}, {Kaltenegger}, {Rowe}, {Espinoza}, {Dang}, \& {Darveau-Bernier}}]{Lim2023}
{Lim}, O., {Benneke}, B., {Doyon}, R., {et~al.} 2023, \bibinfo{title}{{Atmospheric Reconnaissance of TRAPPIST-1 b with JWST/NIRISS: Evidence for Strong Stellar Contamination in the Transmission Spectra},} \apjl, 955, L22, \dodoi{10.3847/2041-8213/acf7c4}

\bibitem[{C.~L. {Lin} {et~al.}(2019){Lin}, {Ip}, {Hou}, {Huang}, \& {Chang}}]{Lin2019}
{Lin}, C.~L., {Ip}, W.~H., {Hou}, W.~C., {Huang}, L.~C., \& {Chang}, H.~Y. 2019, \bibinfo{title}{{A Comparative Study of the Magnetic Activities of Low-mass Stars from M-type to G-type},} \apj, 873, 97, \dodoi{10.3847/1538-4357/ab041c}

\bibitem[{A.~P. {Lincowski} {et~al.}(2023){Lincowski}, {Meadows}, {Zieba}, {Kreidberg}, {Morley}, {Gillon}, {Selsis}, {Agol}, {Bolmont}, {Ducrot}, {Hu}, {Koll}, {Lyu}, {Mandell}, {Suissa}, \& {Tamburo}}]{lincowski23}
{Lincowski}, A.~P., {Meadows}, V.~S., {Zieba}, S., {et~al.} 2023, \bibinfo{title}{{Potential Atmospheric Compositions of TRAPPIST-1 c Constrained by JWST/MIRI Observations at 15 {\ensuremath{\mu}}m},} \apjl, 955, L7, \dodoi{10.3847/2041-8213/acee02}

\bibitem[{H. {Luo} {et~al.}(2024){Luo}, {Dorn}, \& {Deng}}]{Luo2024}
{Luo}, H., {Dorn}, C., \& {Deng}, J. 2024, \bibinfo{title}{{The interior as the dominant water reservoir in super-Earths and sub-Neptunes},} Nature Astronomy, 8, 1399, \dodoi{10.1038/s41550-024-02347-z}

\bibitem[{R. {Luque} \& E. {Pall{\'e}}(2022){Luque} \& {Pall{\'e}}}]{Luque2022}
{Luque}, R., \& {Pall{\'e}}, E. 2022, \bibinfo{title}{{Density, not radius, separates rocky and water-rich small planets orbiting M dwarf stars},} Science, 377, 1211, \dodoi{10.1126/science.abl7164}

\bibitem[{J. {Lustig-Yaeger} {et~al.}(2023){Lustig-Yaeger}, {Fu}, {May}, {Ceballos}, {Moran}, {Peacock}, {Stevenson}, {Kirk}, {L{\'o}pez-Morales}, {MacDonald}, {Mayorga}, {Sing}, {Sotzen}, {Valenti}, {Redai}, {Alam}, {Batalha}, {Bennett}, {Gonzalez-Quiles}, {Kruse}, {Lothringer}, {Rustamkulov}, \& {Wakeford}}]{Lustig-Yaeger2023}
{Lustig-Yaeger}, J., {Fu}, G., {May}, E.~M., {et~al.} 2023, \bibinfo{title}{{A JWST transmission spectrum of the nearby Earth-sized exoplanet LHS 475 b},} Nature Astronomy, 7, 1317, \dodoi{10.1038/s41550-023-02064-z}

\bibitem[{J. {Maldonado} {et~al.}(2020){Maldonado}, {Micela}, {Baratella}, {D'Orazi}, {Affer}, {Biazzo}, {Lanza}, {Maggio}, {Gonz{\'a}lez Hern{\'a}ndez}, {Perger}, {Pinamonti}, {Scandariato}, {Sozzetti}, {Locci}, {Di Maio}, {Bignamini}, {Claudi}, {Molinari}, {Rebolo}, {Ribas}, {Toledo-Padr{\'o}n}, {Covino}, {Desidera}, {Herrero}, {Morales}, {Su{\'a}rez-Mascare{\~n}o}, {Pagano}, {Petralia}, {Piotto}, \& {Poretti}}]{Maldonado2020}
{Maldonado}, J., {Micela}, G., {Baratella}, M., {et~al.} 2020, \bibinfo{title}{{HADES RV programme with HARPS-N at TNG. XII. The abundance signature of M dwarf stars with planets},} \aap, 644, A68, \dodoi{10.1051/0004-6361/202039478}

\bibitem[{U. {Marboeuf} {et~al.}(2014){Marboeuf}, {Thiabaud}, {Alibert}, {Cabral}, \& {Benz}}]{Marboef2014}
{Marboeuf}, U., {Thiabaud}, A., {Alibert}, Y., {Cabral}, N., \& {Benz}, W. 2014, \bibinfo{title}{{From planetesimals to planets: volatile molecules},} \aap, 570, A36, \dodoi{10.1051/0004-6361/201423431}

\bibitem[{V.~S. {Meadows} {et~al.}(2023){Meadows}, {Lincowski}, \& {Lustig-Yaeger}}]{Meadows2023}
{Meadows}, V.~S., {Lincowski}, A.~P., \& {Lustig-Yaeger}, J. 2023, \bibinfo{title}{{The Feasibility of Detecting Biosignatures in the TRAPPIST-1 Planetary System with JWST},} \psj, 4, 192, \dodoi{10.3847/PSJ/acf488}

\bibitem[{W.~B. Moore(2003)Moore}]{Moore2003}
Moore, W.~B. 2003, \bibinfo{title}{Tidal heating and convection in {Io},} Journal of Geophysical Research (Planets), 108, 5096, \dodoi{10.1029/2002JE001943}

\bibitem[{C.~V. {Morley} {et~al.}(2015){Morley}, {Fortney}, {Marley}, {Zahnle}, {Line}, {Kempton}, {Lewis}, \& {Cahoy}}]{Morley2015}
{Morley}, C.~V., {Fortney}, J.~J., {Marley}, M.~S., {et~al.} 2015, \bibinfo{title}{{Thermal Emission and Reflected Light Spectra of Super Earths with Flat Transmission Spectra},} \apj, 815, 110, \dodoi{10.1088/0004-637X/815/2/110}

\bibitem[{C.~V. {Morley} {et~al.}(2017){Morley}, {Kreidberg}, {Rustamkulov}, {Robinson}, \& {Fortney}}]{Morley2017}
{Morley}, C.~V., {Kreidberg}, L., {Rustamkulov}, Z., {Robinson}, T., \& {Fortney}, J.~J. 2017, \bibinfo{title}{{Observing the Atmospheres of Known Temperate Earth-sized Planets with JWST},} \apj, 850, 121, \dodoi{10.3847/1538-4357/aa927b}

\bibitem[{A. {Mortier} {et~al.}(2020){Mortier}, {Zapatero Osorio}, {Malavolta}, {Alibert}, {Rice}, {Lillo-Box}, {Vanderburg}, {Oshagh}, {Buchhave}, {Adibekyan}, {Delgado Mena}, {Lopez-Morales}, {Charbonneau}, {Sousa}, {Lovis}, {Affer}, {Allende Prieto}, {Barros}, {Benatti}, {Bonomo}, {Boschin}, {Bouchy}, {Cabral}, {Collier Cameron}, {Cosentino}, {Cristiani}, {Demangeon}, {Di Marcantonio}, {D'Odorico}, {Dumusque}, {Ehrenreich}, {Figueira}, {Fiorenzano}, {Ghedina}, {Gonz{\'a}lez Hern{\'a}ndez}, {Haldemann}, {Harutyunyan}, {Haywood}, {Latham}, {Lavie}, {Lo Curto}, {Maldonado}, {Manescau}, {Martins}, {Mayor}, {M{\'e}gevand}, {Mehner}, {Micela}, {Molaro}, {Molinari}, {Nunes}, {Pepe}, {Palle}, {Phillips}, {Piotto}, {Pinamonti}, {Poretti}, {Riva}, {Rebolo}, {Santos}, {Sasselov}, {Sozzetti}, {Su{\'a}rez Mascare{\~n}o}, {Udry}, {West}, {Watson}, \& {Wilson}}]{Mortier2020}
{Mortier}, A., {Zapatero Osorio}, M.~R., {Malavolta}, L., {et~al.} 2020, \bibinfo{title}{{K2-111: an old system with two planets in near-resonance},} \mnras, 499, 5004, \dodoi{10.1093/mnras/staa3144}

\bibitem[{E.~R. {Newton} {et~al.}(2014){Newton}, {Charbonneau}, {Irwin}, {Berta-Thompson}, {Rojas-Ayala}, {Covey}, \& {Lloyd}}]{Newton2014}
{Newton}, E.~R., {Charbonneau}, D., {Irwin}, J., {et~al.} 2014, \bibinfo{title}{{Near-infrared Metallicities, Radial Velocities, and Spectral Types for 447 Nearby M Dwarfs},} \aj, 147, 20, \dodoi{10.1088/0004-6256/147/1/20}

\bibitem[{E.~R. {Newton} {et~al.}(2016){Newton}, {Irwin}, {Charbonneau}, {Berta-Thompson}, {Dittmann}, \& {West}}]{Newton2016}
{Newton}, E.~R., {Irwin}, J., {Charbonneau}, D., {et~al.} 2016, \bibinfo{title}{{The Rotation and Galactic Kinematics of Mid M Dwarfs in the Solar Neighborhood},} \apj, 821, 93, \dodoi{10.3847/0004-637X/821/2/93}

\bibitem[{M.~C. {Nixon} \& N. {Madhusudhan}(2021){Nixon} \& {Madhusudhan}}]{Nixon2021}
{Nixon}, M.~C., \& {Madhusudhan}, N. 2021, \bibinfo{title}{{How deep is the ocean? Exploring the phase structure of water-rich sub-Neptunes},} \mnras, 505, 3414, \dodoi{10.1093/mnras/stab1500}

\bibitem[{M.~C. {Nixon} {et~al.}(2024){Nixon}, {Piette}, {Kempton}, {Gao}, {Bean}, {Steinrueck}, {Mahajan}, {Eastman}, {Zhang}, \& {Rogers}}]{Nixon2024}
{Nixon}, M.~C., {Piette}, A. A.~A., {Kempton}, E. M.~R., {et~al.} 2024, \bibinfo{title}{{New Insights into the Internal Structure of GJ 1214 b Informed by JWST},} \apjl, 970, L28, \dodoi{10.3847/2041-8213/ad615b}

\bibitem[{M. {Paegert} {et~al.}(2021){Paegert}, {Stassun}, {Collins}, {Pepper}, {Torres}, {Jenkins}, {Twicken}, \& {Latham}}]{Paegert2021}
{Paegert}, M., {Stassun}, K.~G., {Collins}, K.~A., {et~al.} 2021, \bibinfo{title}{{TESS Input Catalog versions 8.1 and 8.2: Phantoms in the 8.0 Catalog and How to Handle Them},} arXiv e-prints, arXiv:2108.04778, \dodoi{10.48550/arXiv.2108.04778}

\bibitem[{E.~K. {Pass} {et~al.}(2025){Pass}, {Charbonneau}, \& {Vanderburg}}]{Pass2025}
{Pass}, E.~K., {Charbonneau}, D., \& {Vanderburg}, A. 2025, \bibinfo{title}{{The Receding Cosmic Shoreline of Mid-to-Late M Dwarfs: Measurements of Active Lifetimes Worsen Challenges for Atmosphere Retention by Rocky Exoplanets},} arXiv e-prints, arXiv:2504.01182, \dodoi{10.48550/arXiv.2504.01182}

\bibitem[{M.~S. {Peterson} {et~al.}(2023){Peterson}, {Benneke}, {Collins}, {Piaulet}, {Crossfield}, {Ali-Dib}, {Christiansen}, {Gagn{\'e}}, {Faherty}, {Kite}, {Dressing}, {Charbonneau}, {Murgas}, {Cointepas}, {Almenara}, {Bonfils}, {Kane}, {Werner}, {Gorjian}, {Roy}, {Shporer}, {Pozuelos}, {Socia}, {Cloutier}, {Dietrich}, {Irwin}, {Weiss}, {Waalkes}, {Berta-Thomson}, {Evans}, {Apai}, {Parviainen}, {Pall{\'e}}, {Narita}, {Howard}, {Dragomir}, {Barkaoui}, {Gillon}, {Jehin}, {Ducrot}, {Benkhaldoun}, {Fukui}, {Mori}, {Nishiumi}, {Kawauchi}, {Ricker}, {Latham}, {Winn}, {Seager}, {Isaacson}, {Bixel}, {Gibbs}, {Jenkins}, {Smith}, {Chavez}, {Rackham}, {Henning}, {Gabor}, {Chen}, {Espinoza}, {Jensen}, {Collins}, {Schwarz}, {Conti}, {Wang}, {Kielkopf}, {Mao}, {Horne}, {Sefako}, {Quinn}, {Moldovan}, {Fausnaugh}, {F{\.z}{\.z}r{\'e}sz}, \& {Barclay}}]{Peterson2023}
{Peterson}, M.~S., {Benneke}, B., {Collins}, K., {et~al.} 2023, \bibinfo{title}{{A temperate Earth-sized planet with tidal heating transiting an M6 star},} \nat, 617, 701, \dodoi{10.1038/s41586-023-05934-8}

\bibitem[{C. {Piaulet} {et~al.}(2021){Piaulet}, {Benneke}, {Rubenzahl}, {Howard}, {Lee}, {Thorngren}, {Angus}, {Peterson}, {Schlieder}, {Werner}, {Kreidberg}, {Jaouni}, {Crossfield}, {Ciardi}, {Petigura}, {Livingston}, {Dressing}, {Fulton}, {Beichman}, {Christiansen}, {Gorjian}, {Hardegree-Ullman}, {Krick}, \& {Sinukoff}}]{Piaulet2021}
{Piaulet}, C., {Benneke}, B., {Rubenzahl}, R.~A., {et~al.} 2021, \bibinfo{title}{{WASP-107b's Density Is Even Lower: A Case Study for the Physics of Planetary Gas Envelope Accretion and Orbital Migration},} \aj, 161, 70, \dodoi{10.3847/1538-3881/abcd3c}

\bibitem[{C. {Piaulet-Ghorayeb} {et~al.}(2024){Piaulet-Ghorayeb}, {Benneke}, {Radica}, {Raul}, {Coulombe}, {Ahrer}, {Kubyshkina}, {Howard}, {Krissansen-Totton}, {MacDonald}, {Roy}, {Louca}, {Christie}, {Fournier-Tondreau}, {Allart}, {Miguel}, {Schlichting}, {Welbanks}, {Cadieux}, {Dorn}, {Evans-Soma}, {Fortney}, {Pierrehumbert}, {Lafreni{\`e}re}, {Acu{\~n}a}, {Komacek}, {Innes}, {Beatty}, {Cloutier}, {Doyon}, {Gagnebin}, {Gapp}, \& {Knutson}}]{Piaulet-Ghorayeb2024}
{Piaulet-Ghorayeb}, C., {Benneke}, B., {Radica}, M., {et~al.} 2024, \bibinfo{title}{{JWST/NIRISS Reveals the Water-rich ``Steam World'' Atmosphere of GJ 9827 d},} \apjl, 974, L10, \dodoi{10.3847/2041-8213/ad6f00}

\bibitem[{D.~L. {Pollacco} {et~al.}(2006){Pollacco}, {Skillen}, {Collier Cameron}, {Christian}, {Hellier}, {Irwin}, {Lister}, {Street}, {West}, {Anderson}, {Clarkson}, {Deeg}, {Enoch}, {Evans}, {Fitzsimmons}, {Haswell}, {Hodgkin}, {Horne}, {Kane}, {Keenan}, {Maxted}, {Norton}, {Osborne}, {Parley}, {Ryans}, {Smalley}, {Wheatley}, \& {Wilson}}]{Pollacco2006}
{Pollacco}, D.~L., {Skillen}, I., {Collier Cameron}, A., {et~al.} 2006, \bibinfo{title}{{The WASP Project and the SuperWASP Cameras},} \pasp, 118, 1407, \dodoi{10.1086/508556}

\bibitem[{B.~V. {Rackham} {et~al.}(2018){Rackham}, {Apai}, \& {Giampapa}}]{Rackham2018}
{Rackham}, B.~V., {Apai}, D., \& {Giampapa}, M.~S. 2018, \bibinfo{title}{{The Transit Light Source Effect: False Spectral Features and Incorrect Densities for M-dwarf Transiting Planets},} \apj, 853, 122, \dodoi{10.3847/1538-4357/aaa08c}

\bibitem[{M. {Radica} {et~al.}(2025){Radica}, {Piaulet-Ghorayeb}, {Taylor}, {Coulombe}, {Benneke}, {Albert}, {Artigau}, {Cowan}, {Doyon}, {Lafreni{\`e}re}, {L'Heureux}, \& {Lim}}]{Radica2025}
{Radica}, M., {Piaulet-Ghorayeb}, C., {Taylor}, J., {et~al.} 2025, \bibinfo{title}{{Promise and Peril: Stellar Contamination and Strict Limits on the Atmosphere Composition of TRAPPIST-1 c from JWST NIRISS Transmission Spectra},} \apjl, 979, L5, \dodoi{10.3847/2041-8213/ada381}

\bibitem[{A.~D. {Rathcke} {et~al.}(2025){Rathcke}, {Buchhave}, {de Wit}, {Rackham}, {August}, {Diamond-Lowe}, {Mendon{\c{C}}a}, {Bello-Arufe}, {L{\'o}pez-Morales}, {Kitzmann}, \& {Heng}}]{Rathcke2025}
{Rathcke}, A.~D., {Buchhave}, L.~A., {de Wit}, J., {et~al.} 2025, \bibinfo{title}{{Stellar Contamination Correction Using Back-to-back Transits of TRAPPIST-1 b and c},} \apjl, 979, L19, \dodoi{10.3847/2041-8213/ada5c7}

\bibitem[{H. {Rein} \& S.~F. {Liu}(2012){Rein} \& {Liu}}]{rebound}
{Rein}, H., \& {Liu}, S.~F. 2012, \bibinfo{title}{{REBOUND: an open-source multi-purpose N-body code for collisional dynamics},} \aap, 537, A128, \dodoi{10.1051/0004-6361/201118085}

\bibitem[{J. Renner {et~al.}(2000)Renner, Evans, \& Hirth}]{Renner2000}
Renner, J., Evans, B., \& Hirth, G. 2000, \bibinfo{title}{On the rheologically critical melt fraction,} Earth and Planetary Science Letters, 181, 585, \dodoi{10.1016/S0012-821X(00)00222-3}

\bibitem[{J.~G. {Rogers} {et~al.}(2021){Rogers}, {Gupta}, {Owen}, \& {Schlichting}}]{Rogers2021}
{Rogers}, J.~G., {Gupta}, A., {Owen}, J.~E., \& {Schlichting}, H.~E. 2021, \bibinfo{title}{{Photoevaporation versus core-powered mass-loss: model comparison with the 3D radius gap},} \mnras, 508, 5886, \dodoi{10.1093/mnras/stab2897}

\bibitem[{L.~J. {Rosenthal} {et~al.}(2021){Rosenthal}, {Fulton}, {Hirsch}, {Isaacson}, {Howard}, {Dedrick}, {Sherstyuk}, {Blunt}, {Petigura}, {Knutson}, {Behmard}, {Chontos}, {Crepp}, {Crossfield}, {Dalba}, {Fischer}, {Henry}, {Kane}, {Kosiarek}, {Marcy}, {Rubenzahl}, {Weiss}, \& {Wright}}]{Rosenthal2021}
{Rosenthal}, L.~J., {Fulton}, B.~J., {Hirsch}, L.~A., {et~al.} 2021, \bibinfo{title}{{The California Legacy Survey. I. A Catalog of 178 Planets from Precision Radial Velocity Monitoring of 719 Nearby Stars over Three Decades},} \apjs, 255, 8, \dodoi{10.3847/1538-4365/abe23c}

\bibitem[{S. {Seager} {et~al.}(2025){Seager}, {Welbanks}, {Ellerbroek}, {Bains}, \& {Petkowski}}]{seager25}
{Seager}, S., {Welbanks}, L., {Ellerbroek}, L., {Bains}, W., \& {Petkowski}, J.~J. 2025, \bibinfo{title}{{Prospects for Detecting Signs of Life on Exoplanets in the JWST Era},} arXiv e-prints, arXiv:2504.12946, \dodoi{10.48550/arXiv.2504.12946}

\bibitem[{M. Segatz {et~al.}(1988)Segatz, Spohn, Ross, \& Schubert}]{Segatz1988}
Segatz, M., Spohn, T., Ross, M.~N., \& Schubert, G. 1988, \bibinfo{title}{Tidal dissipation, surface heat flow, and figure of viscoelastic models of {Io},} Icarus, 75, 187, \dodoi{10.1016/0019-1035(88)90001-2}

\bibitem[{A. {Seifahrt} {et~al.}(2016){Seifahrt}, {Bean}, {St{\"u}rmer}, {Gers}, {Grobler}, {Reed}, \& {Jones}}]{seifahrt16}
{Seifahrt}, A., {Bean}, J.~L., {St{\"u}rmer}, J., {et~al.} 2016, in Society of Photo-Optical Instrumentation Engineers (SPIE) Conference Series, Vol. 9908, Ground-based and Airborne Instrumentation for Astronomy VI, ed. C.~J. {Evans}, L.~{Simard}, \& H.~{Takami}, 990818, \dodoi{10.1117/12.2232069}

\bibitem[{A. {Seifahrt} {et~al.}(2018){Seifahrt}, {St{\"u}rmer}, {Bean}, \& {Schwab}}]{Seifahrt18}
{Seifahrt}, A., {St{\"u}rmer}, J., {Bean}, J.~L., \& {Schwab}, C. 2018, in Society of Photo-Optical Instrumentation Engineers (SPIE) Conference Series, Vol. 10702, Ground-based and Airborne Instrumentation for Astronomy VII, ed. C.~J. {Evans}, L.~{Simard}, \& H.~{Takami}, 107026D, \dodoi{10.1117/12.2312936}

\bibitem[{A. {Seifahrt} {et~al.}(2020){Seifahrt}, {Bean}, {St{\"u}rmer}, {Kasper}, {Gers}, {Schwab}, {Zechmeister}, {Stef{\'a}nsson}, {Montet}, {Dos Santos}, {Peck}, {White}, \& {Tapia}}]{seifahrt20}
{Seifahrt}, A., {Bean}, J.~L., {St{\"u}rmer}, J., {et~al.} 2020, in Society of Photo-Optical Instrumentation Engineers (SPIE) Conference Series, Vol. 11447, Ground-based and Airborne Instrumentation for Astronomy VIII, ed. C.~J. {Evans}, J.~J. {Bryant}, \& K.~{Motohara}, 114471F, \dodoi{10.1117/12.2561564}

\bibitem[{A. {Seifahrt} {et~al.}(2022){Seifahrt}, {Bean}, {Kasper}, {St{\"u}rmer}, {Brady}, {Liu}, {Zechmeister}, {Stef{\'a}nsson}, {Montet}, {White}, {Tapia}, {Mocnik}, {Xu}, \& {Schwab}}]{Seifahrt22}
{Seifahrt}, A., {Bean}, J.~L., {Kasper}, D., {et~al.} 2022, in Society of Photo-Optical Instrumentation Engineers (SPIE) Conference Series, Vol. 12184, Ground-based and Airborne Instrumentation for Astronomy IX, ed. C.~J. {Evans}, J.~J. {Bryant}, \& K.~{Motohara}, 121841G, \dodoi{10.1117/12.2629428}

\bibitem[{B.~J. {Shappee} {et~al.}(2014){Shappee}, {Prieto}, {Grupe}, {Kochanek}, {Stanek}, {De Rosa}, {Mathur}, {Zu}, {Peterson}, {Pogge}, {Komossa}, {Im}, {Jencson}, {Holoien}, {Basu}, {Beacom}, {Szczygie{\l}}, {Brimacombe}, {Adams}, {Campillay}, {Choi}, {Contreras}, {Dietrich}, {Dubberley}, {Elphick}, {Foale}, {Giustini}, {Gonzalez}, {Hawkins}, {Howell}, {Hsiao}, {Koss}, {Leighly}, {Morrell}, {Mudd}, {Mullins}, {Nugent}, {Parrent}, {Phillips}, {Pojmanski}, {Rosing}, {Ross}, {Sand}, {Terndrup}, {Valenti}, {Walker}, \& {Yoon}}]{Shappee2014}
{Shappee}, B.~J., {Prieto}, J.~L., {Grupe}, D., {et~al.} 2014, \bibinfo{title}{{The Man behind the Curtain: X-Rays Drive the UV through NIR Variability in the 2013 Active Galactic Nucleus Outburst in NGC 2617},} \apj, 788, 48, \dodoi{10.1088/0004-637X/788/1/48}

\bibitem[{J. {Skilling}(2004){Skilling}}]{Skilling2004}
{Skilling}, J. 2004, in American Institute of Physics Conference Series, Vol. 735, Bayesian Inference and Maximum Entropy Methods in Science and Engineering: 24th International Workshop on Bayesian Inference and Maximum Entropy Methods in Science and Engineering, ed. R.~{Fischer}, R.~{Preuss}, \& U.~V. {Toussaint} (AIP), 395--405, \dodoi{10.1063/1.1835238}

\bibitem[{J. Skilling(2006)Skilling}]{Skilling2006}
Skilling, J. 2006, \bibinfo{title}{{Nested sampling for general Bayesian computation},} Bayesian Analysis, 1, 833 , \dodoi{10.1214/06-BA127}

\bibitem[{V.~S. Solomatov \& L.-N. Moresi(2000)Solomatov \& Moresi}]{Solomatov2000}
Solomatov, V.~S., \& Moresi, L.-N. 2000, \bibinfo{title}{Scaling of time-dependent stagnant lid convection: {Application} to small-scale convection on {Earth} and other terrestrial planets,} Journal of Geophysical Research, 105, 21795, \dodoi{10.1029/2000JB900197}

\bibitem[{J.~S. {Speagle}(2020){Speagle}}]{Speagle2020}
{Speagle}, J.~S. 2020, \bibinfo{title}{{DYNESTY: a dynamic nested sampling package for estimating Bayesian posteriors and evidences},} \mnras, 493, 3132, \dodoi{10.1093/mnras/staa278}

\bibitem[{A.~T. {Stevenson} {et~al.}(2025){Stevenson}, {Haswell}, {Faria}, {Barnes}, {Barstow}, {Dickinson}, \& {Standing}}]{Stevenson2025}
{Stevenson}, A.~T., {Haswell}, C.~A., {Faria}, J.~P., {et~al.} 2025, \bibinfo{title}{{RV-exoplanet eccentricities: Good, Beta, and Best},} \mnras, 539, 727, \dodoi{10.1093/mnras/staf502}

\bibitem[{A. {Su{\'a}rez Mascare{\~n}o} {et~al.}(2018){Su{\'a}rez Mascare{\~n}o}, {Rebolo}, {Gonz{\'a}lez Hern{\'a}ndez}, {Toledo-Padr{\'o}n}, {Perger}, {Ribas}, {Affer}, {Micela}, {Damasso}, {Maldonado}, {Gonz{\'a}lez-Alvarez}, {Leto}, {Pagano}, {Scandariato}, {Sozzetti}, {Lanza}, {Malavolta}, {Claudi}, {Cosentino}, {Desidera}, {Giacobbe}, {Maggio}, {Rainer}, {Esposito}, {Benatti}, {Pedani}, {Morales}, {Herrero}, {Lafarga}, {Rosich}, \& {Pinamonti}}]{SuarezMascareno2018}
{Su{\'a}rez Mascare{\~n}o}, A., {Rebolo}, R., {Gonz{\'a}lez Hern{\'a}ndez}, J.~I., {et~al.} 2018, \bibinfo{title}{{HADES RV programme with HARPS-N at TNG. VII. Rotation and activity of M-dwarfs from time-series high-resolution spectroscopy of chromospheric indicators},} \aap, 612, A89, \dodoi{10.1051/0004-6361/201732143}

\bibitem[{M. {Szurgot}(2015){Szurgot}}]{Szurgot2015}
{Szurgot}, M. 2015, in LPI Contributions, Vol. 1839, Comparative Tectonic and Geodynamics of Venus, Earth and Rocky Exoplanets, ed. {LPI Editorial Board}, 5001

\bibitem[{A. {Thiabaud} {et~al.}(2014){Thiabaud}, {Marboeuf}, {Alibert}, {Cabral}, {Leya}, \& {Mezger}}]{Thiabaud2014}
{Thiabaud}, A., {Marboeuf}, U., {Alibert}, Y., {et~al.} 2014, \bibinfo{title}{{From stellar nebula to planets: The refractory components},} \aap, 562, A27, \dodoi{10.1051/0004-6361/201322208}

\bibitem[{T. {Trifonov} {et~al.}(2020){Trifonov}, {Tal-Or}, {Zechmeister}, {Kaminski}, {Zucker}, \& {Mazeh}}]{Trifonov2020}
{Trifonov}, T., {Tal-Or}, L., {Zechmeister}, M., {et~al.} 2020, \bibinfo{title}{{Public HARPS radial velocity database corrected for systematic errors},} \aap, 636, A74, \dodoi{10.1051/0004-6361/201936686}

\bibitem[{M. {Turbet} {et~al.}(2020){Turbet}, {Bolmont}, {Ehrenreich}, {Gratier}, {Leconte}, {Selsis}, {Hara}, \& {Lovis}}]{Turbet2020}
{Turbet}, M., {Bolmont}, E., {Ehrenreich}, D., {et~al.} 2020, \bibinfo{title}{{Revised mass-radius relationships for water-rich rocky planets more irradiated than the runaway greenhouse limit},} \aap, 638, A41, \dodoi{10.1051/0004-6361/201937151}

\bibitem[{G. {Van Looveren} {et~al.}(2024){Van Looveren}, {G{\"u}del}, {Boro Saikia}, \& {Kislyakova}}]{VanLooveren2024}
{Van Looveren}, G., {G{\"u}del}, M., {Boro Saikia}, S., \& {Kislyakova}, K. 2024, \bibinfo{title}{{Airy worlds or barren rocks? On the survivability of secondary atmospheres around the TRAPPIST-1 planets},} \aap, 683, A153, \dodoi{10.1051/0004-6361/202348079}

\bibitem[{G.~J. {Veeder} {et~al.}(2012){Veeder}, {Davies}, {Matson}, {Johnson}, {Williams}, \& {Radebaugh}}]{Veeder2012}
{Veeder}, G.~J., {Davies}, A.~G., {Matson}, D.~L., {et~al.} 2012, \bibinfo{title}{{Io: Volcanic thermal sources and global heat flow},} \icarus, 219, 701, \dodoi{10.1016/j.icarus.2012.04.004}

\bibitem[{P. Virtanen {et~al.}(2020)Virtanen, Gommers, Oliphant, Haberland, Reddy, Cournapeau, Burovski, Peterson, Weckesser, Bright, {van der Walt}, Brett, Wilson, Millman, Mayorov, Nelson, Jones, Kern, Larson, Carey, Polat, Feng, Moore, {VanderPlas}, Laxalde, Perktold, Cimrman, Henriksen, Quintero, Harris, Archibald, Ribeiro, Pedregosa, {van Mulbregt}, \& {SciPy 1.0 Contributors}}]{2020SciPy-NMeth}
Virtanen, P., Gommers, R., Oliphant, T.~E., {et~al.} 2020, \bibinfo{title}{{{SciPy} 1.0: Fundamental Algorithms for Scientific Computing in Python},} Nature Methods, 17, 261, \dodoi{10.1038/s41592-019-0686-2}

\bibitem[{J.~N. {Winn}(2010){Winn}}]{Winn2010}
{Winn}, J.~N. 2010, in Exoplanets, ed. S.~{Seager}, 55--77, \dodoi{10.48550/arXiv.1001.2010}

\bibitem[{J.~G. {Winters} {et~al.}(2021){Winters}, {Charbonneau}, {Henry}, {Irwin}, {Jao}, {Riedel}, \& {Slatten}}]{Winters2021}
{Winters}, J.~G., {Charbonneau}, D., {Henry}, T.~J., {et~al.} 2021, \bibinfo{title}{{The Volume-complete Sample of M Dwarfs with Masses 0.1 {\ensuremath{\leq}} M/M$_{{\ensuremath{\odot}}}$ {\ensuremath{\leq}} 0.3 within 15 Parsecs},} \aj, 161, 63, \dodoi{10.3847/1538-3881/abcc74}

\bibitem[{J. {Yang} {et~al.}(2014){Yang}, {Bou{\'e}}, {Fabrycky}, \& {Abbot}}]{Yang2014}
{Yang}, J., {Bou{\'e}}, G., {Fabrycky}, D.~C., \& {Abbot}, D.~S. 2014, \bibinfo{title}{{Strong Dependence of the Inner Edge of the Habitable Zone on Planetary Rotation Rate},} \apjl, 787, L2, \dodoi{10.1088/2041-8205/787/1/L2}

\bibitem[{K.~J. {Zahnle} \& D.~C. {Catling}(2017){Zahnle} \& {Catling}}]{Zahnle2017}
{Zahnle}, K.~J., \& {Catling}, D.~C. 2017, \bibinfo{title}{{The Cosmic Shoreline: The Evidence that Escape Determines which Planets Have Atmospheres, and what this May Mean for Proxima Centauri B},} \apj, 843, 122, \dodoi{10.3847/1538-4357/aa7846}

\bibitem[{M. {Zechmeister} {et~al.}(2018){Zechmeister}, {Reiners}, {Amado}, {Azzaro}, {Bauer}, {B{\'e}jar}, {Caballero}, {Guenther}, {Hagen}, {Jeffers}, {Kaminski}, {K{\"u}rster}, {Launhardt}, {Montes}, {Morales}, {Quirrenbach}, {Reffert}, {Ribas}, {Seifert}, {Tal-Or}, \& {Wolthoff}}]{Zechmeister2018}
{Zechmeister}, M., {Reiners}, A., {Amado}, P.~J., {et~al.} 2018, \bibinfo{title}{{Spectrum radial velocity analyser (SERVAL). High-precision radial velocities and two alternative spectral indicators},} \aap, 609, A12, \dodoi{10.1051/0004-6361/201731483}

\bibitem[{L. {Zeng} {et~al.}(2016){Zeng}, {Sasselov}, \& {Jacobsen}}]{Zeng2016}
{Zeng}, L., {Sasselov}, D.~D., \& {Jacobsen}, S.~B. 2016, \bibinfo{title}{{Mass-Radius Relation for Rocky Planets Based on PREM},} \apj, 819, 127, \dodoi{10.3847/0004-637X/819/2/127}

\bibitem[{L. {Zeng} {et~al.}(2019){Zeng}, {Jacobsen}, {Sasselov}, {Petaev}, {Vanderburg}, {Lopez-Morales}, {Perez-Mercader}, {Mattsson}, {Li}, {Heising}, {Bonomo}, {Damasso}, {Berger}, {Cao}, {Levi}, \& {Wordsworth}}]{Zeng2019}
{Zeng}, L., {Jacobsen}, S.~B., {Sasselov}, D.~D., {et~al.} 2019, \bibinfo{title}{{Growth model interpretation of planet size distribution},} PNAS, 116, 9723, \dodoi{10.1073/pnas.1812905116}

\bibitem[{S. {Zieba} {et~al.}(2023){Zieba}, {Kreidberg}, {Ducrot}, {Gillon}, {Morley}, {Schaefer}, {Tamburo}, {Koll}, {Lyu}, {Acu{\~n}a}, {Agol}, {Iyer}, {Hu}, {Lincowski}, {Meadows}, {Selsis}, {Bolmont}, {Mandell}, \& {Suissa}}]{Zieba2023}
{Zieba}, S., {Kreidberg}, L., {Ducrot}, E., {et~al.} 2023, \bibinfo{title}{{No thick carbon dioxide atmosphere on the rocky exoplanet TRAPPIST-1 c},} \nat, 620, 746, \dodoi{10.1038/s41586-023-06232-z}

\end{thebibliography}
\bibliographystyle{aasjournal}

\appendix
\setcounter{table}{0}
\renewcommand{\thetable}{A\arabic{table}}

\section{MAROON-X data}
\label{appendix:rvs}


\begin{table}[h]
\centering
\begin{tabular}{|c|c|}
\hline
Column Name & Description                                              \\ \hline
channel     & Channel of observation (``red'' or ``blue'')             \\
flag        & Flag if excluded from RV analysis ('0' if used in RV analysis, '1' otherwise) \\
bjd         & Date of observation (BJD)                                \\
rv          & Radial velocity (m/s)                                    \\
e\_rv         & Radial velocity error (m/s)                              \\
sn\_peak     & Peak SNR of spectrum                                     \\
exptime     & Exposure time (s)                                        \\
berv        & Barycentric radial velocity (m/s)                        \\
airmass     & Airmass of observation                                   \\
dLW         & Differential line width (1000 m$^2$/s$^2$)               \\
e\_dLW       & Differential line width error  (1000 m$^2$/s$^2$0        \\
crx         & Chromatic index (m/s/Np)                                 \\
e\_crx       & Chromatic index error (m/s/Np)                           \\
irt\_ind1    & Calcium infrared triplet (8500.4 \AA \,line) index       \\
irt\_ind1\_e  & Calcium infrared triplet (8500.4 \AA\, line) index error \\
irt\_ind2    & Calcium infrared triplet (8544.4 \AA\, line) index       \\
irt\_ind2\_e  & Calcium infrared triplet (8544.4 \AA\, line) index error \\
irt\_ind3    & Calcium infrared triplet (8664.5 \AA\, line) index       \\
irt\_ind3\_e  & Calcium infrared triplet (8664.5 \AA\, line) index error \\
halpha\_v    & H$\alpha$ index                                          \\
halpha\_e    & H$\alpha$ index Error                                    \\
nad1\_v      & Sodium doublet (5891.6 \AA\, line) index                 \\
nad1\_e      & Sodium doublet (5891.6 \AA\, line) index error           \\
nad2\_v      & Sodium doublet (5897.6 \AA\, line) index                 \\
nad2\_e      & Sodium doublet (5897.6 \AA\, line) index error          \\
\hline
\end{tabular}
\caption{A description of the information contained in our online dataset on GJ\,12\,b.}
\label{tab:RVs}
\end{table}

\end{document}